\documentclass[a4paper,11pt]{article} 

\pdfoutput=1
\usepackage{jheppub,AkashandJaydefs,relsize,caption,afterpage}

\usepackage{jheppub}

\makeatletter
\def\@fpheader{\relax}
\makeatother

\def\sec#1{\S\ref{#1}}

\definecolor{green}{rgb}{0.1,0.8,0.2}

\def\galsigma{\sigma}

\title{On the surface of superfluids}
 
\author[a]{Jay Armas,}
\author[b]{Jyotirmoy Bhattacharya,}
\author[b]{Akash Jain}
\author[c]{and Nilay Kundu}
\affiliation[a]{Physique Th\'{e}orique et Math\'{e}matique Universit\'{e} Libre de Bruxelles and
  International Solvay Institutes ULB-Campus Plaine CP231, B-1050 Brussels, Belgium}
\affiliation[b]{Centre for Particle Theory \& Department of Mathematical Sciences, Durham
  University, South Road, Durham DH1 3LE, United Kingdom}
\affiliation[c]{Center for Gravitational Physics, Yukawa Institute for Theoretical Physics (YITP),
Kyoto University, Kyoto 606-8502, Japan}

\emailAdd{jarmas@ulb.ac.be, jyotirmoy.bhattacharya@durham.ac.uk, akash.jain@durham.ac.uk, nilay.tifr@gmail.com}

\abstract{Developing on a recent work on localized bubbles of ordinary relativistic fluids, we study
  the comparatively richer leading order surface physics of relativistic superfluids, coupled to an
  arbitrary stationary background metric and gauge field in $3+1$ and $2+1$ dimensions.  The
  analysis is performed with the help of a Euclidean effective action in one lower dimension,
  written in terms of the superfluid Goldstone mode, the shape-field (characterizing the surface of
  the superfluid bubble) and the background fields.  We find new terms in the ideal order
  constitutive relations of the superfluid surface, in both the parity-even and parity-odd sectors,
  with the corresponding transport coefficients entirely fixed in terms of the first order bulk
  transport coefficients. Some bulk transport coefficients even enter and modify the surface
  thermodynamics.  In the process, we also evaluate the stationary first order parity-odd bulk
  currents in $2+1$ dimensions, which follows from four independent terms in the superfluid
  effective action in that sector.  In the second part of the paper, we extend our analysis to
  stationary surfaces in $3+1$ dimensional Galilean superfluids via the null reduction of null
  superfluids in $4+1$ dimensions. The ideal order constitutive relations in the Galilean case also
  exhibit some new terms similar to their relativistic counterparts.  Finally, in the relativistic
  context, we turn on slow but arbitrary time dependence and answer some of the key questions
  regarding the time-dependent dynamics of the shape-field using the second law of thermodynamics.
  A linearized fluctuation analysis in $2+1$ dimensions about a toy equilibrium configuration
  reveals some new surface modes, including parity-odd ones.  Our framework can be easily applied to
  model more general interfaces between distinct fluid-phases.}

\begin{document}

\begin{flushright}
\small{DCPT-16/53 \\ YITP-16-145}
\end{flushright}

\maketitle 
%

\section{Introduction and summary}\label{sec:intro}

Matter in the universe exists in diverse forms and very often its collective behaviour is so complex
that its detailed microscopic description becomes intractable. Fortunately, in many situations of
interest, the low-energy collective behaviour can be captured by an effective theory with a few
degrees of freedom. A prominent example of such a finite temperature effective theory is
hydrodynamics, where the description is provided in terms of a few fluid variables in the
long-wavelength approximation.  In this effective description, the relevant microscopic information
is conveniently packaged into the parameters of the theory, referred to as the transport
coefficients. 

The universal nature of this description has lead to its applications in a diverse range of physical
situations, ranging from neutron-stars, quark-gluon plasma to numerous condensed matter
systems. Hence, this subject has had a long history and has been extremely well studied in the past.
However, quite recently there has been a renewed interest in this area, particularly following the
realization that there are some important lacunae in the structural aspects of the fluid equations
that have been considered so far.  It was understood that new transport coefficients must be
incorporated in the effective theory in order to adequately describe certain physical
situations\footnote{For instance, in situations where a symmetry of the underlying theory suffers
  from an anomaly. This was first realized in the study of the map between long-wavelength
  fluctuation of black holes and relativistic hydrodynamics \cite{Erdmenger:2008rm,
    Banerjee:2008th}. Subsequently, the existence of such new coefficients has been inferred
  employing pure hydrodynamic techniques \cite{Son:2009tf}.}. In fact, one of the most interesting
aspects of some of these newly discovered coefficients is their parity-odd nature -- a possibility
that has been largely ignored in the rich and classic literature on the subject.

In regimes where the hydrodynamic approximation is applicable, it is often observed that the same
underlying microscopic theory can exist in distinct macroscopic phases. In situations where two such
phases coexist, they are separated by a dynamical interface (or surface).  If we wish to provide an
effective description for such scenarios, then the hydrodynamic description must be appropriately
generalized in order to include the effects specific to such surfaces. Our main goal in this paper
is to explore new surface properties, especially in the context of superfluids, focusing on the
parity-odd effects.

For the case of ordinary relativistic space-filling fluids, the degrees of freedom include the fluid
velocity $u^\mu$, temperature $T$ and chemical potential(s) $\mu$ corresponding to any global
symmetries that the fluid may enjoy. In this paper, we will assume this global symmetry to be a
$\rmU(1)$ symmetry. The equation of motion for these fluid fields are simply the conservation of the
energy-momentum tensor and charge current, which in turn are expressed in terms of the fluid
variables subjected to constitutive relations. The structure of the constitutive relations is
determined based on symmetry principles and is severely constrained by the second law of
thermodynamics \cite{Landau:1987gn}.

In the case of superfluids, the $\rmU(1)$ symmetry is spontaneously broken and the phase of the
order parameter $\phi$ serves as a massless Goldstone boson, which must be included in the
low-energy effective description in addition to the ordinary fluid fields. In order to preserve
gauge invariance, $\phi$ enters into the constitutive relations only through its gauge-covariant
derivative, referred to as the superfluid velocity $\xi_\mu$ (see \cite{Landau:1987gn} for more
details on the basics of superfluid dynamics). In the case of space-filling superfluids, the most
general constitutive relations consistent with the second law of thermodynamics up to first order in
the derivative expansion have been worked out more recently in \cite{Bhattacharya:2011tra}.

If we wish to provide a unified description of two (super)fluid phases separated by a dynamical
surface, we need to include a new field $f$ in the hydrodynamic description, which keeps track of
the shape of the surface. The surface is considered to be located at $f=0$. This shape-field $f$ is
quite analogous to the Goldstone boson $\phi$ in the case of superfluids. In fact, $f$ may be
considered to be the Goldstone boson corresponding to the spontaneous breaking of translational
invariance in the direction normal to the fluid surface\footnote{See \cite{Bhattacharya:2009gm,
    Armas:2015ssd} for a relevant recent discussion in the stationary case and \cite{Kovtun:2016lfw}
  for an application of similar ideas to the study of polarization effects on surface currents in
  the context of magnetohydrodynamics.}. The guiding symmetry principle for incorporating this
shape-field into the constitutive relations is the reparametrization invariance, i.e. the fluid
must be invariant under arbitrary redefinitions of $f$ as long as its zeroes are unchanged.  This
essentially implies that the dependence of the fluid currents on $f$ happens primarily\footnote{As
  we shall explain in more detail below, another way in which $f$ may enter the constitutive
  relations is via the distribution function $\theta(f)$ and its reparametrization invariant
  derivatives.}  through $n_\mu$, the normal vector to the surface, and its derivatives.

Now, for a superfluid bubble placed inside an ordinary fluid, there is a rich interplay between the
Goldstone boson $\phi$ and the shape-field $f$ on the surface of the superfluid bubble. In this
paper, we study these surface effects and work out the ideal order surface currents for a
superfluid.

This paper is organized as follows: in the remaining of this section we will give a detailed summary
of the main points and techniques used in this paper. In \S \ref{sec:3plus1} we discuss stationary
superfluid bubbles suspended in ordinary fluids in $3+1$ dimensions, and extend it to $2+1$
dimensions in \S \ref{sec:2plus1} (see the summary in \S\ref{ssec:staintro}). Then in \S
\ref{sec:nulsf}, we discuss stationary Galilean superfluid bubbles using the technique of null
superfluids \cite{Banerjee:2016qxf}, and use it to understand about the non-relativistic limit of
surface phenomenon in superfluids (see the summary in \S\ref{ssec:intronull}). Later in
\S\ref{sec:surdyn}, we turn on slow but arbitrary time dependence and study time-dependent dynamics
of the shape-field $f$ using the second law of thermodynamics as well as linearized fluctuations
about an equilibrium configuration. We finish with some discussion in \S \ref{sec:disco}. The paper
has three appendices. In appendix \ref{app:2+1thermo} we discuss surface thermodynamics for $2+1$
dimensional superfluid bubbles. Then in appendix \ref{app:YLeomf}, we give a generic derivation of
the Young-Laplace equation for stationary superfluid bubbles, that determines the shape of the
surface. Finally, in appendix \ref{app:formula} we collect some useful formulae and notations.

\subsection{Stationary superfluid bubbles}\label{ssec:staintro}

To begin with, following \cite{Armas:2015ssd}, we shall mainly focus on stationary relativistic
superfluid bubbles, which will enable us to employ the partition function techniques discussed in
\cite{Banerjee:2012iz, Bhattacharyya:2012xi, Jensen:2012jh}.
Our main objective here is to write down a Euclidean
effective action for the Goldstone boson $\phi$ and the shape-field $f$ in one lower dimension, from
which the surface currents can be easily read off using a variational principle. One of our primary
focuses in this analysis will be the parity violating terms. Therefore, we will separately discuss
the cases of $3+1$ and $2+1$ dimensions\footnote{Note that there is a subtlety in the discussion of
  finite temperature superfluidity in $2+1$ dimensions.  At finite temperature, the low-energy
  physics is blind to the time-like direction and therefore the dynamics is effectively two
  dimensional. In our context, this is clearly reflected by the fact that in $2+1$ dimensions we
  write down a two dimensional Euclidean action for the massless Goldstone boson. This brings us
  within the purview of the Mermin-Wagner theorem implying that superfluidity in these
  dimensions may be destroyed by strong quantum fluctuations. However, this conclusion is rendered
  invalid in the large-N limit. In fact, much of our discussions here might be relevant for $3+1$
  dimensional hairy black holes in AdS, via the AdS/CFT correspondence \cite{Sonner:2010yx}.  Also,
  our discussion in $2+1$ dimensions may be relevant for other microscopic mechanisms for $2+1$
  dimensional superfluidity like the BKT transition.  It would be definitely interesting to make
  this connection more precise.}, which have significantly different parity-odd structures.

We will consider stationary bubbles of a superfluid in the most general background spacetime metric
and gauge field which admits a time-like Killing vector $\dow_t$
\begin{align}\label{relmet}
  \df s^2 &= \cG_{\mu\nu} \df x^\mu \df x^\nu = - e^{2 \sigma\left(\vec{x}\right)} \left(\df t +
            a_i\left(\vec{x}\right) \df x^i \right)^2 +
            g_{ij}\left(\vec{x}\right) \df x^i \df x^j~, \nn\\
  \mathcal A &= \mathcal A_0 \left(\vec{x}\right) \df t + \mathcal A_{i} \left(\vec{x}\right) \df
               x^i~. 
\end{align}
Here, the $i$-index runs over the spatial coordinates. We will denote the covariant derivative
associated with $\cG_{\mu\nu}$ by $\nabla_\mu$, while the one associated with $g_{ij}$ by
$\Df_i$. For later use, we also define respective surface derivatives by
$\tilde\nabla_\mu(\cdots)=1/\sqrt{\N^\nu f \N_\nu f}~\nabla_\mu(\sqrt{\N^\nu f \N_\nu f}\cdots)$ and
the correspondent one associated to $g_{ij}$ by
$\tilde \Df_i(\cdots)=1/\sqrt{\Df^j f \Df_j f}~\Df_i(\sqrt{\Df^j f \Df_j f}\cdots)$.  
Now, since we wish to provide a finite temperature partition function\footnote{Here by partition
  function we refer to (the exponential of) the Euclidean effective action in the presence of
  arbitrary background sources.}  description of our system, we will Wick-rotate to Euclidean time
and compactify this direction, with an inverse radius $T_0$.  Thus, the set of all background data
comprises of (see \cite{Banerjee:2012iz, Armas:2015ssd} for more details)
\begin{equation}\label{bakdata}
\{ \sigma\left(\vec{x}\right),a_i\left(\vec{x}\right),g_{ij}\left(\vec{x}\right), \mathcal A_0 \left(\vec{x}\right), 
 \mathcal A_{i} \left(\vec{x}\right), T_0\}~.
\end{equation}
Apart from $T_0$, there is another length scale in the problem corresponding to the chemical
potential $\mu_0$. However, it is always possible to absorb this into the time component of the
arbitrary gauge field $\mathcal A_0$. Therefore, we will not make $\mu_0$ explicit in our
discussions.

In addition to the background data \eqref{bakdata}, there are two fields which must be included in
the partition function if we wish to describe superfluid bubbles. One of them is the phase of the
scalar operator responsible for the spontaneous breaking of the $\rmU(1)$ symmetry, which we denote
by $\phi\left(\vec{x}\right)$ (see \cite{Bhattacharyya:2012xi} for more details).  The other is the
shape-field $f \left(\vec{x}\right)$, where $f\left(\vec{x}\right) = 0$ denotes the location of the
interface between the superfluid and the ordinary charged fluid (see \cite{Armas:2015ssd} for more
details). In the superfluid description, the first derivative of the Goldstone boson
$\phi\left(\vec{x}\right)$ is treated as a quantity which is zeroth order in derivatives, and is
referred to as the superfluid velocity $\xi_\mu = -\partial_{\mu} \phi + \mathcal A_{\mu}$ \footnote{
  Here we follow the conventions of \cite{Bhattacharyya:2012xi}.  See also
  \cite{Bhattacharya:2011tra, Bhattacharya:2011eea} for an out of equilibrium discussion of
  relativistic superfluids.}.  In the reduced language, that is, using the KK decomposition \eqref{relmet}, 
  since $\phi$ is time independent, we have that $\xi_{\mu} = \{\xi_0 = \cA_0, \xi_i = -\partial_i \phi + \mathcal A_i \}$.

As has been explained in detail in \cite{Banerjee:2012iz, Armas:2015ssd}, the partition function
function must be constructed in terms of quantities that are invariant under spatial
diffeomorphisms, Kaluza-Klein (KK) gauge transformations (redefinitions of time,
$t \ra t + \vq^t(\vec x)$) and $\rmU(1)$ gauge transformations. Therefore, following
\cite{Banerjee:2012iz}, we first define a KK invariant gauge field
\begin{equation}
  A = A_0 \df t + A_i \df x^i, \quad\text{where}, \quad
  A_0 \equiv \mathcal A_0, \quad\text{and},\quad
  A_i \equiv \mathcal A_i - \mathcal A_0 a_i~. 
\end{equation}
In the context of superfluids, it is convenient to redefine the spatial components of the superfluid
velocity so that they are invariant under both the U(1) and KK gauge transformations (the time
component is automatically invariant) \cite{Bhattacharyya:2012xi}
\begin{equation}
 \zeta_i \equiv \xi_i - A_0 a_i = -\partial_i \phi + A_i~.  
\end{equation}

The dependence of the partition function on the shape-field $f(\vec{x})$ follows exactly the same
form as described in \cite{Armas:2015ssd}. This dependence is primarily constrained by the
reparametrization invariance of the surface $f \rightarrow g(f)$ with $g(0)=0$. The elementary
reparametrization invariant building block made out of $f$ is the normal vector to the surface
$n_\mu$, which for stationary configurations takes the form
\begin{equation}\label{normalvec}
  n_{\mu} = -\frac{\partial_\mu f}{\sqrt{\N_{\nu} f \N^{\nu} f}} = \{0, n_i \}, \quad\text{where},\quad
  n_i = -\frac{\partial_i f}{\sqrt{\Df_{j} f \Df^{j} f}} ~.
\end{equation}
We would again like to emphasize the remarkable similarity in the way the fields $\phi$ and $f$
enter the partition function. In fact following the analogy, we will consider the normal vector
$n_\mu$, just like the superfluid velocity $\xi_\mu$, as a zero derivative order quantity.

Now, we wish to describe a stationary bubble of a superfluid inside an ordinary charged fluid.  The
entire set of data which constitutes the building blocks of the partition function for ordinary
charged fluids away from the interface are
\begin{equation}
 \tilde {\mathbb B} =  \{ \sigma\left(\vec{x}\right),a_i\left(\vec{x}\right),g_{ij}\left(\vec{x}\right),  A_0 \left(\vec{x}\right), 
  A_{i} \left(\vec{x}\right), T_0\}~~.
\end{equation}
On the superfluid side, away from the interface, there is an additional ingredient
\begin{equation}
   \mathbb B =  \tilde{\mathbb B} \cup \{ \zeta_i\left(\vec{x}\right) \}~~.
\end{equation}
On the surface, this set must also include the normal vector to the interface
\begin{equation}\label{surfieldset}
  \mathbb B_{(s)} =  \mathbb B \cup \{ n_i(\vec x) \}~~.
\end{equation}
The structure of the Euclidean effective action
for a bubble of a superfluid inside a charged fluid will take
the most general form
\begin{equation}\label{fulpartintoform}
  W = \int \{\df \vec x\} \sqrt{g}\ \frac{\E{\sigma}}{T_0} \bigg( \theta(f) ~\mathcal
  S_{(b)} \left( \mathbb B, \partial \mathbb B,\dots \right) + \tilde \delta(f) ~\mathcal S_{(s)}
  \left( \mathbb B_{(s)}, \dots\right) + \theta(-f) ~\mathcal S_{(e)} \left( \tilde{\mathbb
      B}, \partial \tilde{\mathbb B}, \dots \right) \bigg) ~~,
\end{equation}
where $\mathcal S_{(b)}$ and $\mathcal S_{(e)}$ are the partition functions of space-filling
superfluids and ordinary charged fluids respectively, while $ \mathcal S_{(s)}$ is the partition
function of the interface. $\q(f)$ is a distribution which captures the thickness of the wall. For
an infinitely thin wall, $\q(f)$ can be taken to be the Heaviside theta function. Furthermore,
$\tilde\d(f) = - n^\mu \dow_\mu \q(f) = \sqrt{\N^\mu f \N_\mu f} \d(f)$ (with
$\d(f) = \df\q(f)/\df f$) is the reparametrization invariant derivative of $\q(f)$. Here,
$\mathcal S_{(b)}, \mathcal S_{(e)}, \mathcal S_{(s)}$ are expanded in a derivative expansion as in
ordinary fluid dynamics. In addition, one must also consider terms containing reparametrization
invariant derivatives of $\tilde \delta(f)$ (i.e. terms with two or higher derivatives of
$\theta(f)$). In this way, there are, in fact, two dimensionless small parameters in the effective
theories studied in this paper. One is the usual fluid expansion parameter $\omega/T \ll 1$
($\omega$ being the typical frequency of fluctuations), which allows us to make the usual derivative
expansion in fluid dynamics. The other small parameter is $\tau T\ll 1$ ($\tau$ being the length scale
associated with the thickness of the surface (see \cite{Armas:2015ssd} for more details)). The
derivatives of $\theta(f)$ keep track of this second parameter. Thus, \eqref{fulpartintoform} should
be thought of as a double expansion in both these parameters.

The energy-momentum tensor and charge current that follows from the partition function
\eqref{persuppart}, have the structural form
\begin{equation}\label{curststr}
 \begin{split}
  T^{\mu \nu} &= \theta(f) ~T^{\mu \nu}_{(b)} + \tilde \delta(f) ~T_{(s)}^{\mu \nu} + \theta(-f) ~T^{\mu \nu}_{(e)} + \dots~~,\\
  J^{\mu} &= \theta(f) ~J^{\mu}_{(b)} + \tilde \delta(f) ~J_{(s)}^{\mu} + \theta(-f) ~J^{\mu}_{(e)} + \dots~~, \\
 \end{split}
\end{equation}
where the ellipsis denotes terms with higher derivatives of $\q(f)$.  We will refer to
$\mathcal S_{(b)}$ as the bulk of the superfluid bubble, $\mathcal S_{(e)}$ as the exterior and
$\mathcal S_{(s)}$ as the surface. Correspondingly, $T^{\mu \nu}_{(b)}$, $J^\mu_{(b)}$ are bulk
superfluid currents, $T^{\mu \nu}_{(e)}$, $J^\mu_{(e)}$ are exterior fluid currents, and
$T^{\mu \nu}_{(s)}$, $J^\mu_{(s)}$ are surface currents. The former two have been well explored in
the literature (see e.g. \cite{Banerjee:2012iz,Bhattacharyya:2012xi}), so our main focus here will
be on the surface currents, and how the bulk/exterior of the bubble affects the surface.

In this paper, we will obtain the surface currents in \eqref{curststr} in a special hydrodynamic
frame\footnote{See \cite{Armas:2015ssd} for a detailed and complete description of issues on the
  choice of frames.}, which is the frame that follows directly from equilibrium partition
functions. In this frame, defining $\tilde K=\partial_t$ as the time-like Killing vector field of
the background, the usual ordinary fluid variables are given by
\begin{equation}\label{equisolpfframe}
  T = e^{-\sigma} T_0~~,\qquad
  u^{\mu}= \E{-\sigma} \tilde K^{\mu}~~,\qquad
  \mu = e^{-\sigma} A_0 ~~,
\end{equation}
to all orders in the derivative expansion. We will refer to this frame as the partition function
frame. Furthermore, the surface equations of motion that follow from the conservation of the currents
\eqref{curststr} can be thought of as equations which constrain the boundary conditions that should
be imposed when solving the bulk equations.  If we work in a regime where $\tilde \delta'(f)$ terms
can be neglected, then the exercise of finding new configurations reduces to a boundary value
problem from the bulk point of view. This problem should be solved  with
the boundary conditions themselves being determined by the surface conservation
equations.\footnote{We would like to emphasize that using solutions of the surface equations as
  boundary conditions for the bulk equations is clearly consistent at least in equilibrium, where
  there exists a continuous solution \eqref{equisolpfframe} of the fluid variables for the combined
  set of equations, following from the conservation of currents in 
  \eqref{curststr} (see \cite{Armas:2015ssd} for more detailed discussion of this issue). In \S
  \ref{sec:surdyn}, we also show that this method may be applied in time-dependent situations as well at leading order in derivatives.}

Before summarizing our results regarding the detailed  structure of the partition function, we would
like to justify the distribution function  $\theta(f)$ appearing in \eqref{fulpartintoform} in terms
of  the  Landau-Ginzburg  paradigm.  Here,  we  are  describing  an  interface between  two  phases,
distinguished by  the status of a  $\rmU(1)$ symmetry, which  is spontaneously broken in  one phase,
while is intact  in the other. The two  phases, therefore, are distinguished by  an order parameter,
with the help of which it is possible to write down a Landau-Ginzburg theory
\begin{equation}\label{LGsimp}
  \mathcal A_{LG} = \int |\Df_{\mu}\Phi|^2 + a |\Phi|^2 + b |\Phi|^4~~, 
\end{equation}
where $\Phi = \psi e^{i \phi}$ is a complex scalar field. Here $\psi$ is the order parameter, which
is 1 in the superfluid phase and is 0 outside, and smoothly interpolates between 1 and 0 on the
interface separating the two phases. The hydrodynamic degrees of freedom can be seen as small
fluctuations about the profile of the condensate.  In such situations, the profile of $\psi$ itself
provides us with a smooth distribution function $\theta(f)$ required in the partition function
\eqref{fulpartintoform}.  The terms that are proportional to the derivatives of $\psi$ are localized
on the interface and contribute to $ \mathcal S_{(s)}$ in \eqref{fulpartintoform}.  Now, as we have
discussed before, the derivative of the phase $\phi$, referred to as the superfluid velocity, enters
the superfluid dynamics. 
We would like to point out that $\psi$ starts decreasing from 1, as we approach the interface from the superfluid 
side, and goes to zero with the onset of the ordinary charged fluid. 
This implies that it is possible to have a non-trivial profile of $\phi$ at the interface.  This prompts us to
include a dependence of the interface partition function $\mathcal S_{(s)}$ on the superfluid
velocity. In this context, it is worthwhile pointing out that expanding around the background
interpolating profile of $\psi$, and keeping terms up to the quadratic order in $\phi$, we see that
$\mathcal S_{(s)}$ can depend on the magnitude of the superfluid velocity, as well as on its
component along the direction normal to the surface, both being Lorentz scalars from the interface
point of view. Following the analogy with ordinary fluids\footnote{For ordinary fluids, the normal
  component of the fluid velocity at the surface vanishes in equilibrium,
  i.e. $u^{\mu}n_\mu|_{f=0}=0$.  See \cite{Armas:2015ssd} and \S \ref{sec:surdyn} for further
  details.}, it is tempting to anticipate that the component of the superfluid velocity normal to
the surface should vanish in the stationary case. However, we were unable to obtain any rigorous
justification why this should be the case, and hence we will perform all our analyses keeping this
component non-zero and arbitrary. In fact, an entropy current analysis at leading order, performed
in \S \ref{sssec:jent3p10ord} for situations away from equilibrium, also allows for a non-zero
component of the superfluid velocity normal to the surface.

It may also be noted that, while describing superfluids where there is a normal fluid component, the
usual fluid fields ($u^\mu$, $T$ and $\mu$) are also present along with $\Phi$ in the
Landau-Ginzburg setting.  All these fields, including $\Phi$, may be composite or effective fields
constituted out of the more fundamental degrees of freedom. In such situations, we may consider
interaction terms between $\Phi$ and other fluid variables in the effective action
\eqref{LGsimp}. In 2+1 dimensions, in particular, it is possible to write down such an interesting
parity-odd interaction term of the form
\begin{equation}\label{LGsimppo}
  \mathcal A_{LG} = \int \dots +  i \epsilon^{\mu \nu \rho} u_{\mu} (\Df_{\nu} \Phi)  (\Df_{\rho} \Phi)^{\star}~~. 
\end{equation}
Again, considering fluctuations about the background interpolating $\psi$, it is evident that
\eqref{LGsimppo} generates a term of the form $\epsilon^{\mu \nu \rho} u_{\mu} n_\nu \xi_{\rho}$
localized on the interface. In a time-independent context, in the reduced language, this would imply
that in general the surface partition function $\mathcal S_{(s)}$ can depend on
$\bar \lambda = \epsilon^{ij} n_i \zeta_j$. As we will see later, this fact has an important and
non-trivial consequence on the surface thermodynamics of $2+1$ dimensional superfluid bubbles.

The construction of $S_{(b)}$, up to first order in derivative expansion in $3+1$ dimensions was
presented in \cite{Bhattacharyya:2012xi} and is given by\footnote{Our convention for the spatial
  Levi-Civita tensor in $2+1$ dimensions is $\epsilon^{12} = 1/\sqrt{g}$ while in $3+1$ dimensions
  is $\epsilon^{123} = 1/\sqrt{g}$. Thus, we have the following reductions on the time circle
\begin{equation}
  u_\mu \epsilon^{\mu \nu \sigma}  \rightarrow - \epsilon^{ij} ~, 
  \qquad u_\mu \epsilon^{\nu \sigma \rho \mu}  \rightarrow \epsilon^{ijk} ~,
\end{equation}
in $2+1$ and $3+1$ dimensions respectively.}
\begin{equation}\label{partfn1stord3+1intro}
 \begin{split}
  W&=  W_{\text{even}} + W_{\text{odd}}, ~~\text{where}\\
  W_{\text{even}} &= \int \df^3x \sqrt{g} \ \frac{\E{\sigma}}{T_0} ~\theta(f) \left( P_{(b)}
    - \E{-\sigma} \alpha_1~ \z^i \partial_i \sigma
    + \frac{\alpha_2}{T_o}~ \z^i \partial_i A_0
    - \alpha_3~ \Df_i \left( \E{\sigma} \mathcal F \z^i \right) \right)~~, \\
  W_{\text{odd}} &= \int \df^3x \sqrt{g} \ \frac{\E{\sigma}}{T_0} ~ \theta(f) \left( T_0 \alpha_4~ \epsilon^{ijk} \zeta_i \partial_{j} a_k 
  + \alpha_5~ \epsilon^{ijk} \zeta_i \partial_{j} A_k \right)~~.
 \end{split}
\end{equation}
Note that, as demonstrated in \cite{Bhattacharyya:2012xi}, the term proportional to $\alpha_3$ is
the leading order equation of motion of $\phi$ in the bulk and its effects can be trivially removed
by a shift of $\phi$.\footnote{In the presence of surfaces, the effect of such $\phi$ shifts at the
  surface can be absorbed by a redefinition of the surface partition function.} Therefore, for the
sake of simplicity we set $\alpha_3$ to zero in our analysis.

In this paper, we also construct $S_{(b)}$ in $2+1$ dimensions up to first order in derivatives in
\S \ref{sec:2plus1}. The parity-even sector is identical to that of \eqref{partfn1stord3+1intro},
while the parity-odd sector is richer than its $3+1$ dimensional counterpart 
\footnote{Note that these terms are also the parity-odd first order corrections on the surface of the 
$3+1$ dimensional superfluid bubble. }
\begin{equation}\label{po1stord2p1supintro}
  W_{\text{odd}} = \int \df^2x \sqrt{g} \ \frac{\E{\sigma}}{T_0} \theta(f) \left( m_{\omega} \epsilon^{ij} \partial_i a_j 
    + m_{B} \epsilon^{ij} \partial_i A_j + \beta_1 \epsilon^{ij} \zeta_i \partial_j \sigma 
    + \beta_2 \epsilon^{ij} \zeta_i \partial_j A_0 \right) ~.
\end{equation}
The bulk currents that follow from \eqref{po1stord2p1supintro} have not yet been analyzed in the
literature, to the best of our knowledge. We perform this exercise in \S \ref{ssec:perfsuper2p1}. We
find that there is a total of 35 relations among transport coefficients that are determined in terms
of the four coefficients in \eqref{po1stord2p1supintro} (in addition to the parity-even terms).

Since we are only considering terms up to the first order in derivatives on both the sides far away
from the surface, it suffices to only consider a zeroth order term at the surface for
$\mathcal S_{(s)}$.  This is the surface tension term which was considered in
\cite{Armas:2015ssd}. Since we will be dealing with superfluids on one side of the interface, the
surface tension can now also depend on the superfluid velocity.

In $3+1$ dimensions, we work out the ideal order surface currents in \eqref{persupcovcurs_der} by
varying the partition function, with associated surface thermodynamics given in
\eqref{pesurthermo3p1}. Later in \S \ref{sec:2plus1}, we work out the analogous surface currents in
$2+1$ dimensions in \eqref{persupcovcurs2p1aa}, with respective thermodynamics given in
\eqref{eq:thermo2+1rel}, which also includes parity-odd effects. One of the most interesting
features of our equilibrium analysis is the fact that the equation of motion for the shape-field $f$
(the Young-Laplace equation) is identical to the normal component of the energy-momentum
conservation equation at the surface.  We rigorously argue in Appendix \ref{app:YLeomf} that this
must continue to hold at all orders in the derivative expansion.

\subsection{Non-relativistic stationary superfluid bubbles}\label{ssec:intronull}
%
In order to obtain an understanding of the non-relativistic limits of superfluid surface currents,
in \S\ref{sec:nulsf} we study Galilean\footnote{There is a subtle difference between Galilean and
  non-relativistic systems. As we will explain in more detail below, in the context of fluid
  dynamics, non-relativistic fluids are only a special class of the Galilean ones. Moreover, there
  can be other non-relativistic systems, such as Lifshitz systems with a dynamical exponent
  $z\neq 1,2$, which are not Galilean.} superfluids in $3+1$ dimensions. For this analysis, we use
the technique of null (super)fluids developed in
\cite{Banerjee:2015hra,Banerjee:2015uta,Banerjee:2016qxf}, where it was realized that transport
properties of a Galilean (super)fluid are in one-to-one correspondence with that of a relativistic
system - null (super)fluid in one higher dimension. Here, the basic idea is that in order to obtain
the most generic Galilean (super)fluid currents in $3+1$ dimensions, we can start with a null
(super)fluid on a null background\footnote{A null background is one which admits a null Killing
  vector $V^\sM$ such that a component of the gauge field is fixed as $V^\sM \cA_\sM = -1$. A null
  fluid is a fluid which couples to such a null background and the respective fluid velocity $u^\sM$
  is null instead of time-like. It is normalized so that $u^\sM V_\sM = -1$.} in $4+1$ dimensions,
and then perform a null reduction on it \cite{Duval:1984cj,Julia:1994bs} (also see
\cite{Rangamani:2008gi,Hassaine:1999hn} for some earlier application of null reductions in the
context of fluid dynamics).  The null reduction reduces the underlying Poincar\'{e} symmetry algebra
of a null (super)fluid to the Bargmann symmetry algebra (Galilean algebra with a central extension
with the mass operator) of a Galilean (super)fluid. Though we find the null-reduction prescription
more useful for our purposes, it is worth mentioning that these Galilean results can also be
obtained directly in a $3+1$ dimensional Newton-Cartan setting following
\cite{Jensen:2014aia,Jensen:2014ama} (see also \cite{Geracie:2015xfa}).

The equilibrium currents of a null (super)fluid can be obtained from a partition function written in
terms of the background fields, Goldstone boson and the shape-field, in very much the same way as
for the relativistic fluids discussed in \S\ref{ssec:staintro}.  There is however, one crucial new
ingredient for null backgrounds: in addition to the time-like Killing vector $\tilde K$ as in
\eqref{relmet}, null backgrounds also have a null Killing vector $V$. Choosing a set of coordinates
$\{x^\sM\} = \{x^-,t,x^i\}$ such that $\tilde K = \partial_t$ and $V=\partial_{-}$, the most general metric
and gauge field configurations respecting both Killing vectors are given as
\begin{align}\label{nulmetnA}
 \df s^2 &= \mathcal G_{\sM\sN} \df x^\sM \df x^\sM = - 2 e^{\galsigma(\vec x)} \left(\df t+a_i(\vec x)
        \df x^i \right)\left( \df x^- - \mathcal B_0(\vec x) \df t - \mathcal B_i(\vec x) \df
        x^i\right) + g_{ij}(\vec x) \df x^i \df x^j, \nn\\
 \mathcal A &= -\df x^- + \mathcal A_0(\vec x) \df t + \mathcal A_i(\vec x) \df x^i,
\end{align}
where all the introduced quantities are independent of the $x^-$ and $t$ coordinates.  In
torsionless Galilean/null spacetimes, in equilibrium, we must also have that
$\dow_i \galsigma= \dow_{[i} a_{j]} = 0$.  However, while writing an equilibrium partition function,
we will not require our background to be torsionless and will only impose it at the end of the
computation (see \cite{Banerjee:2015hra} for details).  As in the relativistic case, we would like
to construct the partition function in terms of all the background data which are manifestly
invariant under diffeomorphisms on the null background and gauge transformations.  In order to do
so, we need to consider the following invariant combinations (we refer the reader to
\cite{Banerjee:2016qxf} for more details regarding the transformation properties)
\begin{equation}
  B_i = \mathcal B_i - a_i \mathcal B_0, \qquad
  B_0 = \mathcal B_0, \qquad
  A_i = \mathcal A_i - a_i \mathcal A_0 - B_i, \qquad
  A_0 = \mathcal A_0.
\end{equation}
Since we are interested in superfluids, we also have the Goldstone boson $\phi$, and as in
\S\ref{ssec:staintro} its only gauge invariant combination is $\zeta_i = -\partial_i \phi + A_i$.
The full superfluid velocity thus takes the form
$\xi_{\sM} = \dow_\sM \f + \cA_\sM = \{-1,A_0, \zeta_i + a_i A_0 + B_i \}$.

Compared to \cite{Banerjee:2016qxf}, the additional ingredient in our discussion is the shape-field
$f$, since eventually we are interested in the non-relativistic limit of the superfluid surface. The
surface of the null superfluid needs to respect both the Killing vectors $V$ and $\tilde K$,
rendering it independent of $x^-$ and $t$ coordinates.  Again, since $f$ can only appear in the
partition function in a reparametrization invariant fashion, the primary dependence on $f$ comes through
the normal vector $n_\sM = \{0,0,n_i\}$ with $n_i$ being again given by \eqref{normalvec}.

The background data invariant under all the required symmetries, in terms of which the partition
function for bubbles of a null superfluid should be constructed, is given by ($1/T_0$ is the radius
of the Euclidean time circle)
\begin{equation}\label{nullpfdata}
  \mathbb B
  =  \{ \galsigma(\vec{x}),a_i(\vec{x}),g_{ij}(\vec{x}), B_0 (\vec{x}), 
  B_{i} (\vec{x}), A_0 (\vec{x}), A_{i} (\vec{x}),\zeta_i(\vec{x}), T_0\}, \quad
  {\mathbb B}_{(s)} = \mathbb B \cup \{ n_i(\vec x) \}.
\end{equation}
Note that in this case the background data is clearly larger compared to the relativistic case,
leading to more terms in the partition function at any given derivative order. This in turn implies
that the Galilean fluid obtained after null reduction will in general have more transport
coefficients than its relativistic counterpart. This is to be expected for a non-relativistic fluid
as well, e.g. in the non-relativistic limit the energy of a relativistic fluid splits into a rest
mass density part and the residual internal energy, hence increasing the count. Though this counting
accounts for the extra coefficients at ideal order, there is no reason to believe that at higher
orders as well such splitting will account for all the extra transport coefficients of a Galilean
fluid\footnote{Furthermore, transport coefficients of a Galilean fluid have dependence on an extra
  scalar as opposed to a relativistic fluid, namely, the mass chemical potential. However, for any
  non-relativistic fluid obtained as a limit of a relativistic fluid, this dependence must be
  trivial.}. Therefore, the most generic non-relativistic fluid is, at best, a subset of the
Galilean fluid discussed in this paper, exploration of which we leave for future work.

Finally, the equilibrium partition function for a $4+1$ dimensional null superfluid bubble immersed
in an ordinary fluid, up to first derivative order in the bulk and ideal order on the surface, can
be written as
\begin{multline}\label{galpf}
  W = \int \df^3 x\sqrt{g} \ \q(f) \frac{\E{\galsigma}}{T_0}  \bigg(
        P_{(b)}
  - f_1 \z^i\dow_i \galsigma
  + \E{-\galsigma} f_2 \z^i\dow_i A_0
  + \E{-\galsigma} f_3 \z^i\dow_i B_0 \\
  + T_0\E{-\galsigma} (g_1 + g_2) \e^{ijk} \z_i \dow_j B_k
  + T_0\E{-\galsigma} g_2 \e^{ijk} \z_i \dow_j A_k
  + T_0(g_1 \E{-\galsigma} B_0 + g_2 \E{-\galsigma} A_0 - g_3 ) \e^{ijk} \z_i \dow_j a_k \bigg) \\
  + \int \df^3 x \sqrt{g} \ \tilde\d(f) \frac{\E{\galsigma}}{T_0} \cC
  + \int \df^3 x \sqrt{g} \ \q(-f) \frac{\E{\galsigma}}{T_0} P_{(e)}.
\end{multline}
Note that there are no possible first order terms that we can write on the ordinary fluid side outside
the bubble. All the transport coefficients are functions of the zeroth order background scalar data,
while those on the surface have an additional dependence on $n_i \zeta^i$ as in the relativistic
case.  Note that in writing \eqref{galpf}, we have ignored a total derivative term in the bulk,
which can be absorbed in the surface term and, similarly to the relativistic case, we have not
considered a bulk term proportional to the zeroth order $\phi$ equation of motion.

Using the partition function \eqref{galpf} and the variational formulae \eqref{null_variational}, we
can work out the currents for a $4+1$ dimensional null superfluid bubble, which we report in
\eqref{nullconrel}. Given this, it is straightforward to exploit the null isometry to perform a null
reduction and get the surface currents for a Galilean superfluid \eqref{galleadcurs}. Even in this
case, we find that the ideal order surface currents receive contributions from the bulk transport
coefficients leading to different thermodynamics compared to the bulk.

\subsection{Time dependent fluctuations of the surface}\label{ssec:introtime}

Having understood the nature of the surface currents in equilibrium, we proceeded and introduced a
slow but arbitrary time dependence.  Away from equilibrium, there is no variational principle that
can help us in deducing the structure of surface currents\footnote{Given some of the latest
  developments in writing down actions in terms of fluid variables in non-equilibrium situations
  \cite{Haehl:2015uoc, Crossley:2015evo}, it would be interesting to understand if this setup can be
  suitably generalized to describe out of equilibrium fluid surfaces as well.}. Therefore, we have
to resort back to the second law of thermodynamics in order to constrain the transport coefficients.

The surface of the fluid interacts freely with the bulk. In order to account for this exchange of degrees of freedom between the bulk and the surface, the local form of the second law at the surface needs to be suitably modified. This modification takes the following form
\begin{equation}\label{surseclaw}
 \tilde \nabla_\mu J^\mu_{(s)\text{ent}} - n_\mu J^\mu_{(b)\text{ent}} \geq 0~~,
\end{equation}
where $J^\mu_{(s)\text{ent}}$ and $J^\mu_{(b)\text{ent}}$ represent the local surface and bulk
entropy currents respectively. Eq.~\eqref{surseclaw} corresponds to the $\tilde \delta(f)$ equation
obtained from the divergence of the total entropy current which is of the form\footnote{The reader
  may wonder, since the second law is expressed as an inequality for the divergence of the total
  entropy current, whether it is legitimate to implement the inequality separately for terms
  proportional to $\theta(f)$ and $\delta(f)$.  This is, however justified, since there can be fluid
  configurations where a non-trivial bulk entropy current is divergence free and the second law
  inequality must be valid for all fluid configurations.}
\begin{equation}
 J^\mu_{\text{ent}} = J^\mu_{(b)\text{ent}} \theta(f) + J^\mu_{(s)\text{ent}} \tilde \delta(f) + \dots~~.
\end{equation}

There are a few important aspects of out of equilibrium dynamics that are a priori unclear, even in
the context of ordinary fluid surfaces. One of the key aspects that needs to be understood is the
nature of the normal component of the fluid velocity $u_\mu n^\mu$ at the surface.  In equilibrium,
$u_\mu n^\mu$ vanishes by construction but once the location of the surface becomes time dependent,
this component may become non-trivial. Drawing from the analogy between $f$ and $\phi$, this problem
is analogous to the problem of understanding how the Josephson equation
$u^\mu \xi_\mu = \mu + \mu_\text{diss}$ is determined.  In a recent paper \cite{Jain:2016rlz}, it
was observed that the Josephson equation, even at ideal order, followed from the second law of
thermodynamics, when the $\phi$ field is considered off-shell. While working in an appropriate
generalization of the partition function frame, $\mu_\text{diss}$ reduces to the equation of motion
of $\phi$ in equilibrium. This leads to the interpretation of the Josephson condition as the
equation of motion of $\phi$ away from equilibrium.

Following this analogy, we consider $u^\mu n_\mu = \gamma + \gamma_\text{diss}$.  In
\S\ref{sssec:udotnzero}, we demonstrate that the local form of the second law of thermodynamics on
the surface sets $\gamma$ to zero. The form of $\gamma_{\text{diss}}$ is frame dependent like the
Josephson condition. We derive $\g_{\text{diss}}$ in a frame which is the appropriate generalization
of the partition function frame in \eqref{equisolpfframe}. In equilibrium, it reduces to the
equation of motion of $f$ (or equivalently to the Young-Laplace equation, which is the component of
the energy-momentum conservation equation normal to the surface).  Also, it is noteworthy that in
out of equilibrium situations, the equation $u^\mu n_\mu = \gamma_{\text{diss}}$ is distinct from
the corresponding Young-Laplace equation. They together determine two scalar degrees of freedom at
the boundary: $u^\mu n_\mu$ and $f$, the former of which turns out to be trivial in equilibrium.

Proceeding to the superfluid case in out of equilibrium scenarios, we tackle the corresponding problem for the normal component of the superfluid velocity at the surface $n^\mu \xi_\mu = \lambda + \lambda_{\text{diss}}$. In equilibrium, the $\phi$ equation of motion at  $\tilde \delta'(f)$ order imposes the condition that $\partial\mathcal C/\partial \lambda=0$, where $\cal C$ is (minus) the surface tension. Given a particular dependence of $\mathcal C$ on $\lambda$, implied by some microscopic description,  the condition $\partial \mathcal C / \partial \lambda = 0$ should be seen as the equation determining the value of $\lambda$ at the surface.  However, in the special case for which the surface tension does not depend on $\lambda$, the effective action  does not impose any restriction on $\lambda$. The equation of motion for $\phi$ may be obtained by  an off-shell implementation of the second law, as in \cite{Jain:2016rlz}. However, in the case of $\lambda$, we have shown in \S \ref{sssec:jent3p10ord} that an entropy current analysis does not impose any constraints on $\lambda$, once the leading order entropy density is modified by terms involving $\lambda$. This modification to the entropy density is identical to what is obtained from the equilibrium partition function in \S\ref{ssec:1stord3p1}. Since none of the physical constrains is able to set $\lambda$ to zero,  we report all our results keeping $\lambda$ arbitrary. Note that there can of course be many configurations with $\lambda=0$ but our analysis suggests that these will only be a subset of all possible configurations. 

Also, as explained previously, the surface equations may also be interpreted as determining the
possible set of boundary conditions that are allowed for the bulk fluid equations. Clearly, in the
equilibrium case, there are consistent solutions to the full set of bulk and surface equations. In
the partition function frame, such a solution corresponded to the one where the fluid velocity is
aligned with a Killing vector field of the background. However, away from equilibrium, even with a
judicious choice of frame, such a solution may be considerably complicated. In order to obtain some
idea of the nature of such solutions in time-dependent cases, we study the linearized fluctuations
around a toy equilibrium configuration, only considering the perfect fluid equations of motion.

In \S \ref{ssec:linfluc}, we work with $2+1$ dimensional ordinary fluids in flat space and consider
the background equilibrium configuration to be one in which a static fluid fills half space. At
first, we set the surface entropy to zero, recovering the standard dispersion relation of surface
capillary waves $\omega \sim \pm k^{3/2}$. If the amplitude of the surface ripples is much larger
than the surface thickness, then ignoring the surface degrees of freedom is a perfectly legitimate
approximation. However, as soon as we allow the surface tension to be a function of $T$, thus
introducing some non-trivial surface entropy, our surface equations predict a dispersion relation of
the form $\omega \sim \pm k$. We are then able to solve the bulk equations with such sound-like
boundary conditions. This new kind of surface sound wave for ordinary fluids is expected to be
visible if the amplitude of the waves is comparable or less than the surface thickness. These waves
are very reminiscent of the third sound mode for superfluids.  We perform a similar analysis for
$2+1$ dimensional superfluids, for which the leading order surface equations contain parity-odd
terms. We find that parity violation leaves its imprint on the spectrum of linearized fluctuations,
which contains a sound mode with $\omega \sim k$ while its partner under a parity transformation
$k \rightarrow -k$ is absent.

\section{Stationary superfluid bubbles in 3+1 dimensions}\label{sec:3plus1}

In this section we study stationary bubbles of a $3+1$ dimensional relativistic superfluid immersed in
an ordinary fluid. We work out the respective constitutive relations up to first derivative order in
the bulk and ideal order at the interface using equilibrium partition functions.

\subsection{Perfect superfluid bubbles (d+1 dimensions, d $\geq$ 3)}\label{ssec:perfsuper}

A discussion of the surface properties in perfect superfluids was initiated in
\cite{Armas:2015ssd}. Here we will elaborate and extend upon that discussion.  As explained
in \S \ref{ssec:staintro}, the equilibrium partition function for superfluids takes the form given
in \eqref{fulpartintoform}. If the partition function does not contain any derivatives, the
respective superfluid is called a perfect superfluid. It is of course a fictitious simplified system
just like a perfect fluid, nevertheless it is an instructive toy system to study before moving to
more complicated generalizations.

For a perfect superfluid bubble with an ordinary charged fluid outside, the most generic partition
function takes the form
\begin{equation}\label{persuppart}
  W = \int \df^3 x \sqrt{g} \ \frac{\E{-\sigma}}{T_0}
 \left( \theta(f)  P_{(b)} \left( T, \mu, \chi \right) +
  \tilde \delta(f) \mathcal C \left( T, \mu, \tilde\chi, \lambda \right)
  + \theta(-f)  P_{(e)} \left( T, \mu \right) \right)~~,
\end{equation}
where we have defined $T = T_0 e^{-\sigma}$ and $\mu = A_0 e^{-\sigma}$ suggestively for later
identification with the temperature and chemical potential respectively, while
$\lambda = n_\mu \xi^\mu = n_i \zeta^i$, $\chi = - \xi^\mu \xi_\mu = \mu^2 - \zeta^i \zeta_i$,
$\tilde\xi^\mu = - (\cG^{\mu\nu} - n^\mu n^\nu)\xi_\nu$ and
$\tilde\chi = - \tilde\xi^\mu \tilde\xi_\mu = \mu^2 - \zeta^i \zeta_i + \l^2$.  The discussion in
this subsection is immediately applicable to perfect superfluid bubbles in all dimensions, except in
2+1 dimensions\footnote{Even in 2+1 dimensions, if we restrict only to the parity even sector, then
  the discussion of this section is applicable.}  where there can be parity-odd effects at
ideal order, and will be treated separately in \S \ref{sec:2plus1}.

We start by varying the partition function \eqref{persuppart} with respect to the Goldstone boson
$\phi$, and work out the respective equations of motion
\begin{equation}\label{persurphieq}
 \begin{split}
   \theta(f) &\left[ \Df_i \left( \frac{2}{T} \frac{\partial \mathcal P_{(b)} }{\partial \chi} \z^i
     \right) \right] = 0~~, \\
   \tilde \delta(f) &\left[ T \tilde\Df_i \left( \frac{1}{T} \frac{\partial \mathcal C}{\partial \lambda} n^i
       - \frac{2}{T} \frac{\partial \mathcal C}{\partial \tilde\chi} \tilde\z^i
     \right)
     + 2 \l \frac{\partial  P_{(b)}}{\partial \chi}\right] =0~~, \\
   \tilde \delta'(f) &\left[\frac{\partial \mathcal C}{\partial \lambda}\right] = 0~~,
 \end{split}
\end{equation}
where $\Df_i$ denotes the spatial covariant derivative associated with $g_{ij}$, while $\tilde\Df_i$
denotes the spatial covariant derivative on the surface defined in \S\ref{ssec:staintro}. The last
line of this equation is particularly interesting, as it tells us that on-shell, the boundary
function $\cC$ is independent of the component of the superfluid velocity along $n_\mu$, i.e. $\l = n^\mu \xi_\mu$, \footnote{Since the surface tension function $\cC$ is given a priory, derived
  from the microscopics, the condition $\dow \cC/\dow \l = 0$ should be thought of as a condition
  determining $\lambda$ and not as a consistency condition on $\mathcal C$.} and is only dependent
on the projected components $\tilde\xi_\mu$ through $\tilde\chi$.

The first line in \eqref{persurphieq} is a non-linear second order differential equation, which
yields the profile of the Goldstone mode $\phi$ in the bulk of the superfluid bubble. For cases
where the superfluid velocity can be taken to be small, this equation may be linearized and
converted into a second order linear partial differential equation. This equation must be solved
with suitable boundary conditions at the interface, which are provided by the solutions to the
second and third lines in \eqref{persurphieq}. The third equation provides the derivative of $\phi$
normal to the interface, while the second equation provides the initial condition necessary to
evolve the first equation away from the interface. Note that as we move to higher orders, we will
have an additional condition at the surface, and correspondingly, the order of the first
differential equation will increase by one.

Varying the partition function \eqref{persuppart} and using the variational formulae
\eqref{rel_variation}, we can read out the bulk and boundary currents. The form of the
energy-momentum tensor and charge current inside the bubble takes the usual perfect superfluid form
and has been thoroughly discussed in \cite{Bhattacharyya:2012xi}, while outside the bubble it is
just an ordinary perfect charged fluid. The new ingredients in our discussion however are the
currents at the interface, found via variation as (upon using the $\tilde\d'(f)$ order $\phi$
equation of motion)
\begin{align}\label{persupcurs}
  {T_{(s)}}_{00} = e^{2\sigma}\left(
  - \mathcal C 
  + T \frac{\partial \mathcal C}{\partial T}
  + \mu \frac{\partial \mathcal C}{\partial \mu} 
  \right)
  &+ 2 \xi_0^2 \frac{\partial \mathcal C}{\partial \tilde\chi}~,~~
    {T_{(s)}}_{\ 0}^{i} = \xi_0 2 \frac{\partial \mathcal C}{\partial \tilde \chi}\tilde\z^i~,~~
    {T_{(s)}}^{ij} = \mathcal C h^{ij}
    + 2 \frac{\partial \mathcal C}{\partial \tilde\chi} \tilde\z^i \tilde\z^j~, \nn\\ 
  {J_{(s)}}_{0} = - e^{\sigma}
  &\frac{\partial \mathcal C }{\partial \mu} - 2 \xi_0 \frac{\partial
    \mathcal C}{\partial \tilde\chi}~, \qquad
    {J_{(s)}}^{i} = - 2  \frac{\partial \mathcal C}{\partial \tilde\chi} \tilde\z^i~~,
\end{align}
where $h^{ij} = g^{ij} - n^i n^j$ and $\tilde\z^{i} = h^{ij}\z_j$. It is further instructive to
write down the equation for the shape-field $f$ that follows from the partition function
\eqref{persuppart} (upon using the $\tilde\d'(f)$ order $\phi$ equation of motion)
\begin{equation}
  P_{(b)} -  P_{(e)} + T \Df_{i} \left( \frac{1}{T} \mathcal C n^i
    + \frac{2\l}{T} \frac{\partial \mathcal C}{\partial \tilde\chi} \tilde\z^i\right)  = 0~~.
\end{equation}
This is the modified Young-Laplace equation in the present case. As argued in Appendix
\ref{app:YLeomf}, this equation is simply the normal component of the energy-momentum conservation
equation on the surface.

Let us now study the implications of this analysis on the covariant form of the charge current and
energy-momentum tensor.  We would like to work in a hydrodynamic frame most suitable for the
analysis using the partition function. It is a frame where we have
\begin{equation} \label{frame}
  u^{\mu} = e^{-\sigma} ( 1,0,0,0 ),
  \qquad T = e^{- \sigma} T_0 ,
  \qquad \mu = e^{- \sigma} A_0~~,
\end{equation}
everywhere to all derivative orders, including at the interface. Such frame choice should be
always possible to make as long as we are in equilibrium. The most general ideal order surface
currents with the conditions that $T^{\mu\nu}_{(s)} n_\nu = J^\mu_{(s)} n_\mu = 0$ \footnote{The
  tangentiality conditions on the surface energy-momentum tensor and currents are a direct
  consequence of their conservation equations to leading order in the surface thickness. If
  thickness corrections are taken into account (by including $\tilde\delta'(f)$ terms or of higher
  derivative order in the energy-momentum tensor or currents) then these tangentiality conditions
  will be modified by extra terms on the right hand side \cite{Vasilic:2007wp, Armas:2012ac, Armas:2013aka}.}, can be written as
\begin{align}\label{persupcovcurs}
  T^{\mu \nu}_{(s)} &= (\cE - \cY) u^{\mu} u^\nu - \mathcal Y (\cG^{\mu\nu} - n^\mu n^\nu) + \mathcal F \tilde
  \xi^{\mu} \tilde \xi^{\nu} + 2\l_1 u^{(\mu}\tilde\xi^{\nu)} + 2\l_2 \xi^{(\mu} \bar n^{\nu)} +
  2\cU u^{(\mu} \bar n^{\nu)}~~, \nn\\
  J^{\mu}_{(s)} &= \mathcal Q u^{\mu} + \mathcal F' \tilde \xi^{\mu} + \cV \bar n^\mu~~,
\end{align}
where we have defined $\bar n^\mu = \epsilon^{\mu\nu\r\sigma} u_{\nu}\xi_{\r} n_{\sigma}$ as the
only parity-odd ideal order data.  Now reducing \eqref{persupcovcurs} on the time circle and
comparing it with \eqref{persupcurs}, we obtain\footnote{Note that we are defining
  $\mathcal{Y}=-\mathcal{C}$ in order to have the same sign convention for the surface tension
  $\mathcal{Y}$ as in classical literature.}
\begin{equation}\label{0thorderiden}
  \mathcal E = - \mathcal C + T \frac{\partial \mathcal C}{\partial T} + \mu \frac{\partial \mathcal
    C}{\partial \mu}, \quad
  \mathcal Y = -\mathcal C, \quad
  \mathcal Q =  \frac{\partial \mathcal C}{\partial \mu}, \quad
  \mathcal F = - \mathcal \cF' = 2  \frac{\partial \mathcal C}{\partial \tilde\chi}, \quad
    \cS = \frac{\partial \mathcal C}{\partial T}~~,
\end{equation}
and $\l_1 = \l_2 = \cU= \cV = 0$. We will see, however, that the coefficients $\cU$, $\cV$
get non-zero values when we introduce first order terms in the bulk. The coefficient $\cS$
introduced in \eqref{0thorderiden} is the surface entropy density and enters in the respective entropy
current as $J^\mu_{(s)\text{ent}} = \cS u^\mu$.  From \eqref{0thorderiden}, we can now recover the
Euler relation and the Gibbs-Duhem relation of thermodynamics respectively on the surface (upon
using the $\tilde\d'(f)$ order $\phi$ equation of motion)
\begin{equation}\label{persupthermocom}
  \mathcal E - \mathcal Y = T\cS + \mu \mathcal Q~~, \qquad
  \df \cY = - \cS \df T - \cQ \df \mu - \frac{1}{2} \mathcal F \df\tilde\chi~~.
\end{equation}
The first law of thermodynamics trivially follows from here as
\begin{equation}
  \df \cE = T\df \cS + \mu \df \cQ - \frac{1}{2} \mathcal F \df\tilde\chi~~.
\end{equation}
These thermodynamic relations are exactly the same as their bulk counterparts. However, as we will show
in the next subsection, the surface thermodynamics will modify upon including first order
corrections in the bulk.

\subsection{First order corrections away from the interface}\label{ssec:1stord3p1}

Since the surface currents sit on a boundary separating two phases of a fluid, transport
coefficients at a particular derivative order in the bulk can affect the surface currents at lower
orders via an ``inflow'' (via a differentiation by parts in the partition function
language). Therefore, we expect the ideal order surface currents to get contributions from first
order terms in the bulk. In order to do so, we consider first order corrections to the bulk
superfluid partition function (discussed in \cite{Bhattacharyya:2012xi})
\begin{equation}
  W^{(1)} =  W^{(1)}_{\text{even}} + W^{(1)}_{\text{odd}}~,
\end{equation}
where
\begin{equation}\label{partfn1stord3+1}
 \begin{split}
   W^{(1)}_{\text{even}} &= \int \df^3 x \sqrt{g} \ \frac{\E{\sigma}}{T_0} \theta(f) \left( \alpha_1
     \z^i \partial_i T + \alpha_2 \z^i \partial_i \nu
     - \alpha_3 \Df_i \left( \frac{1}{T} \frac{\dow \cP_{(b)}}{\dow\chi} \z^i \right) \right)~, \\
   W^{(1)}_{\text{odd}} &= \int \df^3 x \sqrt{g} \ \frac{\E{\sigma}}{T_0} \theta(f) \left( T_0
     \alpha_4 \epsilon^{ijk} \zeta_i \partial_{j} a_k + \alpha_5 \epsilon^{ijk} \zeta_i \partial_{j}
     A_k \right)~.
 \end{split}
\end{equation}
As discussed in \S \ref{ssec:staintro}, while working up to first order in derivatives in the bulk
of the superfluid, it is consistent to consider only the ideal order surface tension term at the
surface, which was considered in \eqref{persuppart}. Also, far outside the superfluid bubble, the
ordinary charged fluid does not receive any first order corrections, as there are no possible terms
that can be written in the partition function. Consequently, $W^{(1)}$ in \eqref{partfn1stord3+1}
constitutes the entire first order corrections to the perfect fluid partition function in
\eqref{persuppart}.

The bulk energy-momentum tensor and charge current that follow form \eqref{partfn1stord3+1} have
been thoroughly examined in \cite{Bhattacharyya:2012xi}. In particular, it was pointed out in
\cite{Bhattacharyya:2012xi} that the term proportional to $\alpha_3$ enters the constitutive
relations in a trivial fashion. The reason is that, since $\alpha_3$ multiplies the lower order
equation of motion of $\phi$, it can be shifted to zero by a suitable field redefinition of $\phi$
\footnote{We need to shift $\phi$ by a term proportional to $\alpha_3$. Note that in the
  natural way of counting derivatives for superfluids, $\phi$ is $-1$ order, while
  $\alpha_3(T, \nu, \chi)$ is zeroth order. Therefore, this entails shifting $\phi$ (and hence the
  superfluid velocity $\xi_\mu$) by a higher order term.}. In the presence of a surface, such a
shift would also involve surface quantities. However, at the level of the partition function for
instance, we can always redefine the surface tension to absorb these terms and ignore any higher
order terms.

The surface energy-momentum tensor and charge current, in addition to \eqref{persupcurs}, will now
also have the following contributions from \eqref{partfn1stord3+1} (after setting $\a_3 = 0$)
\begin{equation}\label{stcur3p1}
 \begin{split}
   {T_{(s)}}_{00} = e^{2 \sigma} \lambda T \alpha_1, &\qquad {T_{(s)}}_{\ 0}^{i} = \E{\sigma}\left(
      T \alpha_4 - \mu \alpha_5 \right) \bar n^i,
   \qquad {T_{(s)}}^{ij} = 0~, \\
   & {J_{(s)}}_{0} = - e^{\sigma} \lambda \frac{\alpha_2}{T}~,
   \qquad {J_{(s)}}^{i} = \alpha_5 \bar n^i~,
\end{split}
\end{equation}
where $\bar n^i = \epsilon^{ijk} \zeta_j n_k$.  The equation of motion of $\phi$ is modified
to
\begin{align}\label{eomphi1stord3p1}
  \q(f)
  &\bigg[ \Df_i \bigg( \frac{2}{T}\frac{\dow \cP_{(b)}}{\dow \chi} \z^i
    + \frac{2}{T} \frac{\dow \a_1}{\dow \chi} \z^i\z^j \dow_j T
    - \frac{\a_1}{T} g^{ij} \dow_j T
    + \frac{2}{T} \frac{\dow \a_2}{\dow \chi} \z^i\z^j \dow_j \nu
    - \frac{\a_2}{T} g^{ij} \dow_j \nu \nn\\
  &\quad + 2\E{\sigma} \frac{\dow\a_4}{\dow\chi} \z^i \epsilon^{ajk} \zeta_a \partial_{j} a_k
    - \E{\sigma} \a_4 \epsilon^{ijk} \partial_{j} a_k
    + \frac{2}{T}\frac{\dow\alpha_5}{\dow\chi} \z^i\epsilon^{ajk} \zeta_a \partial_{j}A_k
    - \frac{\alpha_5}{T} \epsilon^{ijk} \partial_{j}A_k
    \bigg) \bigg] = 0, \nn\\
  \tilde\d(f)
  &\bigg[ 
    T \tilde\Df_i \left( 
    \frac{1}{T} \frac{\partial \mathcal C}{\partial \lambda} n^i
    - \frac{2}{T} \frac{\partial \mathcal C}{\partial \tilde\chi} \tilde\z^i
    \right) 
                + 2 \l \frac{\partial  P_{(b)}}{\partial \chi}
                + 2 \l \frac{\dow \a_1}{\dow \chi} \z^j \dow_j T
  - \a_1 n^j \dow_j T
  + 2\l\frac{\dow \a_2}{\dow \chi} \z^j \dow_j \nu
  - \a_2 n^j \dow_j \nu \nn\\
  &\quad + 2\l T_0 \frac{\dow\a_4}{\dow\chi} \epsilon^{ajk} \zeta_a \partial_{j} a_k
  - T_0 \a_4 \epsilon^{ijk} n_i \partial_{j} a_k
  + 2\l\frac{\dow\alpha_5}{\dow\chi} \epsilon^{ajk} \zeta_a \partial_{j}A_k
  - \alpha_5 \epsilon^{ijk} n_i \partial_{j}A_k
                \bigg] =0, \nn\\
  \tilde \delta'(f) &\left[\frac{\partial \mathcal C}{\partial \lambda}\right] = 0~,
\end{align}
while the modified $f$ equation of motion (Young-Laplace equation) is given as (see Appendix
\ref{app:YLeomf} for a detailed discussion on Young-Laplace equations in the generic case)
\begin{multline}
  \cP_{(b)} - \cP_{(e)} + \a_1\z^i \partial_i T + \alpha_2 \z^i \partial_i \nu
  + T_0 \a_4 \e^{ijk} \z_i \partial_{j} a_k + \a_5 \e^{ijk} \z_i \partial_{j} A_k \\
  + T \Df_{i} \left( \frac{1}{T} \mathcal C n^i
    + \frac{2\l}{T} \frac{\partial \mathcal C}{\partial \tilde\chi} \tilde\xi^i\right) = 0~~.
\end{multline}
It is worth pointing out that, instead of the partition function $W$ in \eqref{partfn1stord3+1},
we could have started with a covariant version (ignoring the $\a_3$ term), i.e.,
\begin{equation}\label{partfn1stord3+1_cov}
 \begin{split}
   W^{(1)}_{\text{even}} &= \int \df^4 x \sqrt{-\cG} ~\theta(f) \left( \frac{f_1}{T}
     \z^\mu \partial_\mu T + T f_2 \z^\mu \partial_\mu \nu \right)~, \\
   W^{(1)}_{\text{odd}} &= \int \df^4 x \sqrt{-\cG} ~\theta(f) \left(
     g_1 \epsilon^{\mu\nu\r\sigma} \z_\mu u_\nu \o_{\r\sigma}
     + \half g_2 \epsilon^{\mu\nu\r\sigma} \z_\mu u_\nu F_{\r\sigma} \right)~.
 \end{split}
\end{equation}
Comparing it to \eqref{partfn1stord3+1}, we can simply read out the respective coefficients
\begin{equation}
  \a_1 = \frac{f_1}{T}~~, \qquad
  \a_2 = f_2 T~~, \qquad
  - T(\a_4 - \nu \a_5) = g_1~~, \qquad
  \alpha_5 = g_2~~.
\end{equation}

Now, the covariant form of the energy-momentum tensor and charge current, after imposing the
equation of motion for $\phi$, are modified from that in \eqref{persupcovcurs} to
\begin{equation}\label{persupcovcurs_der}
\begin{split}
 & T^{\mu \nu}_{(s)} = \cE~u^{\mu} u^\nu + \cY~\mathcal P^{\mu \nu} + \mathcal F \tilde \xi^{\mu} \tilde \xi^{\nu} 
 + 2 \cU u^{(\mu}\bar n^{\nu)},\\
 & J^{\mu}_{(s)} = \cQ ~u^{\mu} - \cF \tilde \xi^{\mu} + \cV \bar n^\mu~,
\end{split}
\end{equation}
where $\bar n^\mu = \epsilon^{\mu\nu\r\sigma}  u_{\nu}\xi_{\r} n_{\sigma}$.
Most notably, the coefficients of the surface currents now receive contributions from the first
order transport coefficients and we have
\begin{align}\label{para1stord3p1}
   \mathcal E &= - \mathcal C + T \frac{\partial \mathcal C}{\partial T} + \mu \frac{\partial
     \mathcal C}{\partial\mu} + \lambda f_1~~,
   \qquad \mathcal Y =- \mathcal C~~ , \qquad
   \cS = \frac{\partial \mathcal C}{\partial T} + \frac{\lambda}{T} \lb f_1 - \mu f_2\rb~~, \nn\\
   \mathcal Q &= \frac{\partial \mathcal C}{\partial \mu} + \lambda f_2 ~~,
   \qquad
   \mathcal F = 2 \frac{\partial \mathcal C}{\partial \chi}~~, \qquad
          \mathcal U = g_1 ~~, \qquad
            \mathcal V = g_2 ~~.
\end{align}
Here we have defined $\cS$ as the surface entropy density with the respective entropy current
given by
\begin{equation}
  J^\mu_{(s)\text{ent}} = \cS u^\mu + \frac{1}{T} \lb \cU - \mu \cV \rb \bar n^\mu.
\end{equation}
The identification \eqref{para1stord3p1} leads to the Euler relation and a modified Gibbs-Duhem
relation of thermodynamics at the surface\footnote{In \eqref{pesurthermo3p1}, $f_1$ and $f_2$ should
  be thought of as first order bulk transport coefficients (see \cite{Bhattacharyya:2012xi}) rather
  than parameters of the partition function.  }
\begin{equation}\label{pesurthermo3p1}
 \mathcal E - \mathcal Y = T \cS + \mu \mathcal Q~~, \qquad
 \df \cY = - \Big( \cS  - \frac{\l}{T} \lb f_1 - \mu f_2 \rb \Big) \df T
 - \Big(\cQ - \l f_2\Big) \df \mu
 - \frac{1}{2} \mathcal F \df\tilde\chi ~~.
\end{equation}
We clearly see that the thermodynamics has changed. The respective modified first law of
thermodynamics now takes the form
\begin{equation}\label{pesurthermo3p1_firstlaw}
  \df \Big( \cE - \l f_1 \Big) = T \df \Big( \cS  - \frac{\l}{T} \lb f_1 - \mu f_2 \rb \Big)
  + \mu \df \Big(\cQ - \l f_2\Big) - \frac{1}{2} \mathcal F \df\tilde\chi~~.
\end{equation}
This modification can be interpreted as follows. The surface densities $\cE$, $\cQ$, $\cS$ and $\cF$
have, in general, two contributions: from the thermodynamics on the surface and from the inflow from the
bulk. If we identify the inflow contributions to $\cE$, $\cQ$, $\cS$ and $\cF$ as $\l f_1$,
$\l f_2$, $\l\lb f_1 - \mu f_2 \rb/T$ and $0$ respectively, the remaining thermodynamic
contributions satisfy the thermodynamics \eqref{pesurthermo3p1}-\eqref{pesurthermo3p1_firstlaw}.

Note that the parity-odd ideal order surface transport coefficients $\mathcal U$ and $ \mathcal V$
(or correspondingly $g_1$ and $g_2$) do not enter the thermodynamics
\eqref{pesurthermo3p1}-\eqref{pesurthermo3p1_firstlaw}. However, since all the first order bulk
transport coefficients $f_1$, $f_2$, $g_1$, $g_2$ do modify the ideal order surface transport, they
can be measured by carefully designing experiments which probe the ideal order surface properties of
superfluids.

\section{Stationary superfluid bubbles in 2+1 dimensions}\label{sec:2plus1}

In this section, we study stationary superfluid bubbles in $2+1$ dimensions and particularly focus
on the parity-odd sector, where there is a significant difference compared to the $3+1$ dimensional
case. In fact, an exhaustive analysis of the first order parity-odd terms in the bulk of $2+1$
dimensional superfluids has not been executed so far, to the best of our knowledge. Therefore, we
also evaluate the stationary bulk currents following from the parity-odd first order bulk partition
function in \S \ref{ssec:perfsuper2p1} before analyzing their surface effects.

\subsection{Parity-odd effects for perfect superfluid bubbles}\label{ssec:perfsuper2p1}

We have discussed perfect superfluids in general dimensions in \S \ref{ssec:perfsuper}. However, as
explained in \S \ref{ssec:staintro}, in $2+1$ dimensions there can be parity-odd terms which may
have a non-trivial effect on the surface tension. Hence, before going into the details of the
surface effects of first order corrections in the bulk of $2+1$ dimensional superfluids, we will
revisit the zeroth order case once more.

In $2+1$ dimensions, apart from $\lambda$, it is also possible to define a parity-odd zeroth order
scalar on the surface
$\bar{\lambda} = \epsilon^{\mu \nu \sigma} n_{\mu} u_\nu \xi_\sigma = \epsilon^{ij} n_i \zeta_j$. As
explained in \S \ref{ssec:staintro}, due to the possible presence of a term like \eqref{LGsimppo} in
a Landau-Ginzburg effective theory, the surface tension will, in general, depend on $\bar \lambda$
and we can write\footnote{Note that once we have assumed that $\mathcal C$ depends both on $\lambda$
  and $\bar \lambda$, a further dependence on $\tilde\chi$ is redundant, since it is no longer an
  independent variable and is given by
\begin{equation}\label{chiexp}
  \tilde\chi = \mu^2 - \bar \lambda^2~~.
\end{equation}
This is due to the fact that, on the interface there are only two independent components of the
superfluid velocity, the one along the surface and the one orthogonal to the surface, which are
denoted by $\bar \lambda$ and $\lambda$ respectively. We should also note that, due to the relation
\eqref{chiexp}, the surface term in \eqref{persuppart2p1} may also be written as
\begin{equation}\label{Cspl}
  \mathcal C \left( T, \mu, \lambda, \bar{\lambda} \right) = 
  \mathcal C_0 \left( T, \mu, \tilde\chi, \lambda \right) + 
  \bar{\lambda}  \mathcal C_1 \left( T, \mu, \tilde\chi, \lambda \right)~~,
\end{equation}
however in our discussion we choose to proceed with the form \eqref{persuppart2p1}. }
\begin{equation}\label{persuppart2p1}
  W = \int \df^2x \sqrt{g} \ \frac{\E{\sigma}}{T_0}\left(
    \theta(f)  P_{(b)} \left(T, \mu, \chi \right)+
    \tilde \delta(f) \mathcal C \left( T, \mu, \lambda, \bar\lambda \right)
    + \theta(-f)  P_{(e)} \left( T, \mu \right) \right).
\end{equation}
We start with the $\phi$ equations of motion following from the partition function
\eqref{persuppart2p1}
\begin{equation}\label{persurphieq2p1}
 \begin{split}
   \theta(f) &\left[ \Df_i \left( \frac{2}{T} \frac{\partial  P_{(b)}}{\partial \chi}
       \z^i \right) \right] = 0~~, \\
   \tilde \delta(f) &\left[ T \tilde\Df_i \left(
       \frac{1}{T\bar\l} \frac{\partial \mathcal C}{\partial \bar{\lambda}} \tilde\z^i
       + \frac{1}{T} \frac{\partial \mathcal C}{\partial \lambda} n^i \right)
     + 2 \l \frac{\partial  P_{(b)}}{\partial \chi} \right] =0~~, \\
   \tilde \delta'(f) &\left[ \frac{\partial \mathcal C}{\partial \lambda} \right] = 0~~,
 \end{split}
\end{equation}
where again, $\tilde\Df_i$ denotes the covariant derivative on the surface. The energy-momentum
tensor and charge current far away from the interface is exactly the same as in the $3+1$
dimensional case. At the interface however, we can use the formulae in appendix \ref{app:formula} in
order to determine the energy-momentum tensor and charge current as following (after imposing the
$\tilde\d'(f)$ part of the $\phi$ equation of motion)
\begin{align}\label{persupcurs2p1}
  {T_{(s)}}_{00} = e^{2\sigma}\left( -\cC + T \frac{\partial \mathcal C}{\partial T} + \mu
    \frac{\partial \mathcal C}{\partial \mu}
  \right)~,\quad
  &{T_{(s)}}_{\ 0}^{i} =
  - \xi_0 \frac{1}{\bar\l}\frac{\partial \mathcal C}{\partial \bar \lambda} \tilde\z^i
  ~, \quad
  {T_{(s)}}^{ij} = \mathcal C h^{ij} - \frac{1}{\bar\l} \frac{\partial \mathcal C}{\partial \bar
    \lambda} \tilde\z^i\tilde\z^j
  ~, \nn\\
  {J_{(s)}}_{0} = - e^{\sigma} \frac{\partial \mathcal C }{\partial \mu}~,&
\qquad
 {J_{(s)}}^{i} = 
 \frac{1}{\bar\l}\frac{\partial \mathcal C}{\partial \bar \lambda} \tilde\z^i~~,
\end{align}
where $h^{ij} = g^{ij} - n^i n^j = \e^{ia}n_a \e^{jb} n_b$ and $\tilde \z^i = h^{ij} \z_j$.  Note that
$\mathcal C$ in \eqref{persupcurs2p1} contains both parity-odd and parity-even contributions and is
given by \eqref{Cspl}. We can also easily obtain the equation of motion for the shape-field $f$, which
now involves parity-odd pieces as well, namely the Young-Laplace equation (after imposing the
$\tilde\d'(f)$ part of the $\phi$ equation of motion)
\begin{equation}
   P_{(b)} -  P_{(e)} + \Df_{i} \left( \frac{1}{T} \mathcal C n^i 
    - \frac{\l}{T\bar\l} \frac{\partial \mathcal C}{\partial \bar \lambda} \tilde\z^i
  \right)  = 0~~.
\end{equation}
We again choose to work in a hydrodynamic frame suitable for the partition function analysis, where
\begin{equation} \label{frame2p1}
  u^{\mu} = e^{-\sigma} ( 1,0,0)~, \qquad
  T = T_0 e^{- \sigma}~, \qquad
  \mu = e^{- \sigma} A_0~,
\end{equation}
everywhere, including at the interface. Using the $\tilde \delta' (f)$ part of the equation of motion
for $\phi$ \eqref{persurphieq2p1}, the covariant form of the energy-momentum tensor and current may be
expressed as\footnote{ In $2+1$ dimensions there can at most be $3$ independent vectors but we
  have at least $6$ on the surface: $u^\mu$, $\tilde\xi^\mu$, $n^\mu$, $\e^{\mu\nu\r}u_\nu n_\r$,
  $\e^{\mu\nu\r}u_\nu \tilde\xi_\r$, $\e^{\mu\nu\r}n_\nu \tilde\xi_\r$. Choosing a basis of any
  three, we can write the others in terms of the chosen basis, for example choosing $u^\mu$,
  $\tilde\xi^\mu$, $n^\mu$ (as we did in \eqref{persupcovcurs2p1})
  \begin{equation}
    \e^{\mu\nu\r}u_\nu n_\r = -\frac{1}{\bar \l} (\tilde\xi^\mu + \mu u^\mu)~, \qquad
    \e^{\mu\nu\r}u_\nu \tilde\xi_\r = \bar \l n^\mu~, \qquad
    \e^{\mu\nu\r}n_\nu \tilde\xi_\r = - \frac{\mu}{\bar\l} \tilde\xi^\mu - \frac{\mu^2 - \bar\l^2}{\bar\l} u^\mu~~.
  \end{equation}
  This allows us to write the constitutive relations \eqref{persupcovcurs2p1} in many other
  basis. For example, choosing $\e^{\mu\nu\r}u_\nu n_\r$ in favour of $\tilde\xi^\mu$ we find
  \begin{equation}
    \begin{split}
      & T^{\mu \nu}_{(s)} = (\cE' - \cY') u^{\mu} u^\nu
      - \cY' (\cG^{\mu\nu} - n^\mu n^\nu)
      + 2 \cF' \mu u^{(\mu} \e^{\nu)\sigma\r}u_\sigma n_\r,~ \\
      & J^{\mu}_{(s)} = \cQ' u^{\mu} + \cF' \e^{\mu\nu\r}u_\nu n_\r~~,
    \end{split}
  \end{equation}
  where $\cE' = \cE + \mu^2\cF$, $\cY' = \cY - \bar\l^2\cF$, $\cQ' = \mathcal Q + \mu\cF$ and
  $\cF' = \bar\l \cF$ in terms of the coefficients in \eqref{persupcovcurs2p1}.  }
\begin{equation}\label{persupcovcurs2p1}
\begin{split}
 & T^{\mu \nu}_{(s)} = (\cE - \cY )  u^{\mu} u^\nu - \mathcal Y (\cG^{\mu\nu} - n^\mu n^\nu)
  + \cF \tilde\xi^\mu \tilde\xi^\nu~~, \\
 & J^{\mu}_{(s)} = \mathcal Q u^{\mu} -\cF \tilde\xi^\mu~~,
\end{split}
\end{equation}
according to which comparison with \eqref{persupcurs2p1} allows us to obtain
\begin{align} \label{comp2+1}
  \cE = - \cC + T \frac{\partial \mathcal C}{\partial T}
  &+ \mu \frac{\partial \mathcal C}{\partial
  \mu} + \frac{\mu^2}{\bar \l } \frac{\partial \mathcal C}{\partial \bar \lambda}~~, \qquad
    \cY = - \mathcal C~~, \qquad
  \cF = - \frac{1}{\bar \l } \frac{\partial \mathcal C}{\partial \bar \lambda}~~, \nn\\
  &\cQ = \frac{\partial \mathcal C}{\partial \mu} + \frac{\mu}{\bar \l } \frac{\partial
     \mathcal C}{\partial \bar \lambda} ~~, \qquad
   \cS = \frac{\partial \mathcal C}{\partial T}~~.
\end{align}
Here $\partial \mathcal C / \partial \lambda $ is zero on-shell due to the $\tilde\d'(f)$ equation
of motion for $\phi$ \eqref{persurphieq2p1}. $\cS$ is again the surface entropy current, entering
the entropy current as $J^\mu_{(s)\text{ent}} = \cS u^\mu$.  These relations can be summarized as
the surface Euler relation and the first law of thermodynamics specific to $2+1$ dimensions,
respectively as
\begin{equation}\label{persupthermocom2p1}
  \mathcal E - \mathcal Y = T \cS + \mu \mathcal Q~~, \qquad
  \df \cY = - \cS \df T - \cQ \df\mu - \cF \lb \mu \df \mu - \bar\l \df \bar \l \rb~~.
\end{equation}
Note that, if we do not have any parity-odd dependence in the surface tension $\cY$, i.e. if it only
depends on $\bar\l^2$, the final differential becomes
$\mu \df \mu - \bar\l \df \bar \l = \half\df (\mu^2 - \bar\l^2) = \half \df \tilde\chi$ and we get
the familiar perfect superfluid surface first law of thermodynamics as in
\eqref{persupthermocom}. The first law of thermodynamics in this case however is slightly subtle,
and we have discussed it in appendix \ref{app:2+1thermo}.

\subsection{First order corrections away from the interface}

We now wish to extend the results of the previous section by considering first derivative
corrections to the partition function of the bulk superfluid and that of the exterior charged fluid,
which by the same ``inflow'' mechanism explained in \S \ref{ssec:1stord3p1}, will affect the surface
currents in an important way. A significant difference between $2+1$ dimensional superfluid bubbles
and that of its $3+1$ dimensional counterpart, is the fact that the partition function of the
exterior charged fluid also receives non-zero contributions at first order in derivatives.

The total first order partition function $W^{(1)}$ for $2+1$ dimensional superfluid bubbles can be
expressed in terms of the parity-odd corrections to the bulk superfluid partition function
$W^{(1)}_{\text{odd}}$ and the parity-odd corrections to the exterior charged fluid partition
function $\tilde W^{(1)}_{\text{odd}}$ such that
\begin{equation}
W^{(1)}=W^{(1)}_{\text{even}}+W^{(1)}_{\text{odd}}+\tilde W^{(1)}_{\text{odd}}~~,
\end{equation}
where $W^{(1)}_{\text{even}}$ corresponds to the first order corrections to the partition function
of the bulk superfluid in the parity-even sector \eqref{partfn1stord3+1}, since these corrections
are universal irrespective of spacetime dimensions.

The parity-odd first order corrections in the bulk of the $2+1$ dimensional superfluid are
significantly different from the $3+1$ dimensional case. They are given by
\begin{equation}\label{po1stord2p1sup}
 W^{(1)}_{\text{odd}} = \int \df^2x \sqrt{g}\frac{e^{\sigma}}{T_0} ~\theta(f) \left( m_{\omega} ~ \epsilon^{ij} \partial_i a_j 
 + m_{B} ~\epsilon^{ij} \partial_i A_j + \beta_1 ~\epsilon^{ij} \zeta_i \partial_j \sigma 
 + \beta_2 ~\epsilon^{ij} \zeta_i \partial_j A_0 \right) ~~.
\end{equation}
On the other hand, the parity-odd corrections to the partition function of the exterior charged
fluid read \cite{Banerjee:2012iz}
\begin{equation}\label{po1stord2p1ord}
 \tilde{W}^{(1)}_{\text{odd}} = \int \df^2x \sqrt{g}\frac{e^{\sigma}}{T_0} ~\theta(f) \left( M_{\omega} ~\epsilon^{ij} \partial_i a_j 
 + M_{B} ~\epsilon^{ij} \partial_i A_j\right) ~~.
\end{equation}
The coefficients $m_\omega, m_B, \beta_1, \beta_2$ in \eqref{po1stord2p1sup} depend on the three scalars $T$, $\mu$ and $\chi$, while 
the coefficients $M_\omega, M_B$ parametrizing the charged fluid in \eqref{po1stord2p1ord} only depend on $T$ and $\mu$. 

The bulk currents that follow from direct variation of \eqref{po1stord2p1ord} were studied in
\cite{Banerjee:2012iz}, while the surface effects were recently considered in
\cite{Kovtun:2016lfw}. Therefore, we defer the reader to these references for more details on these
currents.  However, neither the bulk nor the surface effects of \eqref{po1stord2p1sup} have been
previously analyzed in the literature. Below, we explicitly provide the bulk and surface currents
that follow from \eqref{po1stord2p1sup}. In \S \ref{sec:con2+1bulk}, we obtain the constraints among
bulk transport coefficients that $W^{(1)}_{\text{odd}}$ imposes on the $2+1$ dimensional superfluid,
while in \S \ref{sec:con2+1surf} we obtain the constraints imposed by it on the surface transport
coefficients. Finally, in \S \ref{sec:thermo2+1} we study the rich thermodynamic properties of the
interface between the bulk superfluid and the exterior charged fluid by considering all surface
effects arising from $W^{(1)}$.

\subsubsection*{\emph{Bulk currents}}

The bulk energy-momentum tensor and the charge current obtained by varying \eqref{po1stord2p1sup} using the formulae of Appendix \ref{app:formula} take the form 
\begin{equation}\label{po2p1blkstct00}
  \begin{split}
  {T_{(b)}}_{00} =&- e^{2 \sigma}
 \bigg[ \bigg(m_\omega -  T \frac{\partial m_\omega}{\partial T} 
 - 2 e^{-2\sigma}\xi_0^2 \frac{\partial m_\omega}{\partial \chi}\bigg) \epsilon^{ij}  \partial_i a_j \\
 &~~~~~~~~~~+ \bigg( m_B -  T \frac{\partial m_B}{\partial T} 
 -2 e^{-2\sigma}\xi_0^2  \frac{\partial m_B}{\partial \chi} +\beta_1 \bigg) \epsilon^{ij}  \partial_i A_j \\
  &~~~~~~~~~~+ \bigg(\beta_2 - T \frac{\partial \beta_2 }{\partial  T} 
 - 2 e^{-2\sigma}\xi_0^2\frac{\partial \beta_2}{\partial \chi}-\frac{1}{T_0} \frac{\partial \beta_1}{\partial  \nu} \bigg) \epsilon^{ij}  \zeta_i \partial_j A_0 \\
  &~~~~~~~~~~-  2 e^{-2\sigma}\xi_0^2\frac{\partial \beta_1}{\partial T} \epsilon^{ij}  \zeta_i \partial_j \sigma  
 - \frac{\partial \beta_1}{\partial \chi} \epsilon^{ij} \zeta_i  \partial_j \chi \bigg]~~,
 \end{split}
\end{equation}
\begin{equation}\label{po2p1blkstcti0}
  \begin{split}
    &{T_{(b)}}_{\ 0}^{i} = \left(m_{\omega} - A_0 (\beta_1+m_B)-T\left( \frac{\partial
          m_\omega}{\partial T} - A_0 \frac{\partial m_B}{\partial T}\right)\right)
    \epsilon^{ij} \partial_j \sigma \\ &+\left(\frac{1}{T_0} \left( \frac{\partial
          m_\omega}{\partial \nu}
        - A_0 \frac{\partial m_B}{\partial \nu} \right) - A_0 \beta_2 \right) \epsilon^{ij} \partial_j A_0  + \left(\frac{\partial m_\omega}{\partial \chi} - A_0 \frac{\partial m_B}{\partial \chi} \right) \epsilon^{ij} \partial_j \chi \\
    & +2 A_0 \xi^i \left(\frac{\partial m_\omega}{\partial \chi} \epsilon^{jk} \partial_j
      a_k+\frac{\partial m_B}{\partial \chi} \epsilon^{jk} \partial_j A_k+ \frac{\partial
        \beta_1}{\partial \chi} \epsilon^{jk} \zeta_j \partial_k \sigma + \frac{\partial
        \beta_2}{\partial \chi} \epsilon^{jk} \zeta_j \partial_k A_0 \right)~~,
\end{split}
\end{equation}
\begin{equation}\label{po2p1blkstctij}
  \begin{split}
 {T_{(b)}}^{ij} = &2 \xi^i \xi^j \bigg[ \frac{\partial m_{\omega}}{\partial \chi}  ~ \epsilon^{kl} \partial_k a_l 
 + \frac{\partial m_{B}}{\partial \chi} ~\epsilon^{kl} \partial_k A_l + \frac{\partial \beta_1}{\partial \chi} ~\epsilon^{kl} \zeta_k \partial_l \sigma 
 + \frac{\partial \beta_2}{\partial \chi} ~\epsilon^{kl} \zeta_k \partial_l A_0 \bigg]~~,
 \end{split}
\end{equation}
\begin{equation}\label{po2p1blkstcj0} \begin{split}
{J_{(b)}}_{0} =&- e^{2 \sigma} \bigg[ 
 \bigg( \frac{1}{T_0} \frac{\partial m_\omega}{\partial \nu } + 2 A_0 e^{-2 \sigma} \frac{\partial m_\omega}{\partial \chi}
 \bigg) \epsilon^{ij} \partial_i a_j \\ & ~~~~~~~~~
 + \left(  \frac{1}{T_0} \frac{\partial m_B}{\partial \nu } 
 + 2 A_0 e^{-2 \sigma} \frac{\partial m_B}{\partial \chi} + \beta_2 \right) \epsilon^{ij} \partial_i A_j \\
  &  ~~~~~~~~~ +\left(  \frac{1}{T_0} \frac{\partial \beta_1}{\partial \nu } 
 + 2 A_0 e^{-2 \sigma} \frac{\partial \beta_1}{\partial \chi} -\beta_2 +  T \frac{\partial \beta_2 }{\partial T}\right) \epsilon^{ij} \zeta_i \partial_j \sigma \\
  &  ~~~~~~~~~
 + 2 A_0 e^{-2 \sigma} \frac{\partial \beta_2}{\partial \chi} \epsilon^{ij} \zeta_i \partial_j A_0
 - \frac{\partial \beta_2}{\partial \chi} \epsilon^{ij} \zeta_i  \partial_j \chi \bigg]  ~~,
 \end{split}
\end{equation}
\begin{equation}\label{po2p1blkstcji} \begin{split}
 &{J_{(b)}}^{i} = - 2 \xi^i \left( \frac{\partial m_\omega}{\partial \chi} \epsilon^{jk} \partial_j a_k 
 + \frac{\partial m_B}{\partial \chi} \epsilon^{jk} \partial_j A_k  
 + \frac{\partial \beta_1}{\partial \chi} \epsilon^{jk} \zeta_j \partial_k \sigma 
 + \frac{\partial \beta_2}{\partial \chi} \epsilon^{jk} \zeta_j \partial_k A_0  \right) \\ & ~~
 + \left(\beta_1+m_B - T \frac{\partial m_B}{\partial T }\right) \epsilon^{ij} \partial_{j} \sigma +\left(  \beta_2 + \frac{1}{T_0} \frac{\partial m_B}{\partial  \nu }\right) \epsilon^{ij} \partial_{j} A_0 +
  \frac{\partial m_B}{\partial \chi } \epsilon^{ij}\partial_j \chi~~.
\end{split}
\end{equation}
It is important to note that the bulk energy-momentum tensor and current are entirely determined in terms of
the coefficients $m_\omega, m_B, \beta_1, \beta_2$.

\subsubsection*{\emph{Surface currents}}

The surface energy-momentum tensor and charge current obtained by varying \eqref{po1stord2p1sup} take the form
\begin{equation} \label{po2p1surstc}
  \begin{split}
    {T_{(s)}}^{ij} = 0~~, \qquad
    &{T_{(s)}}_{00}=  e^{2 \sigma} \bar \lambda \beta_1~~, \qquad
    {T_{(s)}}_{0}^{i} =- \left( m_\omega - A_0 m_B\right) \epsilon^{ij} n_j~~, \\
    & {J_{(s)}}_{0} =  \bar \lambda \beta_2 e^{2 \sigma}~~, \qquad
    {J_{(s)}}^{i} = - m_B \epsilon^{ij} n_j~~.
\end{split}
\end{equation}
The effect of the coefficients $m_\omega$ and $m_B$ appearing in \eqref{po1stord2p1sup} on the surface currents is essentially the same 
as the effect of the coefficients $M_\omega$ and $M_B$ appearing in \eqref{po1stord2p1ord}. In fact, the contributions to the surface currents 
due to the parity-odd  sector of $W^{(1)}$ is entirely given by \eqref{po2p1surstc} with the replacement $m_\omega\to m_\omega-M_\omega$ and
$m_B\to m_B-M_B$.

\subsubsection{Constraints on the bulk parity-odd constitutive relations} \label{sec:con2+1bulk}

In this section we derive the constraints on the covariant form of the bulk energy-momentum tensor
and charge current that are implied by Eqs.\eqref{po2p1blkstct00}--\eqref{po2p1blkstcji}, which in
turn follow from the partition function \eqref{po1stord2p1sup}.  In order to do so, one must
classify all first order parity-odd terms, which are non-zero in equilibrium, based on their
transformation properties under the spatial rotation group. The possible scalars have been listed in
table \ref{tab:scal}, while the vectors orthogonal to both $u^\mu$ and $\xi^{\mu}$ have been listed
in table \ref{tab:vec}.\footnote{Note that in 2+1 dimensions it is not possible to write any second
  rank symmetric tensor invariant under the spatial rotation group with both indices projected
  orthogonal to both $u^\mu$ and $\xi^{\mu}$.}

\begin{center} 
\begin{tabular}{ |c|c|c| } 
 \hline
 $a$ & $\mathcal S_{(a)}$ & Reduced form $\mathcal S_{(a)}^{\text{red}}$ \rule{0pt}{2.6ex} \\ [0.7ex]
 \hline
 1 & $\epsilon^{\sigma \nu \lambda} u_{\sigma} \omega_{\nu \lambda}$ 
 & $e^{\sigma} \epsilon^{ij}\partial_i a_j$ \rule{0pt}{2.6ex}\\   [0.6ex]
 2 & $\epsilon^{\sigma \nu \lambda} u_{\sigma} \cal{F}_{\nu \lambda}$ 
 & $-2 \epsilon^{ij}(\partial_i A_j+A_0 \partial_i a_j)$ \rule{0pt}{2.6ex} \\  [0.6ex]
 3 & $\epsilon^{\sigma \nu \lambda} \xi_{\sigma} u_{\nu} \partial_{\lambda} T$ 
 & $\epsilon^{ij}\zeta_i \partial_j T = - T \epsilon^{ij}\zeta_i \partial_j \sigma$ \rule{0pt}{2.6ex}\\   [0.6ex]
 4 & $\epsilon^{\sigma \nu \lambda} \xi_{\sigma} u_{\nu} \partial_{\lambda} \nu $
 & $\epsilon^{ij}\zeta_i \partial_j \nu = \frac{1}{T_0} \epsilon^{ij}\zeta_i \partial_j A_0 $ \rule{0pt}{2.6ex} \\  [0.6ex]
 5 & $- \epsilon^{\sigma \nu \lambda} \xi_{\sigma} u_{\nu} \partial_{\lambda} \chi $ 
 & $- \epsilon^{ij}\zeta_i \partial_j \chi $ \rule{0pt}{2.6ex}\\   [0.6ex]
 \hline
\end{tabular}
\captionof{table}{Parity-odd first order scalars in $2+1$ dimensions and their dimensional
  reduction.}\label{tab:scal}
\end{center}
\begin{center} 
\begin{tabular}{ |c|c|c| } 
 \hline
 $a$ & $\mathcal V^{\mu}_{(a)}$ & Reduced form $\mathcal V^{i}_{(a)}$ \rule{0pt}{2.6ex} \\ [0.7ex]
 \hline
 1 & $\tilde P^{\mu}_{~\sigma} \epsilon^{\sigma \nu \lambda}  u_{\nu} \partial_{\lambda}T$  &  
 $\tilde P^{ik} g_{kl}\epsilon^{lj}\partial_j T = -T \tilde P^{ik} g_{kl} \epsilon^{lj}\partial_j \sigma$ \rule{0pt}{2.6ex}\\   [0.6ex]
 2 & $\tilde P^{\mu}_{~\sigma} \epsilon^{\sigma \nu \lambda} u_{\nu} \partial_{\lambda} \nu$  
 &  $\tilde P^{ik} g_{kl}\epsilon^{lj}\partial_j \nu = \frac{1}{T_0} \tilde P^{ik} g_{kl} \epsilon^{lj}\partial_j A_0$ \rule{0pt}{2.6ex} \\  [0.6ex]
 3 & $- \tilde P^{\mu}_{~\sigma} \epsilon^{\sigma \nu \lambda}   u_{\nu} \partial_{\lambda}\chi $  
 &  $- \tilde P^{ik} g_{kl}\epsilon^{lj} \partial_j \chi $ \rule{0pt}{2.6ex}\\   [0.6ex]
 \hline
\end{tabular}
\captionof{table}{Parity-odd vectors in $2+1$ dimensions and their dimensional reduction. Here we
  have defined the projector
  $\tilde P_{\mu \nu} = \mathcal G_{\mu \nu} + u_\mu u_\nu - \frac{\zeta^{\mu}
    \zeta^{\nu}}{|\zeta|^2}$,
  where $\zeta^{\mu} = \xi^{\mu} + (u^\nu \xi_\nu) u^\mu$. After the reduction on the time circle,
  we find $(\mathcal V_{(a)})_0 = 0$ while $\mathcal V^{i}_{(a)}$ is specified in the right
  column. Also, after the reduction, the tangential projector takes the form
  $\tilde P^{ij} = g^{ij} - \frac{\zeta^{i} \zeta^{j}}{(g^{lk} \zeta_l \zeta_k)}, ~\tilde P_{00}=
  P^i_{\ 0}=0 $.}\label{tab:vec}
\end{center}
Using tables \ref{tab:scal} and \ref{tab:vec}, the most general bulk energy-momentum tensor and charge
current allowed by symmetries, describing $2+1$ superfluid bubbles in the parity-odd sector at first
order in derivatives are
\begin{equation}\label{po2p1covstc}
\begin{split}
  T_{(b)}^{\mu\nu} = &\sum_{a=1}^5 \left( \mathfrak{s}^{(1)}_{a} \mathcal S_{(a)} u^{\mu} u^\nu 
+ \mathfrak{s}^{(2)}_{a} \mathcal S_{(a)} \mathcal G^{\mu \nu} 
+ \mathfrak{s}^{(3)}_{a} \mathcal S_{(a)} \zeta^{\mu} \zeta^{\nu} 
+ \mathfrak{s}^{(4)}_{a} \mathcal S_{(a)} \zeta^{(\mu} u^{\nu)} \right) \\
& + \sum_{a=1}^{3} \left( \mathfrak{v}^{(1)}_{a} u^{(\mu} \mathcal V^{\nu)}_{(a)} 
+ \mathfrak{v}^{(2)}_{a} \zeta^{(\mu} \mathcal V^{\nu)}_{(a)} \right)~~, \\
  J_{(b)}^{\mu} = &  \sum_{a=1}^5 \left( \mathfrak{s}^{(5)}_{a} \mathcal S_{(a)} u^{\mu} +
  \mathfrak{s}^{(6)}_{a} \mathcal S_{(a)} \zeta^{\mu} \right) 
  + \sum_{a=1}^{3} \mathfrak{v}^{(3)}_{a}  \mathcal V^{\mu}_{(a)} ~~,
\end{split}
\end{equation}
where $\zeta^\mu = \xi^\mu + (u^\nu \xi_\nu) u^\mu$ is the superfluid velocity projected
orthogonally to the fluid velocity.  The energy-momentum tensor and charge current in
\eqref{po2p1covstcred} are parametrized by a total of 39 transport coefficients, which will
ultimately be constrained in terms of the four parameters $m_\omega, m_B, \beta_1, \beta_2$
appearing in the partition function \eqref{po1stord2p1sup}.  Using the explicit reductions provided
in tables \ref{tab:scal} and \ref{tab:vec}, we can readily reduce the energy-momentum tensor and
current given in \eqref{po2p1covstc} into
\begin{equation} \label{po2p1covstcred}
\begin{split}
{T_{(b)}}_{00} =&  e^{2\sigma} \sum_{a=1}^5\mathcal S^{\text{red}}_{(a)}
(\mathfrak{s}^{(1)}_{a}-\mathfrak{s}^{(2)}_{a})~~, \qquad
{T_{(b)}}_{0}^i = - {e^{\sigma} \over 2} \bigg[ \sum_{a=1}^5\mathcal S^{\text{red}}_{(a)}  \mathfrak{s}^{(4)}_{a}\zeta^i +\sum_{a=1}^3 \mathfrak{v}^{(1)}_{a}\mathcal V^{i}_{(a)} \bigg] ~~,\\
{T_{(b)}}^{ij} =& \sum_{a=1}^5\mathcal S^{\text{red}}_{(a)} \bigg\{\mathfrak{s}^{(2)}_{a} g^{ij}-\mathfrak{s}^{(3)}_{a} \zeta^i \zeta^j  \bigg \} + \sum_{a=1}^3\mathfrak{v}^{(2)}_{a} \zeta^{(i} \mathcal V^{j)}_{(a)}~~, \\
{J_{(b)}}_{0} =& - e^{\sigma} \sum_{a=1}^5 \mathfrak{s}^{(5)}_{a}  \mathcal
S^{\text{red}}_{(a)}~~, \qquad
{J_{(b)}}^{i} =  \sum_{a=1}^5 \mathfrak{s}^{(6)}_{a} \zeta^i  \mathcal S^{\text{red}}_{(a)} + \sum_{a=1}^3 \mathfrak{v}^{(3)}_{a}\mathcal V^{i}_{(a)}~~.\\
\end{split}
\end{equation}

Comparing the components of the energy-momentum tensor and current in Eq.\eqref{po2p1covstcred} with the
corresponding expressions in \eqref{po2p1blkstct00}--\eqref{po2p1blkstcji} that follow from the
partition function, leads to the following four sets of relations among transport coefficients
\begin{equation} \label{po2p1blkrel1}
\begin{split}
&\mathfrak{s}^{(2)}_{a} = 0,~ (a=1,2,3,4,5),~~ \mathfrak{v}^{(2)}_{a} = 0,~ (a=1,2,3), \\
& \mathfrak{v}^{(3)}_{3} =- {\partial m_B \over \partial \chi},~~
\mathfrak{s}^{(6)}_{2} = {\partial m_B \over \partial \chi},~~ \mathfrak{s}^{(3)}_{2} = {\partial m_B \over \partial \chi},~~ \mathfrak{s}^{(4)}_{2} = 2 {\nu} {T}{\partial m_B \over \partial \chi},~~\mathfrak{s}^{(6)}_{5} = -{\mathfrak{v}^{(3)}_{3} \over |\zeta|^2}, \\ 
&\mathfrak{s}^{(3)}_{5} = {\mathfrak{v}^{(3)}_{2} \over |\zeta|^2},~~\mathfrak{s}^{(4)}_{5} = {\mathfrak{v}^{(1)}_{3} \over |\zeta|^2},~~ \mathfrak{s}^{(6)}_{1} = -2 e^{-\sigma}\left({\partial m_{\omega} \over \partial \chi} -A_0{\partial m_B \over \partial \chi}\right),\\
&\mathfrak{s}^{(3)}_{1} = -2 e^{-\sigma}\left({\partial m_{\omega} \over \partial \chi} -A_0{\partial m_B \over \partial \chi}\right), ~~ \mathfrak{v}^{(1)}_{3} = 2 e^{-\sigma}\left({\partial m_{\omega} \over \partial \chi} -A_0{\partial m_B \over \partial \chi}\right)~~,\\
\end{split}
\end{equation}
\begin{equation}\label{po2p1blkrel1_2}
\begin{split}
&\mathfrak{s}^{(4)}_{1} =  4{\nu}{T} e^{-\sigma}\left({\partial m_{\omega} \over \partial \chi} -A_0{\partial m_B \over \partial \chi}\right),
~~ \mathfrak{s}^{(1)}_{5} = {\partial \beta_1 \over \partial \chi},~~ \mathfrak{s}^{(1)}_{3} =- 2 {T} {\nu}^2 {\partial \beta_1 \over \partial \chi}, \\
& \mathfrak{s}^{(3)}_{3} ={2 \over {T}} {\partial \beta_1 \over \partial \chi},~~\mathfrak{s}^{(4)}_{3}-{\mathfrak{v}^{(1)}_{1} \over |\zeta|^2}=4{\nu} {\partial \beta_1 \over \partial \chi}~~,~~ \mathfrak{s}^{(6)}_{3}+{\mathfrak{v}^{(3)}_{1} \over |\zeta|^2}={2 \over {T}} {\partial \beta_1 \over \partial \chi},\\
& \mathfrak{s}^{(5)}_{4} = 2{\nu} {T} {\partial (T_0\beta_2) \over \partial \chi},~~ \mathfrak{s}^{(5)}_{5} = {1 \over {T}} {\partial (T_0\beta_2) \over \partial \chi},~~ \mathfrak{s}^{(3)}_{4} = -2{\partial (T_0\beta_2) \over \partial \chi},\\
& \mathfrak{s}^{(6)}_{4}+{\mathfrak{v}^{(3)}_{2} \over |\zeta|^2}=-2{\partial (T_0\beta_2) \over \partial \chi},~~\mathfrak{s}^{(4)}_{4}+{\mathfrak{v}^{(1)}_{2} \over |\zeta|^2}=4{\nu}{T} {\partial (T_0\beta_2) \over \partial \chi}, \mathfrak{v}^{(3)}_{2} = T_0 \beta_2 + {\partial m_B \over \partial {\nu}}~~,\\
\end{split}
\end{equation}
\begin{equation}\label{po2p1blkrel1_3}
\begin{split}
& \mathfrak{s}^{(5)}_{2} = {1 \over 2{T}} \left[ T_0 \beta_2 + {\partial m_B \over \partial {\nu}}+2 {\nu} {T}^2 {\partial m_B \over \partial \chi}\right],~~\mathfrak{v}^{(3)}_{1} = -{1 \over {T}} \left( \beta_1 +m_B - {T} {\partial m_B \over \partial {T}} \right),\\
& \mathfrak{s}^{(1)}_{2} = {1 \over 2} \left[\beta_1+m_B - {T} {\partial m_B \over \partial {T}}-2 \nu^2 {T}^2 {\partial m_B \over \partial \chi}\right],\\
& \mathfrak{s}^{(5)}_{1} = -{e^{-\sigma} \over {T}} \bigg[ \left({\partial m_{\omega} \over \partial {\nu}} -A_0 {\partial m_{B} \over \partial {\nu}}\right) +2 {\nu} {T}^2\left({\partial m_{\omega} \over \partial \chi} -A_0{\partial m_B \over \partial \chi}\right)\bigg]-{\nu} T_0 \beta_2, \\
&\mathfrak{s}^{(1)}_{1} = {\nu} \beta_1-e^{-\sigma} \bigg[ (m_{\omega} -A_0 m_B)-{T}\left({\partial m_{\omega} \over \partial {T}} -A_0 {\partial m_{B} \over \partial {T}}\right)\\ &~~~~~~~~- 2 {\nu}^2 {T}^2\left({\partial m_{\omega} \over \partial \chi} -A_0{\partial m_B \over \partial \chi}\right)\bigg]~~,\\
\end{split}
\end{equation}
\begin{equation}\label{po2p1blkrel1_4}
\begin{split}
&\mathfrak{s}^{(5)}_{3} =-{1 \over {T}^2} \left[ {\partial\beta_1 \over \partial {{\nu}}} -T_0 \beta_2 +{T} {\partial (T_0 \beta_2) \over \partial {T}} +2{\nu} {T}^2 {\partial \beta_1 \over \partial \chi}\right], \\
&\mathfrak{s}^{(1)}_{4} = {\partial\beta_1 \over \partial {{\nu}}} -T_0 \beta_2 +{T} {\partial (T_0 \beta_2) \over \partial {T}} +2{\nu}^2 {T}^2 {\partial (T_0 \beta_2) \over \partial \chi},\\
&\mathfrak{v}^{(1)}_{1} ={2 \over {T}} \left[(m_{\omega} -A_0 m_B) - {T}\left({\partial m_{\omega} \over \partial {T}} -A_0 {\partial m_{B} \over \partial {T}}\right) \right]-2 {\nu} \beta_1,\\
&\mathfrak{v}^{(1)}_{2} = 2{\nu} {T} (T_0 \beta_2) - 2 e^{-\sigma} \left({\partial m_{\omega} \over \partial {\nu}} -A_0 {\partial m_{B} \over \partial {\nu}}\right)~~.
\end{split}
\end{equation}
Eqs.\eqref{po2p1blkrel1}-\eqref{po2p1blkrel1_4} provide a total of $39$ relations among the transport coefficients. We will now eliminate the coefficients $m_{\omega},~m_B,~\beta_1, ~\beta_2$ from the relations in Eqs.\eqref{po2p1blkrel1}-\eqref{po2p1blkrel1_4} in order to obtain $35$ independent relations among the remaining $35$ transport coefficients present in Eq.\eqref{po2p1covstc}. These relations are listed below
\begin{equation} \label{po2p1blkrel2_1}
\begin{split}
&\mathfrak{s}^{(2)}_{a} = 0~ (a=1,2,3,4,5),~~ \mathfrak{v}^{(2)}_{a} = 0~ (a=1,2,3),~~ \mathfrak{v}^{(3)}_{3} =-\mathfrak{s}^{(6)}_{2},~~\mathfrak{s}^{(3)}_{2} =\mathfrak{s}^{(6)}_{2},\\
& \mathfrak{s}^{(4)}_{2} = 2 {\nu} {T}\mathfrak{s}^{(6)}_{2},~~\mathfrak{s}^{(6)}_{5} =- {\mathfrak{v}^{(3)}_{3} \over |\zeta|^2}, ~~ \mathfrak{s}^{(3)}_{5} = {\mathfrak{v}^{(3)}_{2} \over |\zeta|^2},~~\mathfrak{s}^{(4)}_{5} = {\mathfrak{v}^{(1)}_{3} \over |\zeta|^2},~~\mathfrak{s}^{(3)}_{1} = \mathfrak{s}^{(6)}_{1},\\
&  \mathfrak{v}^{(1)}_{3} = \mathfrak{s}^{(6)}_{1}, ~~\mathfrak{s}^{(4)}_{1} =  -2 {\nu}{T} \mathfrak{s}^{(6)}_{1}, ~~\mathfrak{s}^{(1)}_{3} =-2 {T} {\nu}^2\mathfrak{s}^{(1)}_{5},~~ \mathfrak{s}^{(3)}_{3} ={2 \over {T}}\mathfrak{s}^{(1)}_{5},\\
&\mathfrak{s}^{(4)}_{3}-{\mathfrak{v}^{(1)}_{1} \over |\zeta|^2}=4\nu \mathfrak{s}^{(1)}_{5},~~\mathfrak{s}^{(6)}_{3}+{\mathfrak{v}^{(3)}_{1} \over |\zeta|^2}={2 \over {T}}\mathfrak{s}^{(1)}_{5},~~ \mathfrak{s}^{(5)}_{4} = 2 \nu {T}^2  \mathfrak{s}^{(5)}_{5},~~ \mathfrak{s}^{(3)}_{4} = -2 {T} \mathfrak{s}^{(5)}_{5}~~, \\
\end{split}
\end{equation}
\begin{equation} \label{po2p1blkrel2_2}
\begin{split}
& \mathfrak{s}^{(6)}_{4}+{\mathfrak{v}^{(3)}_{2} \over |\zeta|^2}=-2{T} \mathfrak{s}^{(5)}_{5},~~ \mathfrak{s}^{(4)}_{4}+{\mathfrak{v}^{(1)}_{2} \over |\zeta|^2}=4{\nu}{T}^2 \mathfrak{s}^{(5)}_{5},~~\mathfrak{s}^{(5)}_{2} = {1 \over 2{T}}\left[\mathfrak{v}^{(3)}_{2}+2 {\nu} {T}^2\mathfrak{s}^{(6)}_{2} \right]~~,  \\
& \mathfrak{s}^{(1)}_{2} =- {{T} \over 2} \left[ \mathfrak{v}^{(3)}_{1} +2 {\nu} {T}\mathfrak{s}^{(6)}_{2} \right],~~  \mathfrak{s}^{(1)}_{1} =- {{T} \over 2} \mathfrak{v}^{(1)}_{1}+ {\nu}^2 {T}^2\mathfrak{s}^{(6)}_{1},~~  \mathfrak{s}^{(5)}_{1} = -{\mathfrak{v}^{(1)}_{2}\over 2{T} }  + {\nu} {T}\mathfrak{s}^{(6)}_{1}, \\
& \mathfrak{s}^{(1)}_{4} = {T}^2 \left[2 {\nu}^2 {T}\mathfrak{s}^{(5)}_{5}-\mathfrak{s}^{(5)}_{3}-2{\nu}  \mathfrak{s}^{(1)}_{5}  \right],~~\mathfrak{s}^{(5)}_{5} = {1 \over {T}}\left({\partial \mathfrak{v}^{(3)}_{2}  \over \partial \chi} - {\partial \mathfrak{s}^{(6)}_{2} \over \partial {\nu}}\right)~~, \\
\end{split}
\end{equation}
\begin{equation} \label{po2p1blkrel2_3}
\begin{split}
& \mathfrak{v}^{(3)}_{2} =  {T} {\partial\mathfrak{v}^{(3)}_{2} \over \partial {T}}+ {T}^2 \mathfrak{s}^{(5)}_{3} -{T} {\partial \mathfrak{v}^{(3)}_{1}  \over \partial {\nu}}+2 {\nu} {T}^2 \mathfrak{s}^{(1)}_{5}, \\
& \mathfrak{s}^{(1)}_{5} = -\left[ \mathfrak{s}^{(6)}_{2} - {T} {\partial \mathfrak{s}^{(6)}_{2} \over \partial{T}} + {T} {\partial\mathfrak{v}^{(3)}_{1} \over \partial \chi} \right], ~~ \mathfrak{s}^{(6)}_{2} =  {1 \over 2{T}} \left[{\partial \mathfrak{s}^{(6)}_{1} \over \partial {\nu}} - {\partial\mathfrak{v}^{(1)}_{2} \over \partial \chi} \right]- {\nu} {T} \mathfrak{s}^{(5)}_{5}, \\
& \mathfrak{s}^{(6)}_{1} =-{{T} \over 2} \left[ {\partial\mathfrak{v}^{(1)}_{1} \over \partial \chi} + 2 {\nu}\mathfrak{s}^{(1)}_{5} - {\partial \mathfrak{s}^{(6)}_{1} \over \partial{T}}   \right]~~.
\end{split}
\end{equation}
As expected, the remaing 35 transport coefficients are determined in terms of the four coefficients $m_\omega, m_B, \beta_1, \beta_2$.

\subsubsection{Bulk parity-odd effects on the surface currents} \label{sec:con2+1surf}
Following the same strategy as in the previous section, here we derive constraints on the covariant
form of the surface energy-momentum tensor and charge current. These constrains are implied by
\eqref{po2p1surstc}, which in turn follows from the partition function \eqref{po1stord2p1sup}.

The only parity-odd scalar we can write at the surface is actually
$\bar\l = \e^{\mu\nu\r}n_\mu u_\nu \xi_\r$, which upon dimensional reduction is equal to
$\bar\l = \e^{ij} n_i \z_j$. On the other hand, there are no new independent parity-odd vectors or
tensors. The reason being that at the surface we are supposed to write tensor structures transverse
to all three of $u^\mu$, $\xi^\mu$ and $n^\mu$, and in $2+1$ dimensions, there are no such possible
tensors structures. Another way to see this is that in $2+1$ dimensions, any vector or tensor can be
expressed in terms of a chosen basis of three vectors, which we naturally have at the surface as
$u^\mu$, $\xi^\mu$ and $n^\mu$.  Therefore, there can be no other vectors or tensors in $2+1$
dimensions, which are not their linear combinations. For example,
$\cG^{\mu\nu} = -u^\mu u^\nu + \frac{1}{|\z|^2}\tilde\z^\mu \tilde\z^\nu + n^\mu n^\nu$, where
$\tilde\z^\mu = \xi^\mu + (u^\nu \xi_\mu) u^\mu - (n^\nu \xi_\nu) n^\mu $. Having said that, in this
sector (the sector of parity-odd transport flowing in from the bulk) we work with an alternate
basis, $u^\mu$, $n^\mu$ and $\bar n^\mu = \e^{\mu\nu\r} u_\nu n_\r$, i.e. exchanging $\xi^\mu$ for
$\bar n^\mu$. This basis is more appropriate because it is simultaneously valid for the ``inflow''
from the exterior ordinary fluid, where $\xi^\mu = 0$. Note that now,
$\cG^{\mu\nu} = -u^\mu u^\nu + \bar n^\mu \bar n^\nu + n^\mu n^\nu$.

Following our discussion above, the most general parity-odd ideal order surface energy-momentum
tensor and charge current allowed by symmetries for $2+1$ dimensional superfluid bubbles is given
as
\begin{align}\label{po2p1covstcsur}
  T_{(s)}^{\mu\nu}
  &= \bigg[ \fa_1~ u^{\mu}u^{\nu} + \fa_2 \bar n^\mu \bar n^\nu
    + \fa_3~n^{\mu}n^{\nu}
  + 2\fa_4~ u^{(\mu}n^{\nu)} \bigg] \bar\l
  + 2\fb_1 ~ u^{(\mu} \bar n^{\nu)} + 2\fb_2 ~n^{(\mu} \bar n^{\nu)}~~,\nn\\
  J_{(s)}^{\mu} &= \bigg[ \fa_5~u^{\mu} + \fa_6~ n^{\mu}\bigg]
  \bar\l + \fb_3 ~ \bar n^\mu~~,
\end{align}
where all the transport coefficients are parity-even, i.e. they do not have $\bar\l$ dependence. The
energy-momentum tensor and charge current \eqref{po2p1covstcsur} are parametrized by a total of 9
transport coefficients. As we will see below, the partition function \eqref{po1stord2p1sup} will
give 5 relations among these 9 coefficients, while determining the other 4 in terms of the four
parameters appearing in \eqref{po1stord2p1sup}. It is straightforward to compare the surface
energy-momentum tensor and charge current \eqref{po2p1covstcsur} with that in \eqref{po2p1surstc}
that follow from the partition function. This comparison leads to the following relations 9
relations
\begin{equation} \label{po2p1surrel}
\begin{split}
  & \fa_1 = \beta_1~~, \quad \fa_5 = - {T_0 \over {T}} \b_2 ~~, \quad \fb_1 = \frac{T}{T_0}
  (m_{\omega}-A_0 m_B)~~, \quad \fb_3 = -m_B~~, \\
  & \fa_2 = \fa_3 = \fa_4 = \fa_6 = \fb_2 = 0~~.
\end{split}
\end{equation}
As expected, all surface transport coefficients are determined in terms of the four coefficients
$m_{\omega}$, $m_B$, $\beta_1$ and $\beta_2$ that appear in the partition function.
\eqref{po1stord2p1sup}.

Finally, before concluding this section, we would like to point out that instead of using the partition function \eqref{po2p1covstcsur} written in the reduced language in two dimensions, we could have used  its covariant version in $2+1$ dimensions, which takes the form
\begin{multline}\label{po1stord2p1supcov}
  W^{(1)}_{odd} = \int d^3x \sqrt{-\mathcal G} ~\theta(f) \bigg( \kappa_1 ~
  \epsilon^{\mu\nu\lambda} u_{\mu} \omega_{\nu\lambda}
  + \half \kappa_2 ~\epsilon^{\mu\nu\lambda} u_{\mu} \mathcal{F}_{\nu\lambda} \\
  + \kappa_3 ~ \epsilon^{\mu\nu\lambda} \zeta_{\mu} u_{\nu} \frac{1}{T} \partial_{\lambda} {T} +
  \kappa_4 ~ \epsilon^{\mu\nu\lambda} \zeta_{\mu} u_{\nu} T\partial_{\lambda} {\nu} \bigg)~~,
\end{multline}
where the coefficients $\kappa_i~,~i=1,2,3,4$ are functions of $T,\mu,\chi$.
Once we reduce this covariant partition function on the time circle and compare it with \eqref{po2p1covstcsur} we readily identify
\begin{equation}
  \k_1 = \frac{T}{T_0}(m_{\omega}-A_0 m_B) ,
  \qquad \k_2 = - m_B  ,
  \qquad \kappa_3 = - \beta_1,
  \qquad \kappa_4 = \frac{T_0}{T} \beta_2~~ .
\end{equation}
These relations can be inverted in order to express the surface coefficients \eqref{po2p1surrel} in terms of the coefficients $\kappa_i$ leading to the identifications
\begin{equation} \label{po2p1surrela}
  \fa_1 = - \kappa_3~~, \qquad
  \fa_5 = - \kappa_4 ~~, \qquad
  \fb_1 = \kappa_1~~, \qquad  \fb_3 = \kappa_2~~,
\end{equation}
while the remaining transport coefficients must vanish.

\subsubsection{Surface currents and thermodynamics} \label{sec:thermo2+1}
In this section we combine the surface contributions from both the parity-even sector \eqref{stcur3p1} (by means of the coefficients $\alpha_1$ and $\alpha_2$) and the parity-odd sector \eqref{po2p1surstc}, including the effects of the exterior charged fluid partition function \eqref{po1stord2p1ord}, which are accounted for by replacing $m_\omega\to m_\omega-M_\omega$ and
$m_B\to m_B-M_B$ in \eqref{po2p1surstc}. Using \eqref{po2p1covstcsur}, \eqref{comp2+1} and \eqref{persupcovcurs_der}, the surface energy-momentum tensor, charge current and entropy current for 2+1 superfluid bubbles takes the form
\begin{equation}\label{persupcovcurs2p1aa}
\begin{split}
  T^{\mu \nu}_{(s)} &= (\mathcal E - \cY) ~u^{\mu} u^\nu - \mathcal Y ~(\cG^{\mu\nu} - n^\mu n^\nu)
  +\mathcal{F}\tilde\xi^{\mu}\tilde \xi^{\nu}
  + 2 \mathcal \cU  u^{(\mu}\bar n^{\nu)} ~~,\\
  J^{\mu}_{(s)} &= \mathcal Q ~u^{\mu} - {\mathcal F}\tilde\xi^{\mu}
  + \cV \bar n^\mu ~~,\\
  J^{\mu}_{(s)\text{ent}} &= \mathcal{S} u^{\mu} + \frac{1}{T} \lb \cU - \mu \cV \rb \bar n^\mu~~,
\end{split}
\end{equation}
where $\bar n^\mu = \epsilon^{\mu \rho \lambda} u_{\rho} n_{\lambda}$. After imposing the on-shell
condition $\partial\mathcal{C}/\partial\lambda=0$, the various transport coefficients are given
by\footnote{One can easily obtain the contribution of the coefficients $\k_3$ and $\k_4$ to the
  surface entropy density $\mathcal{S}$ by performing a variation of the integrand of
  \eqref{po1stord2p1supcov} with respect to $T$.}
\begin{align} \label{eq:thermo2+1rel}
    \cE &= - \cC + T \frac{\dow \cC}{\dow T} + \mu \frac{\dow \cC}{\dow \mu}
    + \frac{\mu^2 }{\bar\lambda}\frac{\partial \mathcal{C}}{\partial\bar\lambda}
    + (\l f_1 - \bar\l \k_3) , \qquad
    \cY = -\mathcal{C}, \qquad 
    \cF = -\frac{1}{\bar\lambda}\frac{\partial \mathcal{C}}{\partial\bar\lambda}~~, \nn\\
    \cQ &= \frac{\partial\mathcal{C}}{\partial\mu}
    + \mu \frac{1}{\bar\lambda}\frac{\partial \mathcal{C}}{\partial\bar\lambda}
    + (\l\alpha_2 - \bar\l\k_4), \qquad
    \mathcal{S} = \frac{\partial\mathcal{C}}{\partial T}
    + \frac{\lambda}{T}(\alpha_1-\mu\alpha_2)
    - \frac{\bar\lambda}{T}(\k_3 - \mu \k_4), \nn\\
    \cU &= \frac{T}{T_0}(m_\omega-M_\omega)-\mu (m_B-M_B)= \kappa_1, \qquad
    \cV = -(m_B-M_B)= \kappa_2~~.
\end{align}
These relations in turn imply the Gibbs-Duhem and Euler relations of thermodynamics at the surface
respectively
\begin{align}
  \df \cY
  &= - \Big( \cS - \frac{\l}{T}(f_1 - \mu f_2)
    + \frac{\bar\l}{T} (\k_3 - \mu \k_4) \Big) \df T
    - \Big(\cQ - \l f_2 + \bar\l\k_4 \Big) \df \mu
    - \mathcal{F}(\mu \df\mu-\bar\lambda \df\bar\lambda) ~~, \nn\\
  \mathcal{E}-\mathcal{Y}
  &= T\mathcal{S}+\mu\mathcal{Q}~~.
\end{align}
The respective first law of thermodynamics has been discussed in appendix \ref{app:2+1thermo}.

We see that the surface thermodynamics of $2+1$ dimensional superfluid bubbles is comparatively richer
than their $3+1$ dimensional counterparts. In particular, not only does the parity-even coefficients
$f_1,f_2$ directly affect the surface thermodynamics as in the $3+1$ the dimensional case, but even
the parity-odd coefficients $\k_3,\k_4$ have an effect, in exactly the same way as the coefficients
$f_1,f_2$ do.

\section{Galilean stationary superfluid bubbles in 3+1 dimensions} \label{sec:nulsf}
In this section, we analyze the surface currents for stationary bubbles of a $3+1$ dimensional
Galilean superfluid immersed in an ordinary fluid. Any appropriately defined non-relativistic limit
of the relativistic currents worked out in \S \ref{sec:3plus1}, should be a special case of Galilean
superfluids. In this sense, Galilean superfluids can provide us with a general understanding of the
respective non-relativistic physics.  As in the relativistic case, our primary focus here will be
the surface currents. A complete analysis of the bulk currents in this case has already been
provided in \cite{Banerjee:2016qxf}.

Our basic setup has been thoroughly described in \S \ref{ssec:intronull}. At first, we have to work
out the constitutive relations for a null superfluid in $4+1$ dimensions using an equilibrium
partition function, and then perform a null reduction on it in order to obtain the Galilean results.
We shall report the results of this section in a slightly different notation compared to the
relativistic case, so as to be closer to those usually used in the non-relativistic superfluid
literature.  Let us define the superfluid potential
$\mu_s = -\half \xi^\sM \xi_\sM = -\mu + \mu_n + \hat\mu_s$, where $\hat\mu_s = - \half \z^k \z_k$,
in addition to the usual zero derivative scalars: temperature $T = \E{-\galsigma}T_0$, chemical
potential $\mu = \E{-\galsigma} A_0$ and mass chemical potential $\mu_n = \E{-\galsigma} B_0$. We
will denote $\nu = \mu/T$ and $\nu_n= \mu_n/T$. It will also be useful to define a boundary
superfluid velocity projected on the surface, $\tilde\xi^\sM = (\cG^{\sM\sN} - n^\sM n^\sN)\xi_\sN$
and an associated potential,
$\tilde\mu_s = -\half \tilde\xi^\sM \tilde\xi_\sM = -\mu + \mu_n + \hat\mu_s + \half \l^2$ with
$\l = n_\sM \xi^\sM = n_i\z^i$, as before.

Up to first order in the bulk and ideal order at the surface, the partition function can be written
in terms of the shape-field and background data \eqref{nullpfdata} as follows
\begin{multline}\label{eqbPF_full_null}
  W = \int \df^3 x\sqrt{g} \ \q(f) \frac{\E{\galsigma}}{T_0}  \bigg(
  P_{(b)}
  - f_1 \z^i\dow_i \galsigma
  + \E{-\galsigma} f_2 \z^i\dow_i A_0
  + \E{-\galsigma} f_3 \z^i\dow_i B_0 \\
  + T_0\E{-\galsigma} (g_1 + g_2) \e^{ijk} \z_i \dow_j B_k
  + T_0\E{-\galsigma} g_2 \e^{ijk} \z_i \dow_j A_k
  + T_0(g_1 \E{-\galsigma} B_0 + g_2 \E{-\galsigma} A_0 - g_3 ) \e^{ijk} \z_i \dow_j a_k \bigg)  \\
  + \int \df^3 x \sqrt{g} \ \tilde\d(f) \frac{\E{\galsigma}}{T_0} \cC
  + \int \df^3 x \sqrt{g} \ \q(-f) \frac{\E{\galsigma}}{T_0} P_{(e)}~~.
\end{multline}
In order to obtain the thermodynamics in the conventional notation, we consider
$P = P \left(T,\mu, \mu_n ,\mu_s\right)$, while the rest of the bulk transport coefficients $f_i$
and $g_i$ are considered to be functions of $\{T, \nu, \nu_n , \hat \mu_s\}$.  On the other hand,
for the surface tension we consider
$\mathcal C = \mathcal C \left( T, \mu, \mu_n , \tilde \mu_s, \lambda \right)$.  Since outside the
bubble there is an ordinary fluid, it cannot depend on the superfluid variables $\z_i$ or $\mu_s$,
leading to no possible terms which can be written at first order. Furthermore,
$P_{(e)} = P_{(e)}(T,\mu,\mu_n)$ is independent of $\mu_s$.

We start with the $\phi$ equations of motion obtained by varying the partition function \eqref{eqbPF_full_null} with respect to $\phi$
\begin{align}\label{phi.eom.null_1}
  \q(f)
  &\bigg[
    \Df_i \bigg( \frac{1}{T}\frac{\dow P_{(b)}}{\dow\mu_s} \z^i
  + \frac{1}{T^2}\frac{\dow f_1}{\dow \hat\mu_s} \z^i \z^j\dow_j T
  - \frac{1}{T^2} f_1 g^{ij} \dow_j T 
  + \frac{\dow f_2}{\dow \hat\mu_s} \z^i \z^j\dow_j \nu
  - f_2 g^{ij} \dow_j \nu \nn\\
  &\qquad + \frac{\dow f_3}{\dow \hat\mu_s} \z^i \z^j \dow_j \nu_n
  - f_3 g^{ij} \dow_i\nu_n
  + \frac{\dow (g_1 + g_2)}{\dow \hat\mu_s} \z^i \e^{ajk} \z_a \dow_j B_k
  - (g_1 + g_2) \e^{ijk} \dow_j B_k \nn\\
  &\qquad + \frac{\dow g_2}{\dow \hat\mu_s} \z^i \e^{ajk} \z_a \dow_j A_k
  - g_2 \e^{ijk} \dow_j A_k
  + \E{\galsigma}\lb \frac{\dow g_1}{\dow\hat\mu_s} \mu_n + \frac{\dow g_2}{\dow\hat\mu_s} \mu -
  \frac{\dow g_3}{\dow\hat\mu_s} \rb \z^i \e^{ajk} \z_a \dow_j a_k \nn\\
  &\qquad - \E{\galsigma}(g_1 \mu_n + g_2 \mu - g_3 ) \e^{ijk} \dow_j a_k \bigg)
    \bigg]~,
\end{align}
\begin{align}\label{phi.eom.null_2}    
  \tilde\d(f)
  &\bigg[
    T \tilde\Df_i \lb \frac{1}{T} \frac{\dow \cC}{\dow\l} n^i
    - \frac{1}{T} \frac{\dow \cC}{\dow\tilde\mu_s} \tilde\z^i
    \rb
    + \l \frac{\dow P_{(b)}}{\dow\mu_s}
  + \frac{1}{T}\frac{\dow f_1}{\dow \hat\mu_s} \l \z^i\dow_i T
  - \frac{1}{T} f_1 n^i \dow_i T 
  + T \frac{\dow f_2}{\dow \hat\mu_s} \l \z^i\dow_i \nu  \nn\\
  &\qquad 
    - T f_2 n^i \dow_i \nu
    + T \frac{\dow f_3}{\dow \hat\mu_s} \l \z^i\dow_i \nu_n
    - T f_3 n^i \dow_i\nu_n
  + T \frac{\dow (g_1 + g_2)}{\dow \hat\mu_s} \l \e^{ijk} \z_i \dow_j B_k  \nn\\
  &\qquad - T (g_1 + g_2) \e^{ijk} n_i \dow_j B_k
  + T \frac{\dow g_2}{\dow \hat\mu_s} \l \e^{ijk} \z_i \dow_j A_k
  - T g_2 \e^{ijk} n_i \dow_j A_k  \nn\\
  &\qquad + T_0 \lb \frac{\dow g_1}{\dow\hat\mu_s} \mu_n + \frac{\dow g_2}{\dow\hat\mu_s} \mu -
  \frac{\dow g_3}{\dow\hat\mu_s} \rb \l \e^{ijk} \z_i \dow_j a_k
  - T_0(g_1 \mu_n + g_2 \mu - g_3 ) \e^{ijk} n_i \dow_j a_k
    \bigg], \nn\\
  \tilde\d'(f) & \bigg[ \frac{\dow \cC}{\dow\l} \bigg] = 0~~.
\end{align}
Taking a variation of the partition function \eqref{eqbPF_full_null} and using the variational
formulae \eqref{null_variational}, we can read off the surface currents (for a discussion on the
bulk currents see \cite{Banerjee:2016qxf}), after using the $\tilde\d'(f)$ equation of motion of
$\phi$
\begin{equation}\nn
  {T_{(s)}}_{--} = \cR_{n} + \cR_{s}, \qquad
  {T_{(s)}}^i_{\ -} = - g_1 \bar n^i  - \cR_{s} \tilde\z^i, \qquad
  {T_{(s)}}^{ij} = - h^{ij} \cY + R_{s,\dow} \tilde\z^i\tilde\z^j,
\end{equation}
\begin{equation}\nn
  {T_{(s)}}_{0-} = \E{\galsigma} \lb \cE - \mu_n \cR_{n} \rb  - \xi_0 \cR_{s}, \qquad
  {T_{(s)}}^i_{\ 0} = - \E{\galsigma} \lb g_3 - g_1 \mu_n \rb \bar n^i + \xi_0 \cR_{s} \tilde\z^i,
\end{equation}
\begin{equation}\label{noncov.curr.null}
  {J_{(s)}}_- = - \cQ + \cR_{s}, \qquad
  {J_{(s)}}^i = - \cR_{s} \tilde\z^i + g_2 \bar n^i~~,
\end{equation}
where $h^{ij} = g^{ij} - n^i n^j$, $\tilde\z^i =h^{ij}\z_j$, $\bar n^i = T\e^{ijk} \z_j n_k$, and we
have defined the surface first law of thermodynamics and the Euler relation
\begin{align}\label{thermo_null}
  \df \Big(\cE - \l f_1 \Big)
    &= T \df \Big( \cS - \frac{\l}{T} (f_1 - \mu f_2 - \mu_n f_3) \Big)
      + \mu_n \df\Big( \cR_n - \l f_3 \Big)
      + \mu\df \Big(\cQ - \l f_2 \Big)
      - \cR_{s} \df \tilde\mu_s~~, \nn\\
  \cE - \cY &= T \cS + \mu \cQ + \mu_n \cR_{n}~~,
\end{align}
where $\cY = -\cC$ is the surface tension.  Finally, the equation of motion of $f$ yields the
Young-Laplace equation for Galilean/null superfluids
\begin{multline}
  P_{(b)} - P_{(e)} + \frac{1}{T} f_1 \z^i\dow_i T + T f_2 \z^i\dow_i \nu + T f_3 \z^i\dow_i \nu_n +
  T (g_1 + g_2) \e^{ijk} \z_i \dow_j B_k
  + T g_2 \e^{ijk} \z_i \dow_j A_k \\
  + T_0(\mu_ng_1 + \mu g_2 - g_3 ) \e^{ijk} \z_i \dow_j a_k + T \Df_i \lb \frac{1}{T} \cC n^i +
  \frac{\l}{T} \cR_s \tilde\z^i \rb = 0~~.
\end{multline}
The same equation can also be obtained by projecting the surface energy-momentum conservation
equation along $n_\sM$ (see appendix \ref{app:YLeomf}).  After properly covariantizing the
expressions \eqref{noncov.curr.null}, and using a hydrodynamic frame suitable for the equilibrium
partition function
\begin{equation}
  u^\sM = \E{-\galsigma} (B_0, 1, 0, 0 , 0), \qquad
  T = \E{-\galsigma} T_0, \qquad
  \mu = \E{-\galsigma} A_0, \qquad
  \mu_n = \E{-\galsigma} B_0~~,
\end{equation}
we have the surface currents
\begin{align}\label{nullconrel}
  T^{\sM\sN}_{(s)}
  &= \cR_{n} u^\sM u^\sN + 2 (\cE{-}\cC) u^{(\sM} V^{\sN)} - \cC (\cG^{\sM\sN}{-}n^\sM n^\sN) + \cR_{s}
    \tilde\xi^{\sM} \tilde\xi^{\sN} + 2 g_1 u^{(\sM} \bar n^{\sN)} + 2g_3 V^{(\sM} \bar n^{\sN)},
    \nn\\
  J^\sM_{(s)} &= \cQ u^\sM - \cR_{s} \tilde\xi^\sM + g_2 \bar n^\sM,
\end{align}
where $\bar n^\sM = T\e^{\sM\sN\sR\sS\sT}V_\sN u_\sR \xi_\sS n_\sT$.  The respective thermodynamics
is given by \eqref{thermo_null}.  Note that the most generic form of the constitutive relations at
ideal order (transverse to $n^\sM$) could have contained three more terms proportional to
$u^{(\sM} \tilde\xi^{\sN)}$, $V^{(\sM} \tilde\xi^{\sN)}$ and $\tilde\xi^{(\sM} \bar n^{\sN)}$ in the
energy-momentum tensor, making a total of 12 independent terms. The equilibrium partition function
fixes these 12 coefficients in terms of a boundary function $\cC$ and 6 first order bulk
coefficients $f_i$, $g_i$.

Finally, upon performing the null reduction, the leading order surface currents and densities for a
$3+1$ dimensional Galilean superfluid can be obtained as
\begin{subequations}\label{galleadcurs}
\begin{align}
  \text{Mass Density:}\quad &\r_{(s)} = \cR_{n} + \cR_{s}, \\
  \text{Mass Current:}\quad & \r^i_{(s)} = \cR_{n} u^i + \cR_{s} \tilde\xi^i + g_1 \bar n^i,  \\
  \text{Stress Tensor:}\quad & t^{ij}_{(s)} = \cR_{n} u^i u^j - \cY h^{ij} + \cR_{s} \tilde\xi^i \tilde\xi^j + 2g_1
                               u^{(i}\bar n^{j)}, \\
 \text{Energy Density:}\quad & \e_{(s)} = \cE + \cR_s \tilde \mu_s + \half \cR_{n} u^k u_k + \half \cR_{s} \tilde\xi^k
  \tilde\xi_k + g_1 \bar n^i u_i, \nn \\
 \text{Energy Current:}\quad & \e^i_{(s)} = u^i \lb \cE - \cY + \half \cR_{n} u^k u_k + g_1 \bar n^i u_j \rb
  + \cR_s \tilde \xi^i \lb  \half \tilde\xi^k \tilde\xi_k + \tilde\mu_s \rb \nn\\
  &\qquad + \lb g_3 + \half g_1 u^k u_k \rb  \bar n^i,  \\
 \text{Charge Density:}\quad & q_{(s)} = \cQ - \cR_{s}, \\
 \text{Charge Current:}\quad & q^i_{(s)} = \cQ u^i - \cR_{s} \tilde\xi^i + g_2 \bar n^i.
\end{align}
\end{subequations}
It is interesting to contrast these results with those in the bulk, as reported by
\cite{Banerjee:2016qxf}. Not only there are new terms in the leading order Galilean constitutive
relations, but some of them are parity-odd as well. Furthermore, all these new terms are completely
determined in terms of the first order bulk transport coefficients. In fact, since all the first
order stationary bulk coefficients appear in the surface constitutive relations, they can, in
principle, be measured by performing carefully designed experiments on the surface of the
superfluid.

\section{Surface dynamics}\label{sec:surdyn}
%
In this section, we study the consequences of a non-trivial time dependence of the shape-field
on the surface. Once we relax the assumption of stationarity, we cannot deduce the constitutive
relations of a (super)fluid through an equilibrium partition function, as we did in \S
\ref{sec:3plus1} and \S\ref{sec:2plus1}. Therefore, we have to resort to the second law of
thermodynamics to constrain and understand the full time-dependent dynamics. Hence, we first analyze
the surface entropy current at ideal order in \S \ref{ssec:entLO}, to understand the structure of
the equations governing the surface dynamics. With this understanding, in \S\ref{ssec:linfluc} we
study linearized fluctuations on the surface and its relation with the fluctuations in the bulk,
both for an ordinary fluid and a superfluid.

\subsection{Surface entropy current analysis at zero derivative order}\label{ssec:entLO}
%

\subsubsection{Surface entropy current for ordinary fluids}\label{sssec:udotnzero}
%
Before proceeding to the superfluid case, we study the entropy current and the consequences of the
second law of thermodynamics for ordinary fluids in the presence of a surface. Once we give up the
assumption of stationarity, the first aspect of surface dynamics we would like to understand is what
determines the normal component of the fluid velocity $u^\mu n_\mu$ at the surface. In the
stationary case, this normal component vanishes as $\tilde K^\mu = \E{\sigma} u^{\mu}$ is a Killing
vector field.\footnote{ This simply follows as
  $u^\mu n_\mu \propto \tilde K^\mu \dow_\mu f = \lie_{\tilde K} f = 0 $. Another way to argue this
  is that on the surface we have $d+3$ undetermined variables in $d+1$ dimensions: $T|_{f=0}$,
  $\mu|_{f=0}$, $d$ components of $u^\mu|_{f=0}$ (including $u^\mu n_\mu$) and $f$. Since we only
  have $d+2$ conservation laws, for the system to be solvable, there must be another relation among
  these variables. Later in this section, we will show that the second law of thermodynamics forces
  such a relation to imply $u^\mu n_\mu = 0$ in equilibrium.
  This goes on to show that $u^\mu n_\mu$ should not be treated as an independent thermodynamic
  variable at the surface, as was done in \cite{Kovtun:2016lfw}.} The second aspect of surface
dynamics we would like to understand is what determines the equation of motion for the shape-field
$f$, since it is not clear a priori if the normal component of the surface energy-momentum
conservation continues to serve as a proxy for the equation of motion of $f$ in non-equilibrium
situations. In this section, we will try to answer both these questions and demonstrate that they
are interrelated.

As mentioned above, in the analysis of equilibrium partition functions, $u^\mu n_\mu$ was zero by
construction. In fact, this condition served as one of the boundary conditions for solving the bulk
fluid equations (see \S\ref{sec:intro} and \cite{Armas:2015ssd} for more details). However, as we
move away from stationarity, the status of $u^\mu n_\mu$ is not clear a priori and we need a
principle to determine it. In order to address this problem, it is extremely useful to remember the
analogy between the shape-field $f$ and the superfluid phase $\phi$, both being a consequence of a
spontaneously broken symmetry. Momentarily, if we take this analogy seriously then $u^\mu n_\mu$
would correspond to $u^\mu\xi_\mu$ in the case of superfluids. Now, as we know, $u^\mu \xi_\mu$ is
not an independent variable in superfluid dynamics. In fact, it is given by the chemical potential
$\mu$ \cite{Landau:1987gn} at leading order and receives further corrections at higher orders, as
determined by the second law of thermodynamics \cite{Bhattacharya:2011tra}.  As noted in
\cite{Jain:2016rlz}, the generalized Josephson equation $u^\mu \xi_\mu = \mu + \mu_{diss}$ can be
derived using an entropy current analysis. It was also observed in \cite{Jain:2016rlz} that in
equilibrium, and in a hydrodynamic frame chosen appropriately for equilibrium, the equation
$u^\mu \xi_\mu = \mu + \mu_{diss}$ reduces to $\mu_{diss} = 0$, which can be identified as the
equation of motion for $\phi$ following from the respective equilibrium partition function
\cite{Bhattacharyya:2012xi}. Therefore, the Josephson equation can be thought of as the equation of
motion for $\f$ outside equilibrium. This gives us an important clue for the case of the
shape-field: $u^\mu n_\mu$ should also be determined by the second law of thermodynamics in terms of
other fluid variables, and the respective determining relation should be the equation of motion for
$f$ outside equilibrium. For this purpose, let us define
\begin{equation}
 u^\mu n_\mu = \gamma + \gamma_{\text{diss}}~~,
\end{equation}
where $\gamma$ is the zeroth order value of $u^\mu n_\mu$ and $\gamma_{\text{diss}}$ contains the
higher derivative corrections.  It is definitely possible to choose a hydrodynamic frame where
$ \gamma_{\text{diss}} = 0$, just as it is possible to choose a frame where $\mu_{\text{diss}} = 0$
in the case of superfluids. However, such a frame would not correspond to the more standard frame
choices like the Landau frame, neither would it be a generalization of the equilibrium frame defined
in \S\ref{sec:intro}.

Let us now proceed to analyze the structure of the divergence of the surface entropy current.  The
bulk energy-momentum tensor and entropy current have the well known form
\begin{equation}\label{bulgenordcurs}
\begin{split}
  T^{\mu \nu}_{(b)} &= \left(  E + P \right) u^{\mu} u^{\nu} + P ~\mathcal G^{\mu \nu} + \Pi^{\mu \nu}_{(b)}~~, \\
  J^{\mu}_{(b)\text{ent}} &= S u^\mu + \U^\mu_{(b)\text{ent}}~~, \quad
  \U^\mu_{(b)\text{ent}} = - \frac{u_\nu}{T} \Pi^{\mu \nu}_{(b)} + \U^{\mu}_{(b)\text{new}}~~,
\end{split}
\end{equation}
where $ \Pi^{\mu \nu}_{(b)}$ and $\U^{\mu}_{(b)\text{new}}$ are higher derivative corrections, which
can be found, for example, in \cite{Bhattacharyya:2012nq}.  It is interesting to note that
$\U^{\mu}_{(b)\text{new}}$ does not receive any first order corrections \cite{Bhattacharyya:2012nq}.
On the other hand, the ideal order surface currents are given by \footnote{Note there that we have
  not assumed the tangentiality conditions
  $T^{\mu \nu}_{(s)}n_\mu = {J^{\mu}_{(s)\text{ent}}}n_\mu=0$ on the surface energy-momentum tensor
  and entropy current, since we wish to derive such tangentiality conditions at leading order from
  the entropy current analysis. Furthermore, note that in \eqref{surgenordcurs} we have not
  considered terms of the form $\theta_1u^{(\mu}n^{\nu)}$ and $\theta_2n^{\mu}n^{\mu}$ in the
  surface energy-momentum tensor neither have we consider a term proportional to $\theta_3n^{\mu}$
  in the surface entropy current. In full generality, such terms must be taken into account but for
  clarity of presentation we have not introduced them. In any case, the second law of thermodynamics
  ultimately implies that $\theta_1=\theta_2=\theta_3=0$.}
\begin{equation}\label{surgenordcurs}
\begin{split}
  T^{\mu \nu}_{(s)} &= \left( \cE - \cY \right) u^{\mu} u^{\nu} - \cY \left( \mathcal{G}^{\mu \nu} -n^\mu n^\nu \right) + \Pi^{\mu \nu}_{(s)}~~, \\
  {J^{\mu}_{(s)\text{ent}}} &= \cS u^\mu + \U^\mu_{(s)\text{ent}}~~, \quad
  \U^\mu_{(s)\text{ent}} = - \frac{u_\nu}{T} \Pi^{\mu \nu}_{(s)} + \U^{\mu}_{(s)\text{new}}~~,
\end{split}
\end{equation}
where the $\cY, \cE, \cS$ are the surface tension, energy density and entropy density on the surface
respectively, and $\Pi^{\mu \nu}_{(s)}$, $\U^{\mu}_{(s)\text{new}}$ are higher derivative
corrections. These derivative corrections will not play any significant role in our discussion
below, but we retain them for completeness.  The surface conservation equation projected along the
fluid velocity takes the form
\begin{multline}\label{surconeqproju}
 u_\nu \tilde \nabla_{\mu} T^{\mu \nu}_{(s)} - u_\nu n_\mu T^{\mu \nu}_{(b)} = 
 - u^\mu \partial_\mu \cE - \left( \cE - \cY \right) \tilde
 \nabla_{\mu} u^\mu \\
 + u^\mu n_\mu \Big( \N_\mu (\cY n^\mu) + E  \Big)
 + u_\nu \left( \tilde \nabla_\mu \Pi^{\mu \nu}_{(s)} - n_\mu \Pi^{\mu \nu}_{(b)} \right) = 0~~.
\end{multline}
Now, the divergence of the entropy current on the boundary, including the possible entropy exchange
with the bulk, must be positive semi-definite. This condition upon using the equation of motion
\eqref{surconeqproju} simplifies to
\begin{multline}\label{entcurineqgenordfl}
  \tilde \nabla_\mu J^{\mu}_{(s)\text{ent}} - n_\mu J^\mu_{(b)\text{ent}}
  = \frac{u^\mu n_{\mu}}{T} \Big( \N_\mu( \cY n^\mu) - P \Big)
  - \Pi^{\mu \nu}_{(s)} \nabla_\mu \bfrac{u_\nu}{T} \\
  + \tilde \nabla_\mu \U_{(s)\text{new}}^\mu -
  n_\mu \U^\mu_{(b)\text{new}}\geq 0 ~~,
\end{multline}
where we have made use of the Euler relation $E+P=TS$ and $\cE-\cY=T\cS$, as well as of the first
law $\df E=T\df S$ and $\df\cE=T\df\cS$. Up to first order in the bulk and ideal order at the
surface, \eqref{entcurineqgenordfl} implies that
\begin{equation}\label{leadordOrdflentcond}
  \frac{u^\nu n_\nu}{T} \Big( \N_\mu( \cY n^\mu) - P \Big) \geq 0 ~~.
\end{equation} 
The condition \eqref{leadordOrdflentcond} must hold for an arbitrary fluid configuration, including
the ones for which the term inside the bracket may have a negative sign. This implies that at
leading order $u^\mu n_\mu$ must vanish, that is
\begin{equation}
 \gamma = 0 ~~.
\end{equation}
This is the first important conclusion of this section.

As we move to higher orders, other terms in \eqref{entcurineqgenordfl} become important for this
analysis. An important noteworthy structural feature in \eqref{entcurineqgenordfl} is the fact that
the only term which contains the bulk transport coefficients is the last term
$n_\mu \U^\mu_{(b)\text{new}}$.  This immediately implies that only the transport coefficients that
arise in $\U^\mu_{(b)\text{new}}$ are the ones that may be related to the surface transport
coefficients. An interesting observation can be made, if we focus on perfect fluid bubbles, i.e.
$ \Pi^{\mu \nu}_{(b)}= \U^{\mu}_{(b)\text{new}} = \Pi^{\mu \nu}_{(s)}= \U^{\mu}_{(s)\text{new}} =
0$. For this choice, \eqref{entcurineqgenordfl} simply implies (after setting $\gamma = 0$)
\begin{equation} \label{eq:gammadiss}
  \frac{\gamma_{\text{diss}}}{T} \Big( \N_\mu( \cY n^\mu) - P \Big) \geq 0~~,
\end{equation}
which has a solution
\begin{equation}
  \gamma_{\text{diss}} = \varsigma \Big( \N_\mu( \cY n^\mu) - P \Big), \quad
  \text{with}, \quad
  \varsigma \geq 0~~.
\end{equation}
Here $\varsigma$ has the status of a dissipative transport coefficient. The respective $f$ equation
of motion away from equilibrium is then
\begin{equation}
  u^\mu n_\mu = \varsigma \Big( \N_\mu( \cY n^\mu) - P \Big).
\end{equation}
In equilibrium $u^\mu n_\mu = 0$, and consequently $\g_{diss} = 0$ implies the perfect ordinary
fluid Young-Laplace equation, $\N_\mu( \cY n^\mu) = P$. In order to see this exactly, note that the
Young-Laplace equation, defined as the normal component of the surface energy-momentum conservation
equation is $- T^{\mu\nu}_{(s)}K_{\mu\nu}= T^{\mu\nu}_{b} n_\mu n_\nu$
\cite{Armas:2015ssd}, which at ideal order implies  that
\begin{equation}\label{first.order.YL}
  \cY K - T \frac{\dow \cY}{\dow T} n_\mu \mathfrak a^\mu = P + \cO\lb\g_{diss}^2,\dow\g_{diss}\rb.
\end{equation}
Here $K_{\mu\nu}=\nabla_{(\mu}n_{\mu)}$ is the extrinsic curvature tensor of the surface,
$\nabla_\mu n^{\mu}=\mathcal{G}^{\mu\nu}K_{\mu\nu}=K$ is the mean extrinsic curvature and
$\mathfrak{a}^{\mu} = u^\nu \N_\nu u^\mu$ is the fluid acceleration. Using the fact that in
equilibrium $n^{\mu}\partial_\mu T = - T\mathfrak{a}^{\mu}n_\mu$, the equivalence between
\eqref{eq:gammadiss} and \eqref{first.order.YL} in equilibrium immediately follows.

However, under the assumption of perfect fluid bubbles, for which \eqref{eq:gammadiss} applies, one
may use the fact that, on-shell, the normal component of the vector bulk equation of motion implies
that $n^{\mu}\partial_\mu T=-T\mathfrak{a}^{\mu}n_\mu$ at the surface. Therefore, ignoring higher
order corrections, the Young-Laplace equation implies that on-shell $u^\mu n_\mu = \g_{\text{diss}} = 0$
for perfect fluids, even away from equilibrium\footnote{This equivalence holds on-shell but not
  off-shell in the sense of \cite{Haehl:2015pja}.}.  When we include first order terms in the bulk,
i.e.  $\Pi^{\mu\nu}_{(b)}=-\eta \sigma^{\mu\nu}-\zeta \Q (\mathcal{G}^{\mu\nu}+u^{\mu}u^{\nu})$,
where $\sigma_{\mu\nu}$ and $\Q$ are the fluid shear tensor and expansion respectively,
Young-Laplace equation modifies as
\begin{equation}\label{firstorder.YL}
  \cY K - T \frac{\dow \cY}{\dow T} n_\mu \mathfrak a^\mu  =
  P-\eta\sigma^{\mu\nu}n_{\mu}n_{\mu}-\zeta \Q + \cO\lb\g_{diss}^2,\dow\g_{diss}\rb ~~.
\end{equation}
On the other hand, the $f$ equation of motion in \eqref{eq:gammadiss} remains unchanged, since
$J^{\mu}_{(b)\text{new}}$ is known to be zero at first order for ordinary fluids. Hence for onshell
configurations, we can rewrite \eqref{firstorder.YL} as
\begin{equation}
  u^\mu n_\mu = \g_{\text{diss}} = -\vs\lb \eta\sigma^{\mu\nu}n_{\mu}n_{\mu} + \zeta \Q \rb + \cO\lb\g_{diss}^2,\dow\g_{\text{diss}}\rb ~~.
\end{equation}
We can see that upon including derivative corrections, $u^\mu n_\mu = \g_{\text{diss}} \neq 0$ away
from equilibrium. We would like to note that, upon including further higher order corrections,
either in the bulk or at the surface, and hence moving further away from the simplified case of
perfect fluid bubbles, we might expect \eqref{eq:gammadiss} as well as \eqref{first.order.YL}, to be
modified.

\subsubsection{Surface entropy current for 3+1 dimensional superfluids}\label{sssec:jent3p10ord}
%
Having understood the behaviour of $u^\mu n_\mu$ for neutral fluids, in this subsection we will
explore the similar entropy current analysis for superfluids with a surface.  We will demonstrate
that the first law of thermodynamics in $3+1$ dimensions modifies like \eqref{pesurthermo3p1}, and
includes contributions from the first order bulk transport coefficients $\alpha_1$ and $\alpha_2$.
We shall also exhibit, that the second law of thermodynamics puts no constraints on $n_\mu \xi^\mu$
at the interface, in contrast to the normal component of the fluid velocity $u^\mu n_\mu$ which is
set to zero at ideal order.

For superfluids, the bulk currents take the well known form
\begin{equation}\label{sublkcurf}
\begin{split}
  T^{\mu \nu}_{(b)} &= \left(  E + P \right) u^{\mu} u^{\nu} + P ~\mathcal G^{\mu \nu} + F
  \xi^{\mu}\xi^{\nu} + \Pi^{\mu \nu}_{(b)}~~, \\
  J^\mu_{(b)} &= Q u^\mu - F \xi^\mu + \U^{\mu}_{(b)}~~,\\
  J^{\mu}_{(b)\text{ent}} &= S u^\mu + \U^{\mu}_{(b)\text{ent}}~~, \quad 
  \U^\mu_{(b)\text{ent}} = - \frac{u_\nu}{T} \Pi^{\mu \nu}_{(b)}
  -\frac{\mu}{T}J^\mu_{(b)} + \U^\mu_{(b)\text{new}}~~.
\end{split}
\end{equation}
Here, the leading order coefficients follow the usual superfluid thermodynamics
$E+P=ST+\mu Q, ~dP = S dT + Q d\mu + \frac{1}{2} F d\chi$. 
In our analysis here, the first order corrections to the bulk entropy current
$\U^\mu_{(b)\text{new}}$ will play an important role.  The first order terms in
$\U^\mu_{(b)\text{new}}$ were obtained in \cite{Bhattacharya:2011tra} and the coefficients were
related to those in the partition function \eqref{partfn1stord3+1} in
\cite{Bhattacharyya:2012xi}. Setting $\alpha_3 = 0$ as in \S\ref{ssec:perfsuper},
$\U^\mu_{(b)\text{new}}$ reads (see
\cite{Bhattacharyya:2012xi,Banerjee:2016qxf})\footnote{ \label{general_entropy_correction} At first
  order, the only contribution to $\U_{(b)\text{new}}^\mu$ comes from the equilibrium sector and is
  obtained as follows \cite{Jain:2016rlz}: write down the most general scalar $\cL$ made out of
  first order data that survives in equilibrium (it can be thought of as a covariant version of the
  partition function), and perform a variation keeping the fluid variables constant
  \begin{equation}
    \frac{1}{\sqrt{\cG}}\d (\sqrt {\cG} \cL) = \half T^{\mu\nu}_\cL \d \cG_{\mu\nu} + J^\mu_\cL \d
    \cA_\mu + K_\cL \d \vf + Y_\cL \d \vf + \N_\mu \Q_\cL^\mu,
  \end{equation}
  where $\N_\mu \Q_\cL^\mu$ is a total derivative gained via differentiation by parts. Having done
  that, in an appropriate hydrodynamic frame which is a generalization of the equilibrium frame, we
  have $\U_{(b)\text{new}}^\mu = \frac{1}{T} \cL u^\mu - \Q_\cL^\mu$. In fact, $T^{\mu\nu}_\cL$,
  $J^\mu_\cL$, $K_\cL$ and $Y_\cL$ are the first order equilibrium energy-momentum tensor, charge
  current, $\phi$ variation and $f$ variation respectively in the bulk, gained via the equilibrium
  partition function.  }
\begin{multline}\label{jnew}
  \U^\mu_{(b)\text{new}} =
  \N_\nu \lb c_1 u^{[\mu}\xi^{\nu]} + c_2 \e^{\mu\nu\r\sigma} u_\r \xi_\sigma \rb \\
  + \lb \frac{f_1}{T} 2 u^{[\mu} \xi^{\nu]} \frac{1}{T} \partial_\nu T
  + \frac{u^\mu \xi_\mu - \mu}{T} \cO_\chi(f_1)^{\mu\nu} \frac{1}{T} \dow_\nu T \rb
  + \lb \frac{f_2}{T} 2 u^{[\mu} \xi^{\nu]} T\partial_\nu \nu
  + \frac{u^\mu \xi_\mu - \mu}{T} \cO_\chi(f_2)^{\mu\nu} T \dow_\nu \nu \rb \\
  + \lb \frac{g_1}{T} \e^{\mu\nu\r\sigma} \z_\nu
  \lb \dow_\r u_\sigma + u_\sigma \frac{1}{T}\dow_\r T \rb
  + \frac{u^\mu \xi_\mu - \mu}{T} \cO_\chi(g_1)^\mu_{\ \a} \e^{\a\nu\r\sigma} u_\nu
  \dow_\r u_\sigma \rb
  \\
  + \lb \frac{g_2}{T} \e^{\mu\nu\r\sigma} \z_\nu \lb \half \cF_{\r\sigma} + u_\sigma T\dow_\r \nu \rb
  + \half \frac{u^\mu \xi_\mu - \mu}{T} \cO_\chi(g_2)^\mu_{\ \a} \e^{\a\nu\r\sigma}
  u_\nu\cF_{\r\sigma} \rb
  ~~,
\end{multline}
where $\z_\mu = \xi_\mu + (u_\nu\xi^\nu) u_\mu$ and we have defined an operator
$\cO_\chi(\cdot)^{\mu\nu} = \lb 2 \z^\mu\z^\nu \frac{\dow}{\dow\chi} - P^{\mu\nu} \rb$ for
clarity. Note that the first term here is a total derivative, i.e. its divergence trivially
vanishes, and hence is not important in the bulk. However it might have some non-trivial boundary
effects. Following \eqref{persupcovcurs}, the surface currents have the general form \footnote{As in
  the case of ordinary fluids, one must consider terms proportional to $n^{\mu}$ in the surface
  energy-momentum tensor and currents. However, such terms will be ultimately set to zero by the
  entropy current analysis and hence we did not consider them here for clarity of presentation.}
\begin{equation}\label{surcursgen}
\begin{split}
  T^{\mu \nu}_{(s)} &= (\mathcal E - \cY) ~u^{\mu} u^\nu - \mathcal Y (\cG^{\mu\nu} - n^\mu n^\nu) +
  \mathcal F ~\tilde \xi^{\mu} \tilde \xi^{\nu} + 2\cU u^{(\mu} \bar n^{\nu)}
  + \Pi^{\mu \nu}_{(s)}~~,\\
  J^{\mu}_{(s)} &= \mathcal Q ~u^{\mu} - \mathcal F ~\tilde \xi^{\mu}  + \cV \bar n^\mu + \U^\mu_{(s)}~~, \\
  J^\mu_{(s)\text{ent}} &= \mathcal{S} u^{\mu} + \U^{\mu}_{(s)\text{ent}}~~, \quad
  \U^\mu_{(s)\text{ent}} = - \frac{u_\nu}{T} \Pi^{\mu \nu}_{(s)}-\frac{\mu}{T}J^\mu_{(s)} +
  \U^\mu_{(s)\text{new}}~~.
\end{split}
\end{equation}
where $\bar n^\mu = \e^{\mu\nu\r\sigma} u_\nu \xi_\r n_\sigma$.  Again, we work out the following
two scalar components of the conservation equations
\begin{align}
  u_\nu \tilde \nabla_{\mu} T^{\mu \nu}_{(s)}
  &- u_\nu n_\mu T^{\mu \nu}_{(b)}
    - u_\nu \cF^{\nu\mu} J^{(s)}_{\mu}
    =
    - u^\mu \partial_\mu \cE - \left( \cE - \cY \right) \tilde \nabla_{\mu} u^\mu
    - \half \cF u^\nu \dow_\nu \tilde\chi \nn\\
  &\qquad - T\tilde\N_\mu \lb \frac{1}{T} \cU \bar n^{\mu} \rb
    - 2T \cU u^{(\mu} \bar n^{\nu)} \N_\mu \bfrac{u_\nu}{T}
    - \cV u_\nu \cF^{\nu\mu} \bar n_\mu
    \nn\\
  &\qquad
   + u^\nu \xi_\nu \lb \tilde\N_\mu (\cF \tilde \xi^\mu) - (n^\mu \xi_\mu) F \rb
   + u^\mu n_\mu \Big( \N_\mu (\cY n^\mu - (n^\nu\xi_\nu) \cF \tilde\xi^\mu ) + E  \Big)
   \nn\\
   &\qquad
     + u_\nu \left( \tilde \nabla_\mu \Pi^{\mu \nu}_{(s)} - n_\mu \Pi^{\mu \nu}_{(b)} \right)
     - u_\nu \cF^{\nu\mu} \U_\mu^{(s)} = 0, \nn\\
   \tilde \nabla_\mu J^\mu_{(s)} -& n_\mu J^\mu_{(b)}
   = u^\mu \dow_\mu \cQ + \mathcal Q \tilde \nabla_\mu u^\mu - \tilde\nabla_\mu(\cF \tilde\xi^\mu)
   + n^\mu \xi_\mu F - u^\mu n_\mu Q + \tilde\N_\mu (\cV \bar n^\mu) \nn\\
   &\qquad + \left( \tilde \nabla_\mu \U_{(s)}^\mu - n_\mu \U^\mu_{(b)}\right) = 0~~.
\end{align}
Now, it is possible to show that the divergence of the entropy current conservation at the surface
reduces to
\begin{multline}\label{second-law-general}
  \tilde \nabla_\mu J^\mu_{(s)\text{ent}} - n_\mu J^\mu_{(b)\text{ent}} =
  - \frac{u^\mu}{T} \lb \dow_\mu \cE - T\dow_\mu \cS -\mu \dow_\mu \cQ + \half \cF \dow_\mu \tilde\chi \rb
  - \frac{1}{T}\lb \cE - \cY + T\cS + \mu \cQ\rb \tilde\N_\mu u^\mu \\
  + \frac{u^\nu \xi_\nu - \mu}{T} \lb \tilde\N_\mu (\cF \tilde \xi^\mu) - (n^\mu \xi_\mu) F \rb
  + \frac{u^\mu n_\mu}{T} \Big( \N_\mu (\cY n^\mu - (n^\nu\xi_\nu) \cF \tilde\xi^\mu )
  + E - TS - \mu Q \Big) \\
  - \Pi^{\mu\nu}_{(s)} \N_\mu\bfrac{u_\nu}{T} - \frac{1}{T}\U^\mu_{(s)} \lb T \N_\mu \nu + u^\nu
  \cF_{\nu\mu} \rb
  - 2 \cU u^{(\mu} \bar n^{\nu)} \N_\mu \bfrac{u_\nu}{T} \\
  - \frac{1}{T} \cV \bar n^\mu \lb T \N_\mu \nu + u^\nu \cF_{\nu\mu} \rb
  + \tilde \nabla_\mu \lb \U^{\mu}_{(s)\text{new}} - \frac{1}{T} (\cU - \mu \cV)
  \bar n^\mu \rb
  - n_\mu \U^{\mu}_{(b)\text{new}} \geq 0.
\end{multline}
Note that we have not imposed the thermodynamics yet, as there are first derivative terms in
$J^{\mu}_{(b)\text{new}}$ which might modify it. Restricting ourselves to first order in the bulk
and ideal order at the boundary, this equation modifies to
\begin{multline}
  - \frac{u^\mu}{T} \lb \dow_\mu \Big( \cE - \l f_1 \Big)
  - T\dow_\mu\Big( \cS - \frac{\l}{T} (f_1 - \mu f_2) \Big)
  - \mu \dow_\mu \Big( \cQ - \l f_2 \Big)
  + \half \cF \dow_\mu \tilde\chi
  \rb \\
  - \frac{1}{T}\lb \cE - \cY + T\cS + \mu \cQ\rb \tilde\N_\mu u^\mu + \frac{u^\nu \xi_\nu - \mu}{T}
  \scrE_\f
  + \frac{u^\mu n_\mu}{T} \scrE_f \\
  - 2(\cU - g_1 ) u^{(\mu} \bar n^{\nu)} \N_{\mu} \bfrac{u_{\nu}}{T}
  - \frac{\cV - g_2}{T} \bar n^\mu \lb T\dow_\mu \nu + u^\nu \cF_{\nu\mu} \rb \\
  + \tilde \nabla_\mu \lb \U^{\mu}_{(s)\text{new}} - \frac{1}{T} (\cU - \mu \cV)
  \bar n^\mu - c_1 u^{[\nu}\xi^{\mu]} n_\nu + c_2 \bar n^\mu \rb \geq 0~,
\end{multline}
where we have used the bulk Euler relation $E+P = ST + \mu Q$, and defined
\begin{align}
  \scrE_\f &= \tilde\N_\mu (\cF \tilde \xi^\mu) - (n^\mu \xi_\mu) F
  - \cO_\chi(f_1)^{\mu\nu} n_\mu \frac{1}{T} \dow_\nu T
  - \cO_\chi(f_2)^{\mu\nu} n_\mu T \dow_\nu \nu \nn\\
  &\qquad - \cO_\chi(g_1)^\mu_{\ \a} n_\mu \e^{\a\nu\r\sigma} u_\nu \dow_\r u_\sigma
    - \half \cO_\chi(g_2)^\mu_{\ \a} n_\mu \e^{\a\nu\r\sigma} u_\nu\cF_{\r\sigma}~~, \nn\\
  \scrE_f &= \N_\mu \lb \cY n^\mu - (n^\nu\xi_\nu) \cF \tilde\xi^\mu \rb - P
            - f_1 \xi^{\nu} \frac{1}{T}\partial_\nu T - f_2 \xi^{\nu} T\partial_\nu \nu \nn\\
           &\qquad   + g_1 \e^{\mu\nu\r\sigma} u_\mu \z_\nu \dow_\r u_\sigma
             + g_2 \half \e^{\mu\nu\r\sigma} u_\mu \z_\nu \cF_{\r\sigma}~~.
\end{align}
The condition of positive semi-definiteness implies the surface thermodynamics
\begin{align}
  \df \Big(\cE - \l f_1\Big)
  &= T\df \Big( \cS - \frac{\l}{T} (f_1 - \mu f_2) \Big)
    + \mu \df \Big( \cQ - \l f_2 \Big)
    - \half \cF \df \tilde\chi~~, \nn\\
  \cE - \cY &= T\cS + \mu\cQ~~,
\end{align}
and the relations
\begin{equation}
  \cU = g_1~~, \qquad
  \cV = g_2~~,
\end{equation}
which are exactly the same as the ones found using the equilibrium partition function. The second
law also implies the corrections to the entropy current\footnote{\label{EC.ambiguity} Note that, we
  can always modify the entropy currents as
  \begin{equation}
    J^\mu_{(b)\text{ent}} \ra J^\mu_{(b)\text{ent}} + \N_\nu X^{[\mu\nu]}~, \qquad
    J^\mu_{(s)\text{ent}} \ra J^\mu_{(b)\text{ent}} + n_\nu X^{[\mu\nu]}~,
  \end{equation}
  without changing the second law, hence the entropy currents always have this
  ambiguity. Interestingly, using this ambiguity we can get rid of both the $c_1$ and $c_2$
  contributions from the theory.  }
\begin{equation}
  \U^{\mu}_{(s)\text{new}} = \frac{1}{T} \lb \cU - \mu \cV - Tc_2 \rb \bar n^\mu
  + c_1 u^{[\nu}\xi^{\mu]} n_\nu~~.
\end{equation}
After imposing all of these, the second law of thermodynamics will turn into
\begin{equation}
  \frac{u^\mu \xi_\mu - \mu}{T} \scrE_\f + \frac{u^\mu n_\mu}{T} \scrE_f \geq 0~~,
\end{equation}
which will admit a general solution
\begin{equation}
  u^\mu \xi_\mu - \mu = \a \scrE_\f + (\b + \b') \scrE_f~~, \qquad
  u^\mu n_\mu = (\b - \b') \scrE_\f + \varsigma \scrE_f~~,
\end{equation}
with $\vs \geq 0$, $\a\vs \geq \b^2$ and an arbitrary $\b'$. These are the respective Josephson
equation and the equation of motion for $f$ outside equilibrium which determines $u^\mu \xi_\mu$ and
$u^\mu n_\mu$ respectively. On the other hand, the second law of thermodynamics leaves $n^\mu \xi_\mu$
undetermined. In equilibrium $u^\nu \xi_\nu = \mu$ and $u^\mu n_\mu = 0$, which implies the
equilibrium versions of the Josephson and Young-Laplace equation respectively
\begin{equation}
  \scrE_\f = 0, \qquad \scrE_f = 0~~,
\end{equation}
which are same as the ones derived using an equilibrium partition function. It is worthwhile noting that outside equilibrium, contrary to
the ordinary fluid case discussed in the previous section, the equation of motion of $f$ is not the Young-Laplace equation.

\subsubsection{Surface entropy current for 2+1 dimensional superfluids}\label{sssec:jent2p10ord}

In this subsection we will give the entropy current analysis for 2+1 dimensional superfluids with a
surface. We will only focus on the boundary computation here, for simplicity. As pointed out in the
previous section, the only way in which the bulk interacts with the boundary in the second law
\eqref{second-law-general}, is via the bulk entropy current correction $\U_{(b)\text{new}}^\mu$. In
$2+1$ dimensions, the form of $\U_{(b)\text{new}}^\mu$ is same as in the $3+1$ dimensional case in
the parity-even sector, but is quite different in the parity-sector. It is given by\footnote{We do
  not know of any reference which discusses generic first order corrections to entropy current for
  $2+1$ dimensional superfluids. However, we can use the results of \cite{Jain:2016rlz} to work out
  the generic $\U^\mu_{(b)\text{new}}$ (see footnote \ref{general_entropy_correction}).}
\begin{multline}
  \U_{(b)\text{new}}^\mu =
  \N_\nu \lb c_1 u^{[\mu} \xi^{\nu]} + c_2 \e^{\mu\nu\rho} u_\r + c_3 \e^{\mu\nu\rho} \xi_\r \rb \\
  + \lb \frac{f_1}{T} 2 u^{[\mu} \xi^{\nu]} \frac{1}{T}\partial_\nu T
  + \frac{u^\mu \xi_\mu - \mu}{T} \cO_\chi(f_1)^{\mu\nu} \frac{1}{T} \dow_\nu T \rb
  + \lb \frac{f_2}{T} 2 u^{[\mu} \xi^{\nu]} T \partial_\nu \nu
  + \frac{u^\mu \xi_\mu - \mu}{T} \cO_\chi(f_2)^{\mu\nu} T \dow_\nu \nu \rb \\
  - \lb \frac{\k_1}{T} \e^{\mu\nu\r}  \frac{1}{T} \dow_\nu (Tu_\r)
  - \frac{\k_1}{T} \frac{\dow(m_\o - m_B \nu)}{\dow \chi} \z^\mu \e^{\a\nu\r} u_\a\dow_\nu u_\r \rb \\
  - \lb \frac{\k_2}{T} \e^{\mu\nu\r} \lb \half \cF_{\nu\r} + u_\r T \dow_\nu \nu \rb
  - \frac{u^\mu \xi_\mu - \mu}{T} \frac{\dow \k_2}{\dow\chi} \z^\mu \half \e^{\a\nu\r} u_\a
  \cF_{\nu\r} \rb \\
  + \lb \frac{\k_3}{T} \e^{\mu\nu\r} \z_\nu \frac{1}{T} \dow_\r T
  + \frac{u^\mu \xi_\mu - \mu}{T} \cO_\chi(\k_3)^\mu_{\ \nu} \e^{\nu\r\sigma} u_\r \frac{1}{T} \dow_\sigma T \rb \\
  + \lb \frac{\k_4}{T} \e^{\mu\nu\r} \z_\nu T \dow_\r \nu
  + \frac{u^\mu \xi_\mu - \mu}{T} \cO_\chi(\k_4)^\mu_{\ \nu} \e^{\nu\r\sigma} u_\r T \dow_\sigma \nu \rb.
\end{multline}
On the other hand, the most generic surface currents are given as \footnote{As in the previous
  examples, we have not considered contributions proportional to $n^{\mu}$ in the surface currents
  for clarity of presentation.}
\begin{equation}
\begin{split}
  T^{\mu \nu}_{(s)} &= (\mathcal E - \cY) ~u^{\mu} u^\nu - \mathcal Y (\cG^{\mu\nu} - n^\mu n^\nu)
  + \mathcal F ~\tilde \xi^{\mu} \tilde \xi^{\nu} + 2 \cU u^{(\mu} \bar n^{\nu)}
  + \Pi^{\mu \nu}_{(s)}~~,\\
  J^{\mu}_{(s)} &= \mathcal Q ~u^{\mu} - \mathcal F ~\tilde \xi^{\mu} + \cV \bar n^\mu + \U^\mu_{(s)}~~, \\
  J^\mu_{(s)\text{ent}} &= \mathcal{S} u^{\mu} + \U_{(s)\text{ent}}^\mu~~, \quad
  \U_{(s)\text{ent}}^\mu = - \frac{u_\nu}{T} \Pi^{\mu
    \nu}_{(s)}-\frac{\mu}{T}\U^\mu_{(s)} + \U^{\mu}_{(s)\text{new}}~~,
\end{split}
\end{equation}
where $\bar n^\mu = \e^{\mu\nu\r} u_\nu n_\r$.  It should be noted that in $2+1$ dimensions,
$\bar n^\mu$ can be written in terms of $u^\mu$, $n^\mu$ and $\tilde\xi^\mu$, but we keep it in this
format in hindsight. Up to first order in the bulk and ideal order at the boundary, the second law
\eqref{second-law-general} takes the form
\begin{multline}
  \tilde \nabla_\mu J^\mu_{(s)\text{ent}} - n_\mu J^\mu_{(b)\text{ent}} = \\
  - \frac{u^\mu}{T} \lb \dow_\mu \cE - T\dow_\mu \cS -\mu \dow_\mu \cQ
  + \cF \lb \mu \dow_\mu \mu - \bar\l \dow_\mu \bar \l \rb
  - \lb \l f_1 - \bar\l \k_3 \rb \frac{1}{T} \dow_\mu T
  - \lb \l f_2 - \bar\l \k_4 \rb T \partial_\mu \nu
  \rb \\
  - \frac{1}{T}\lb \cE - \cY + T\cS + \mu \cQ\rb \tilde\N_\mu u^\mu
  + \frac{u^\mu \xi_\mu - \mu}{T} \scrE_\f
  + \frac{u^\mu n_\mu}{T} \scrE_f \\
  - 2 \lb \cU - \k_1 \rb u^{(\mu} \bar n^{\nu)} \N_\mu \bfrac{u_\nu}{T} 
  - \frac{1}{T} \lb \cV -\k_2 \rb \bar n^\mu \lb T \N_\mu \nu + u^\nu \cF_{\nu\mu} \rb \\
  + \tilde \nabla_\mu \lb \U^{\mu}_{(s)\text{new}} - \frac{1}{T} \cU_\xi \bar n^{\mu} + \nu \cV_\xi
  \bar n^\mu + c_1 u^{[\mu} \xi^{\nu]} n_\nu + c_2 \e^{\mu\nu\rho} n_\nu u_\r + c_3
  \e^{\mu\nu\rho} n_\nu \xi_\r
  \rb
  \geq 0~~,
\end{multline}
where, we have used the bulk Euler relation $E+P = TS + \mu Q$, and defined
\begin{align}
  \scrE_\f
  &=
  \tilde\N_\mu (\cF \tilde \xi^\mu) - (n^\mu \xi_\mu) F
  - \cO_\chi(f_1)^{\mu\nu} n_\mu \frac{1}{T} \dow_\nu T
  - \cO_\chi(f_2)^{\mu\nu} n_\mu T \dow_\nu \nu \nn\\
  &\qquad
    - \frac{\dow \k_1}{\dow \chi} \z^\mu n_\mu \e^{\a\nu\r} u_\a \dow_\nu u_\r
    - \frac{\dow \k_2}{\dow\chi} \z^\mu n_\mu \half \e^{\a\nu\r} u_\a \cF_{\nu\r} \nn\\
  &\qquad
    - \cO_\chi(\k_3)^\mu_{\ \nu} n_\mu \e^{\nu\r\sigma} u_\r \frac{1}{T} \dow_\sigma T
    - \cO_\chi(\k_4)^\mu_{\ \nu} n_\mu \e^{\nu\r\sigma} u_\r T \dow_\sigma \nu~~, \nn\\
  \scrE_f
  &= \N_\mu \lb \cY n^\mu - (n^\nu\xi_\nu) \cF \tilde\xi^\mu \rb
  - P
  - f_1 \xi^{\nu} \frac{1}{T} \partial_\nu T
    - f_2 \xi^{\nu} T\partial_\nu \nu
    - \k_2 \e^{\mu\nu\r} u_\mu \half \cF_{\nu\r} \nn\\
  &\qquad - \k_1 \e^{\mu\nu\r} u_\mu \dow_\nu u_\r
  - \k_3 \e^{\mu\nu\r} \z_\mu u_\nu \frac{1}{T} \dow_\r T
  - \k_4 \e^{\mu\nu\r} \z_\mu u_\nu T \dow_\r \nu~~.
\end{align}
Demanding positive definiteness, we can read out the surface thermodynamics
\begin{align}
  \df \cC
  &= - \Big( \cS - \frac{\l}{T}(f_1 - \mu f_2)
  + \frac{\bar\l}{T} (\k_3 - \mu \k_4) \Big) \df T
  - \df \Big( \cQ - \l f_2 + \bar\l\k_4 \Big) \df \mu
  - \mathcal{F}(\mu \df\mu-\bar\lambda \df\bar\lambda)~~, \nn\\
  \cE - \cY &= T\cS + \mu \cQ~~.
\end{align}
and the constraints
\begin{equation}
  \cU = \k_1~~, \qquad
  \cV = \k_2 ~~,
\end{equation}
which are exactly the same as found using the equilibrium partition function. The respective first
law of thermodynamics has been discussed in appendix \ref{app:2+1thermo}.  Furthermore, we get the
correction to the entropy current\footnote{We can remove the $c_1$, $c_2$, $c_3$ dependence of the
  system, by using the entropy current ambiguity (see footnote \ref{EC.ambiguity}). }
\begin{equation}
  \U^{\mu}_{(s)\text{new}} = \frac{1}{T} (\cU - \mu \cV) \bar n^\mu
  - c_1 u^{[\mu} \xi^{\nu]} n_\nu - c_2 \e^{\mu\nu\rho} n_\nu u_\r - c_3
  \e^{\mu\nu\rho} n_\nu\xi_\r~~.
\end{equation}
After implementing all of these constraints, the second law takes the form
\begin{equation}
  \frac{u^\mu \xi_\mu - \mu}{T} \scrE_\f + \frac{u^\mu n_\mu}{T} \scrE_f \geq 0~~,
\end{equation}
which can be solved, just like in the $3+1$ dimensional case, by
\begin{equation}
  u^\mu \xi_\mu - \mu = \a \scrE_\f + (\b + \b') \scrE_f~~, \qquad
  u^\mu n_\mu = (\b - \b') \scrE_\f + \varsigma \scrE_f~~,
\end{equation}
with $\vs \geq 0$, $\a\vs \geq \b^2$ and an arbitrary $\b'$. These are the respective Josephson
equation and equation of motion for $f$ outside equilibrium which determines $u^\mu \xi_\mu$ and
$u^\mu n_\mu$ respectively. Again, the second law of thermodynamics leaves $n^\mu \xi_\mu$
undetermined. In equilibrium, we recover the equilibrium version of the Josephson and Young-Laplace
equations respectively
\begin{equation}
  \scrE_\f = 0~~, \qquad \scrE_f = 0~~,
\end{equation}
which are same as the ones derived using an equilibrium partition function.

\subsection{Ripples on the surface}\label{ssec:linfluc}

After studying the structure of the leading order surface equations away from equilibrium, in this
section we shall study the nature of linearized fluctuations about an equilibrium configuration. For
simplicity, we shall confine ourselves to the discussion in $2+1$ dimensions.

\subsubsection*{\emph{Ordinary fluids}}

Let us first consider the case of ordinary charged fluids in flat spacetime. We choose the
coordinates $\{x^\mu\} = \{t,x,y\}$ with the flat Minkowski metric $\eta = \text{diag}\{-1,1,1\}$.
We will consider one of the simplest equilibrium configurations, where the ordinary fluid fills the
upper half spacetime $y\geq 0$, so that the equilibrium fluid variables take the form
\begin{equation}\label{ordeqconfig}
  T(t,x,y) = T_0~~, \qquad
  u^\mu(t,x,y) = ( 1, 0,0) ~~, \qquad
  f(t,x,y) = y~~. 
\end{equation}
The line $f=y=0$ is the fluid surface. For such a configuration to exist, the equilibrium pressure
must be uniform everywhere. Also, since the extrinsic curvature of the line vanishes, this uniform
equilibrium pressure must vanish as well $P(T_0)=0$ \footnote{Note that the vanishing of the
  extrinsic curvature only implies that the pressure difference at the surface vanishes. If we
  consider a scenario similar to the one in \cite{Aharony:2005bm}, where a plasma fluid is separated
  from the vacuum by a surface, then the surface pressure and hence the equilibrium pressure
  everywhere in the bulk for the configuration \eqref{ordeqconfig} must vanish. This may be achieved
  if the equation of state is of the form $P(T) = A ~T^\alpha - B$. In such system, the
  configuration \eqref{ordeqconfig} can exist as a metastable state at the phase transition
  temperature $T_0$.  }. Note that although the equilibrium pressure vanishes everywhere, the
entropy density $S(T_0) = P'(T_0)$ and the energy density $E(T_0) = T_0S(T_0) - P(T_0) = T_0S(T_0)$
remains uniformly non-zero.  Now, let us consider linearized fluctuations about this configuration
\begin{equation}\label{ordlinflucvars}
 T = T_0 + \epsilon \delta T + \mathcal O (\epsilon^2)~, \quad
 u^\mu = ( 1,\epsilon \ \delta u^x , \epsilon \ \delta u^y ) + \mathcal O (\epsilon^2)~, \quad
 f = y + \delta f + \mathcal O (\epsilon^2)~~. 
\end{equation}
Note that in \eqref{ordlinflucvars}, $u^\mu$ remains unit normalized up to the relevant order, i.e.
$u^\mu u_\mu = - 1 + \mathcal O (\epsilon^2) $.  The linearized equations in the bulk, which follow
from the conservation of the leading order energy-momentum tensor in \eqref{bulgenordcurs}, are given by,
\begin{subequations}\label{ordbullineq}
\begin{align}
  S(T_0) \lb \partial_x \delta u^x +\partial_y \delta u^y \rb
  + S'(T_0 )~ \partial_t \delta T
  &= 0   \label{ordblklineq1}~~,\\
  S(T_0) \lb \partial_t \delta u^x + \frac{1}{T_0} \partial_x \delta T \rb &= 0  \label{ordblklineq2}~~,\\
  S(T_0) \lb \partial_t \delta u^y + \frac{1}{T_0} \partial_y \delta T \rb &= 0 \label{ordblklineq3}~~.
\end{align}
\end{subequations}
As we have argued in \ref{sssec:udotnzero}, $n_\mu u^\mu$ at leading order must vanish due to the
second law, i.e. $\gamma = 0$. This serves as the additional equation required for determining the
additional variable at the surface. In the linearized approximation this equation is given by
\begin{equation}\label{ordlinequdotn}
\partial_t \delta f =-\delta u^y~~.
\end{equation} 
Using this and the leading order surface energy-momentum tensor \eqref{surgenordcurs}, the surface
conservation laws take the form
\begin{subequations}\label{ordsurlineq}
\begin{align}
  \cS(T_0) ~\partial_x \delta u^x + \cS'(T_0) ~ \partial_t \delta T  &= 0
 \label{ordsurlineq2}~~,\\
  \cS(T_0 ) \lb \partial_t \delta u^x + \frac{1}{T_0}\partial_x \delta T \rb &= 0
 \label{ordsurlineq3}~~, \\
  \cE (T_0 ) ~\partial_t^2 \delta f - \cY (T_0 )~ \partial_x^2  \delta f  &= S(T_0 )~ \delta T 
 \label{ordsurlineq1}~~.
\end{align}
\end{subequations}
Now, the procedure for solving these equations as outlined in \S\ref{sec:intro} includes first
solving the 4 surface equations \eqref{ordlinequdotn}, \eqref{ordsurlineq} for $\d u^x$, $\d u^y$,
$\d T$ and $\d f$ at the surface, and then use the solutions as a boundary condition for solving the
remaining 3 bulk equations \eqref{ordbullineq} for $\d u^x$, $\d u^y$ and $\d T$. The boundary
condition should be specified at $f=0$. In the linearized approximation that we are working in, it
suffices to impose the boundary condition at $y=0$.

In the classical computation of capillary waves \cite{safran1994statistical}, the surface entropy
$\cS$ is considered to be zero, or equivalently, a constant surface tension is assumed. In this
limit, \eqref{ordsurlineq2} and \eqref{ordsurlineq3} are automatically satisfied. This implies that
the set of allowed boundary conditions is less constrained compared to the more general case. Thus,
the bulk equations, in that case, may be solved with partially arbitrary boundary conditions, as
long as \eqref{ordsurlineq1} and \eqref{ordlinequdotn} are ensured to be satisfied.

In order to obtain the dispersion relation of capillary waves, in the absence of any external
gravitational field, the equations \eqref{ordblklineq2}, \eqref{ordblklineq3}, \eqref{ordlinequdotn}
and \eqref{ordsurlineq1} are solved by
\begin{align}\label{capwavsol}
  \delta u^x(t,x,y) &= - \delta f_0 ~\frac{k_x \omega}{\kappa}~\cos\left(k_x x + \omega t \right)
  e^{-\kappa y}~~, \quad
  \delta u^y(t,x,y) =  \delta f_0 ~\omega~\sin\left(k_x x + \omega t \right) e^{-\kappa y}~~, \nn\\
  \delta T(t,x,y) &= \delta f_0 ~\frac{T_0 \omega^2}{\kappa}~\cos\left(k_x x + \omega t \right) e^{-\kappa y} ~~, \quad
  \delta f(t,x)= \delta f_0~\cos\left(k_x x + \omega t \right)~~, \nn\\
  & \text{with}~~ \omega = \pm k_x \sqrt{\frac{\kappa \cY}{E + \kappa \cY}}~~,
\end{align}
where $\delta f_0$ is the wave amplitude, $k_x$ is the wavenumber and $\omega$ is the wave frequency
of the linearized fluctuation.  The remaining equation \eqref{ordblklineq1} provides a condition for
determining the damping factor $\kappa \geq 0$
\begin{equation}
  \k \lb \k^2 - k_x^2\rb
  + T_0 \frac{E'}{E} \lb\k - |k_x| \rb k_x^2
  + \frac{E}{\cY} \lb \k^2 - k_x^2\rb
  + T_0 \frac{E'}{E} |k_x|^3 = 0~~.
\end{equation}
For small $|k_x|$, this condition simply sets $\kappa = |k_x| $, which implies the well-known
dispersion relation of the form $\o \approx \pm k_x^{3/2} \sqrt{\cY / E}$.

However, if we take into account a non-zero surface entropy, then the boundary conditions for
solving the bulk equations must satisfy all the equations in \eqref{ordsurlineq} and
\eqref{ordlinequdotn}.  This completely determines the possible set of boundary conditions. In fact,
\eqref{ordsurlineq} and \eqref{ordlinequdotn} admits a sinusoidal solution with the following
dispersion relations
\begin{equation}\label{smlrip}
  \omega = \pm k \sqrt{\frac{\cY}{\cE}} ~~, \qquad
  \omega = \pm k \sqrt{\frac{\cS}{ \partial \cE / \partial T}}~~ ,
\end{equation}

We see that there are two sound-like modes on the surface. 
We can solve the bulk equations \eqref{ordbullineq} with the sound modes as the boundary condition at $y=0$. For instance, the full bulk solution corresponding the to first dispersion relation in 
\eqref{smlrip} takes the form
\begin{equation}\label{ordlinflucsolful}
\begin{split}
\delta u^x(t,x,y) &= \delta f_0~  \frac{k_x\o}{\kappa} \sin \left(\kappa y\right)\cos (k_x  x+\omega t) ~~,\\
\delta u^y(t,x,y) &= \delta f_0~  \o \cos \left(\kappa y\right) \sin (k_x  x+\omega t)~~, \\  
\delta T(t,x,y) &= - \delta f_0~  \frac{\omega^2 T_0}{\k} \sin \left(\kappa y\right) \cos (k_x  x+\omega t)~~, \\
\delta f(t,x) &= \delta f_0 ~ \cos (k_x  x + \omega t)~~, \\
\text{where,}  ~~\kappa &= |k_x|\sqrt{\frac{E '(T_0 )}{P'(T_0 )}\frac{\cY(T_0)}{\cE(T_0)} - 1} \quad
\text{and the dispersion is,} ~
\omega = \pm k_x \sqrt{\frac{\cY(T_0)}{\cE(T_0)}}~~.
\end{split}
\end{equation}
It can be easily checked that \eqref{ordlinflucsolful} solves both the bulk \eqref{ordbullineq} and surface equations \eqref{ordsurlineq} simultanuously. There also exists a similar sinusoidal solution corresponding to the second dispersion in \eqref{smlrip}.

Note that it should be possible to have both, the
capillary waves in \eqref{capwavsol}, as well as the tiny ripples \eqref{ordlinflucsolful} on the
surface of the same fluid. If the amplitude of the waves is large compared to the thickness of the
surface, then neglecting the surface entropy would be a legitimate approximation. Hence, in that
case, we shall have capillary waves as in \eqref{capwavsol}. On the other hand, if the amplitude of
the surface waves is small or comparable to the surface thickness, then waves like
\eqref{ordlinflucsolful} would be generated.\footnote{In this sense, the linearized solution
  \eqref{ordlinflucsolful} is similar to the third sound mode on superfluid surfaces
  \cite{khalatnikov1995introduction}.}

\subsubsection*{\emph{Superfluids}}

We now move on to surface linear fluctuations in a $2+1$ dimensional superfluid. To start with, we
will consider an equilibrium configuration similar to \eqref{ordeqconfig}, with the superfluid phase
filling half spacetime $y \geq 0$
\begin{equation}
\begin{split}
  T(t,x,y) &= T_0~~,\quad
  \mu(t,x,y) = \mu_0~~, \quad
  u^\mu(t,x,y) = ( 1, 0,0)~~, \quad
  f(t,x,y) = y~~, \\
  \f(t,x,y) &= \f_0~~, \quad
  \xi^{\mu}(t,x,y) = \lb - \mu_0,0,0\rb~~, \quad
  \chi(t,x,y) = \mu_0^2 ~~, \quad
  \l = \bar \l = 0~~.
\end{split}
\end{equation}
We consider the following linearized fluctuations about this equilibrium configuration
\begin{align}\label{linflucsf}
  T = T_0 + \e \delta T + \cO (\epsilon^2)~~,\qquad
  \mu &= \mu_0  - \e\partial_t \delta \phi + \mathcal O (\epsilon^2)~~,
   \qquad
    u^\mu = (1,\e\delta u^x , \e\delta u^y ) + \cO (\e^2)~~ , \nn\\
  \f = \f_0 + \e \d \f + \cO(\e^2)~~, &\qquad
  \xi^{\mu} = \{-\mu_0 + \e\partial_t \delta \phi, - \e\partial_x \delta \phi, - \e\partial_y \delta \phi\}
              + \mathcal O(\epsilon^2)~~, \nn\\
  f = y + \e\delta f &+ \mathcal O (\epsilon^2), \qquad
  \chi = \mu_0^2 - 2\e \mu_0 \dow_t \d\f + \cO (\e^2)~~, \nn\\
  \l = \e (\dow_y \d\f - \mu_0 \d u^y) &+ \cO (\e^2)~~, \qquad
       \bar \l = - \e (\dow_x \d\f - \mu_0 \d u^x) + \cO (\e^2)~~,
\end{align}
where we have used $u^\mu n_\mu = 0$ at ideal order. Note that the curl free condition for the
superfluid velocity and the Josephson condition $u^\mu \xi_\mu = \mu$ have already been implemented
in the ansatz \eqref{linflucsf}, up to the relevant order. The surface equations are given by
conservation of currents in \eqref{persupcovcurs2p1}, which includes parity-odd effects.

With the most general analysis of the fluctuation equations, we found that a system with a generic equation of state $\cY = \cY(T,\mu,\l,\bar\l)$, exhibits 6 independent modes at the surface. These modes can further be used as boundary conditions to solve the bulk equations. For simplicity, however, here we consider a
simplified equation of state 
\begin{equation}\label{the-ansatz}
 \cY = T\cY_1 + \bar\l \cY_2~~,
\end{equation}
where $\cY_1$, $\cY_2$ are constants. With this ansatz, the linearized surface conservation
equations following from the leading order currents \eqref{sublkcurf} and \eqref{persupcovcurs2p1},
together with the condition $\gamma=0$, yield
\begin{align}\label{linflucsfsur}
  T_0 \cY_1 \dow_x \d u^x
  + \mu_0 \cY_2 \dow_t \d u^x
  &= \mu_0 F \lb \dow_y \d\f + \mu_0 \dow_t \d f\rb~~, \nn\\
  - \cY_1 \lb \dow_x \d T + T_0 \dow_t \d u^x \rb
  - \cY_2 \lb 2\mu_0 \dow_x \d u^x - \dow_t^2 \d\f \rb
  &= 0~~, \nn\\
  \cY_1 T_0 \dow_x^2 \d f
  + 2 \mu_0 \cY_2 \dow_t\dow_x \d f
  &=
    - S \d T
    + (Q + \mu_0 F) \dow_t \d\f~~, \nn\\
  \cY_2 \dow_t \d u^x
  &= F_0 \lb \dow_y \d\f + \mu_0 \dow_t\d f \rb~~, \nn\\
  \dow_t \d f + \d u^y &= 0~~.
\end{align}
This system of equations exhibits a sinusoidal solution of the form $e^{i (\omega t - k_x x)} e^{- \ell y}$, only if $\omega$ satisfies
\begin{equation}\label{dispersions-sf}
  \mu_0 \o^2 \lb S\cY_2 \o - \cY_1 (Q + \mu_0 F) k_x \rb
  -  \ell k^2_x \cY_1 \lb 2 \mu_0 \cY_2 \o + T_0 \cY_1 k_x \rb
  = 0~~,
\end{equation}
where $k_x$ is the $x$-momentum, and  $\o$ is the frequency.
 This equation leads to 3 modes $\o \propto k$ (implying that 3
out of 6 modes were lifted due to the specific choice  of the equation of state
\footnote{For instance, in ordinary fluids, the sound modes disappear if we consider an equation of state where the pressure is linear in temperature. This is because the velocity of sound becomes infinite in this limit.} ), one of which is a
sound-like mode. We observe that none of these three modes come with a parity conjugate $k\ra -k$,
which can be seen as an imprint of parity-odd effects on the spectrum of linearized fluctuations.

One quick check which one can perform for this phenomenon is by taking $\ell = 0$. In
this limit, equation \eqref{dispersions-sf} boils down to
$\o^2 \lb S\cY_2 \o - \cY_1 (Q + \mu_0 F) k_x \rb = 0$, which implies dispersion relations
$\o^2 = 0$ and $\o = \frac{ \cY_1 (Q + \mu_0 F)}{\cY_2 S} k_x$. Though the first solution
respects parity in this limit, the second clearly breaks it, as is expected for a system with no
parity invariance.

\section{Discussion}\label{sec:disco}

In this paper, we have worked out the leading order surface energy-momentum tensor and charge
current for a finite bubble of superfluid, both in equilibrium and slightly away from it. In
equilibrium, we were able to write down the most general Euclidean effective action for the
Goldstone boson and the shape-field (in one lower dimension), coupled to arbitrary slowly varying
background fields. By appropriately varying this action, we obtained all surface currents. Away from
equilibrium, we used the second law of thermodynamics, implemented via an entropy current with a
positive semi-definite divergence. Our near equilibrium results reduce to those obtained from the
effective action, upon restricting to the stationary sector.

The ideal order surface currents contain new terms, compared to their bulk counterparts, which are
entirely determined by the first order bulk transport coefficients. This exercise has revealed new
parity-even and parity-odd terms in the ideal order surface currents. In the case of the parity-odd
terms, we have shown that they leave an imprint in the spectrum of linearized fluctuations. Such
terms are also present in the surface currents of Galilean superfluids, which we have obtained by a
null-reduction of $4+1$ dimensional null superfluids. Hence, such new effects should also be
relevant in realistic non-relativistic situations.

The parity-odd surface effects that we discussed here are relevant for theories with microscopic
parity violation\footnote{For instance, in theories with anomalies, there may be additional terms in
  the first order bulk currents that are entirely determined by the anomaly coefficient
  \cite{Banerjee:2008th,Son:2009tf,Jensen:2012kj}. Although we have refrained from discussing such
  (non-gauge invariant) terms in this paper, we hope to return to this question in a later work.},
but they may also be present as an emergent parity odd phenomenon. In order to better understand the
nature of the physical systems in which our results would play an important role, it would be
interesting to write down Kubo-like formulae for the first order parity-odd superfluid coefficients,
along the lines of \cite{Bhattacharyya:2013ida}.

The results found here are extremely relevant in the context of black holes via the AdS/CFT correspondence. In this holographic context, the space-filling configurations of the boundary fluid have a one-to-one correspondence with slowly varying black brane configurations in the bulk \cite{Bhattacharyya:2008jc}.  It is also possible to generalize such maps to the context where the plasma of the deconfined phase fills the space partially while the rest of space is occupied by the confined phase \cite{Aharony:2005bm, Marolf:2013ioa}. In the large N limit, such situations may be described by a plasma fluid separated from the vacuum by a surface in the hydrodynamic approximation. The holographic dual of such fluid configurations is a combination of black branes and the AdS-soliton  patched up in a suitable fashion to account for the fluid surface at the boundary \cite{Lahiri:2007ae, Bhardwaj:2008if, Caldarelli:2008mv, Bhattacharya:2009gm}. Similarly, the holographic dual of the space filling superfluid phase are AdS hairy black holes \cite{Bhattacharya:2011eea}. It would be extremely interesting to construct the holographic duals of the superfluid bubbles discussed in this paper, along the lines of \cite{Aharony:2005bm}. Such hairy black holes, besides being new and interesting solutions of the Einstein equations, may provide a suitable microscopic setting for a better understanding of the functional dependence of the surface tension on its arguments. 

\acknowledgments 
We would like to thank Felix Haehl for many useful discussions and collaboration during the initial
stages of this project. We would also like to thank Nabamita Banerjee, Sayantani Bhattacharyya,
Suvankar Dutta, Arthur Lipstein, Mukund Rangamani, Simon Ross, Tadashi Takayanagi and Amos Yarom for
useful correspondences and comments. We are grateful to Suvankar Dutta, Arthur Lipstein and Simon
Ross for valuable comments on the draft of this manuscript. JA would like to thank NBI for
hospitality during the course of this project. JB would like to acknowledge local hospitality at
YITP during the long term workshop on ``Quantum Information in String Theory and Many-body Systems''
where a part of this work was done. JB would also like to thank IISER Pune and TIFR for hospitality
during the final stages of the project. NK would like to acknowledge the support received from his
previous affiliation HRI, Allahabad, during the course of this project. AJ is funded by the Durham
Doctoral Scholarship offered by Durham University. JA acknowledges the current support of the ERC
Starting Grant 335146 HoloBHC. JB is supported by the STFC Consolidated Grant ST/L000407/1. Research
of NK is supported by the JSPS Grant-in-Aid for Scientific Research (A) No.16H02182.

\appendix

\section{Surface thermodynamics in 2+1 dimensions} \label{app:2+1thermo}


In \S\ref{sec:2plus1} we derived the surface Euler relation and Gibbs-Duhem relation for a superfluid
bubble, which take the form
\begin{align}\label{appendix_2+1_thermo}
  - \df \cY
  &= \Big( \cS - \frac{\l}{T}(f_1 - \mu f_2)
    + \frac{\bar\l}{T} (\k_3 - \mu \k_4) \Big) \df T
    + \Big(\cQ - \l f_2 + \bar\l\k_4 \Big) \df \mu
    + \mathcal{F}(\mu \df\mu-\bar\lambda \df\bar\lambda) ~~, \nn\\
  \mathcal{E}-\mathcal{Y}
  &= T\mathcal{S}+\mu\mathcal{Q}~~.
\end{align}
Though these relations are correct, they mix parity-even and parity-odd sectors.
In this appendix, we will write down mutually independent thermodynamics for parity-even and
parity-odd sectors on the surface, and derive the respective first law of thermodynamics.

The thermodynamic ensemble, as used in \eqref{appendix_2+1_thermo}, is described by
$\cY(T,\mu,\bar\l)$. Using the fact that $\bar\l^2 = \mu^2 - \tilde\chi$, and performing a Taylor
expansion in $\bar\l$, we can split this ensemble into a parity-even and a parity-odd sector
\begin{equation}
  \cY(T,\mu,\bar\l) = \cY_+(T,\mu,\tilde\chi) + \bar\l \cY_-(T,\mu,\tilde\chi).
\end{equation}
Note that both the functions $\cY_+$ and $\cY_-$ are purely parity-even. Keeping in mind the inflow
from the bulk, we define mutually independent Gibbs-Duhem and Euler relations in the parity-even and parity-odd
sectors, in terms of $\cY_+$ and $\cY_-$ respectively
\begin{align}\label{appendix_2+1_split.thermo}
  - \df \cY_+ &= \Big(\cS_+ - \frac{\l}{T}(f_1 - \mu f_2)\Big) \df T
  + \Big(\cQ_+ - \l f_2 \Big) \df \mu + \half \cF_+ \df \tilde\chi, \quad
  \cE_+ - \cY_+ = T\cS_+ + \mu \cQ_+, \nn\\
  - \df \cY_- &= \Big(\cS_- + \frac{1}{T} (\k_3 - \mu \k_4) \Big) \df T
                + \Big(\cQ_- + \k_4\Big) \df \mu + \half \cF_- \df \tilde\chi, \quad
                \cE_- - \cY_- = T\cS_- + \mu \cQ_-~.
\end{align}
Comparing these to the parity-mixed expressions, we can read out the parity splitting of energy,
charge, entropy and superfluid density respectively
\begin{equation}\nn
  \cE = \cE_+ + \bar\l \lb \cE_- - \frac{\mu^2}{\mu^2 - \tilde\chi}\cY_- \rb~, \qquad
  \cQ = \cQ_+ + \bar\l \lb \cQ_- - \frac{\mu}{\mu^2 - \tilde\chi} \cY_- \rb~, \quad
\end{equation}
\begin{equation}
  \cS = \cS_+ + \bar\l\cS_-, \qquad
  \cF = \cF_+ + \bar\l \lb \cF_- + \frac{1}{\mu^2 - \tilde\chi} \cY_- \rb~~.
\end{equation}
Using \eqref{appendix_2+1_split.thermo}, it is easy to derive the first law of thermodynamics for
parity-even and parity-odd sectors respectively
\begin{align}
  \df \Big(\cE_+ - \l f_1\Big)
  &= T\df \Big(\cS_+ - \frac{\l}{T}(f_1 - \mu f_2)\Big)
              + \mu\df \Big(\cQ_+ - \l f_2 \Big) - \half \cF_+ \df \tilde\chi, \nn\\
  \df \Big(\cE_- + \k_3 \Big) &= T\df \Big(\cS_- + \frac{1}{T} (\k_3 - \mu \k_4) \Big)
              + \mu\df \Big(\cQ_+ + \k_4 \Big) - \half \cF_- \df \tilde\chi~~.
\end{align}

\section{Equation of motion for the shape-field and the Young-Laplace equation } \label{app:YLeomf}


In this appendix, we rigorously show that in the stationary case, the Young-Laplace equation that
follows by projecting the surface conservation equation along $n_\mu$, is identical to the equation
of motion of $f$ which follows from the equilibrium partition function, up to all orders in
derivatives. Let us start with the most generic partition function variation parametrized as
\begin{multline}
  \d W =
  \int \df^4 x \sqrt{-\cG} \ \q(f) \lb \half T^{\mu\nu}_{(b)} \d \cG_{\mu\nu} + J^\mu_{(b)} \d \cA_\mu
  \rb
  + \int \df^4 x \sqrt{-\cG} \ \q(-f) \lb \half T^{\mu\nu}_{(e)} \d \cG_{\mu\nu} + J^\mu_{(e)} \d \cA_\mu \rb\\
  + \int \df^4 x \sqrt{-\cG}\ \tilde\d(f) \lb \half T^{\mu\nu}_{(s)} \d \cG_{\mu\nu} + J^\mu_{(s)} \d
  \cA_\mu + \frac{Y}{  \sqrt{\N_\mu f \N^\mu f}} \d f \rb,
\end{multline}
where
\begin{equation}
  \tilde\d^{(n)}(f) = (-)^{n+1} (n^\mu \dow_\mu)^{n+1} \q(f).
\end{equation}
The Young-Laplace seen as equation of motion of $f$ is just $Y = 0$. On the other hand, we know that
$W$ is a gauge invariant scalar, so it must be invariant under a diffeomorphism and gauge variation
of the constituent fields, parametrized by $\scrX = \{\vartheta^\mu,\L_\vartheta\}$
\begin{equation}
\begin{split}
  \d_\scrX  \cG_{\mu\nu} = \N_{\mu} \vartheta_\nu &+ \N_\nu \vartheta_\mu, \quad
  \d_\scrX \cA_\mu = \N_\mu (\L_\vartheta + \cA_\mu \vartheta^\mu) + \vartheta^\nu \cF_{\nu\mu}, \quad\\
  & \d_\scrX f = \vartheta^\mu \dow_\mu f = - \sqrt{\N_\mu f \N^\mu f} \vartheta^\mu n_\mu.
 \end{split}
\end{equation}
This leads to a set of identities
\begin{align}
  \q(f) &\bigg[ \N_\mu T^{\mu\nu}_{(b)} - \cF^{\nu\r} J^{(b)}_\r \bigg], \qquad
          \q(f) \bigg[ \N_\mu J^\mu_{(b)}\bigg] = 0, \nn\\
  \q(f) &\bigg[ \N_\mu T^{\mu\nu}_{(e)} - \cF^{\nu\r} J^{(e)}_\r \bigg], \qquad
  \q(f) \bigg[ \N_\mu J^\mu_{(e)}\bigg] = 0, \nn\\
  \tilde\d(f) &\bigg[ \N_\mu T^{\mu\nu}_{(s)} - \cF^{\nu\r} J^{(s)}_{\r} - n_\mu (T^{\mu\nu}_{(b)} -
                T^{\mu\nu}_{(e)}) + n^\nu Y \bigg] = 0, \quad
  \tilde\d(f) \bigg[ \N_\mu J^\mu_{(s)} - n_\mu (J^\mu_{(b)} - J^\mu_{(e)}) \bigg] = 0, \nn\\
  \tilde\d'(f) &\bigg[ T^{\mu\nu}_{(s)} n_\nu\bigg] = 0, \qquad
  \tilde\d'(f) \bigg[ J^{\mu}_{(s)} n_\mu\bigg] = 0~~.
\end{align}
From here it is clear that an analogous representation of the Young-Laplace equation $Y=0$ is the
$n_\mu$ component of the surface energy-momentum conservation equation
\begin{equation}
  n_\nu\N_\mu T^{\mu\nu}_{(s)} = n_\nu\lb \cF^{\nu\r} J^{(s)}_{\r} + n_\mu (T^{\mu\nu}_{(b)} -
  T^{\mu\nu}_{(e)}) \rb.
\end{equation}
This equation can be thought of as extremizing the partition function $W$ under a restricted
variation where we Lie drag $\cG_{\mu\nu}$ and $\cA_\mu$ along $\vartheta^\mu = \vartheta n^\mu$
keeping $f$ fixed. It might be beneficial to see this explicitly. Let the partition function have the
form
\begin{equation}
  W = \int \df^4 x \sqrt{-\cG} \bigg[ \q(f) \cL_{(b)} + \q(-f) \cL_{(e)} + \tilde\d(f) \cL_{(s)} \bigg].
\end{equation}
We use the facts that the bulk Lagrangians $\cL_{(b)}$, $\cL_{(e)}$ do not have any dependence on
the shape-field $f$, and the dependence of $\cL_{(s)}$ only comes via the reparametrization
invariant $n_\mu$. For the sake of simplicity, we further assume that $\cL_{(s)}$ is only dependent
on $n_\mu$ and not on its derivatives, which is true for our analysis in the bulk of the paper. We
can perform a $f$ variation of $W$ to get
\begin{multline}
  \d_f W = \int \df^4 x \sqrt{-\cG}  \tilde\d(f) \bigg[ \cL_{(b)} - \cL_{(e)}
  + \N_\mu \lb \cL_{(s)} n^\mu + (\d_{\mu}^{\ \nu} - n_\mu n^\nu) \frac{\dow \cL_{(s)}}{\dow n_\mu}  \rb \bigg]  \frac{\d f}{\sqrt{\N^\mu f \N_\mu f}}.
\end{multline}
This allows us to write the Young-Laplace equation directly as
\begin{equation}\label{STyleq}
  \cL_{(b)} - \cL_{(e)} + \N_\mu \lb n^\mu \cL_\dow +  (\d^\mu{}_\nu - n^\mu n_\nu) \frac{\dow\cL_\dow}{\dow n_\nu} \rb =0.
\end{equation}
On the other hand if we perform a restricted variation of $W$ along
$\vartheta^\mu = \vartheta n^\mu$ keeping $f$ fixed one can check that we get
\begin{multline}\label{YLapp_varrestr}
  \d_{\vq n^\mu}W = \int \df^4 x \sqrt{-\cG} \bigg[
  \lb \q(f) \cL_{(b)} + \q(-f) \cL_{(e)} + \tilde\d(f) \cL_{(s)} \rb
  \cG^{\mu\nu} \N_\mu (\vq n_\nu) \\
  + \q(f) \vq n^\mu \dow_\mu \cL_{(b)}
  + \q(-f) \vq n^\mu \dow_\mu \cL_{(e)}
  + \tilde\d(f) \vq n^\mu \dow_\mu \cL_{(s)}
  - \tilde\d(f) \frac{\dow\cL_{(s)}}{\dow n_\mu}  \d_{f} n_\mu
  \bigg].
\end{multline}
We have used the fact that $\cL_{(b)}$, $\cL_{(e)}$ and $\cL_{(s)}$ are scalars and transform
accordingly. Note however that $\cL_{(s)}$ also contains $f$ which we are supposed to keep
constant. To balance this we subtract the last term in \eqref{YLapp_varrestr}. We can simplify this
expression as
\begin{equation}
  \d_{\vartheta n^\mu} W = \int \df^4 x \sqrt{-\cG} \vq \tilde\d(f) \bigg[
  \cL_{(b)} - \cL_{(e)}
  + \N_\mu \lb
  \cL_{(s)} n^\mu + (\d_{\nu}^{\ \mu} - n_\nu n^\mu) \frac{\dow \cL_{(s)}}{\dow n_\mu} 
  \rb
  \bigg],
\end{equation}
which leads to the same Young-Laplace equation \eqref{STyleq}.

\section{Useful notations and formulae} \label{app:formula}

In this appendix, we recollect some useful notations and formulae used throughout this paper.

\begin{table}[t] 
  \renewcommand{\arraystretch}{1.3}
  \centering         
  \begin{tabular}{|l|c|}
    \hline
    \multicolumn{2}{|c|}{\textbf{Background Quantities}} \\
    \hline
    Metric \& cov. derivative & $\cG_{\mu\nu}$, \quad ($\cG = \det \cG_{\mu\nu}$), \quad
                                      $\N_\mu$ \\
    Gauge field \& strength & $\cA_\mu$, \quad
                                     $\cF_{\mu\nu} = \dow_\mu \cA_\nu - \dow_\nu \cA_\mu$ \\
    Levi-Civita tensor & $\e^{\mu\nu\r\sigma}$: $\e^{0123} = \frac{1}{\sqrt{-\cG}}$ (3+1), \quad
                                 $\e^{\mu\nu\r}$: $\e^{012} = \frac{1}{\sqrt{-\cG}}$ (2+1) \\
    \hline\hline
    \multicolumn{2}{|c|}{\textbf{Superfluid Quantities}} \\
    \hline
    Fluid velocity & $u^\mu$ with $u^\mu u_\mu = -1$ \\
    Temperature \& chemical pot. & $T$, \quad
                                   $\mu$, \quad ($\nu = \mu/T$) \\
    Superfluid phase \& velocity & $\f$, \quad
                                             $\xi_\mu = -\dow_\mu \f + \cA_\mu$, \quad
                                             ($\zeta_\mu = \xi_\mu + (u^\nu\xi_\nu) u_\mu$) \\ 
    Superfluid potential & $\chi = - \xi^\mu \xi_\mu$, \quad $\hat\chi = - \zeta^\mu \zeta_\mu $ \\
    \hline
    Shape-field \& normal vector & $f$, \quad
                                   $n_\mu = - \frac{1}{\sqrt{\N^\nu f \N_\nu f}} \dow_\mu f$ \\
    Surface metric \& derivative & $\tilde\cG^{\mu\nu} = \cG^{\mu\nu}{-}n^\mu n^\nu$,~
                                   $\tilde\N_\mu (\cdots) = \frac{1}{\sqrt{\N^\nu f \N_\nu f}} \N_\mu (\sqrt{\N^\nu f
                         \N_\nu f} \cdots) $ \\
    Surface fluid velocity & $ \tilde u^\mu = u^\mu - (u^\nu n_\nu) n^\mu $ \\
    Surface superfluid velocity & $ \tilde \xi^\mu = \xi^\mu - (n^\nu \xi_\nu) n^\mu $, \quad
                                  ($\tilde\z^\mu = \z^\mu - (n^\nu \z_\nu) n^\mu $) \\
    Even surface scalars & $u^\mu n_\mu$, \quad $\tilde\chi = \tilde\xi^\mu \tilde\xi_\mu$, \quad
                           $\l = n^\mu \xi_\mu$ \\
    Distribution fn. \& derivatives & $\q(f)$, \quad
    $\tilde\d(f) = - n^\mu \dow_\mu \q(f)$, \quad
    $\tilde\d'(f) = -n^\mu \dow_\mu \tilde\d(f)$\\
    \hline
    Odd surface velocity & $\bar n^\mu = \e^{\mu\nu\r\sigma} u_\nu \xi_\r n_\sigma$ (3+1), \quad
                           $\bar n^\mu = \e^{\mu\nu\r} u_\nu n_\r$ (2+1) \\
    Odd surface scalar ($2+1$) & $\bar\l = \e^{\mu\nu\r}n_\mu u_\nu \xi_\r$ \\
    \hline
    Energy-momentum tensor & $T^{\mu\nu}_{(b)}$ (bulk),~~$T^{\mu\nu}_{(s)}$ (surface),~~
                             $T^{\mu\nu}_{(e)}$ (exterior),~~$T^{\mu\nu}$ (full)\\
    Charge current & $J^{\mu}_{(b)}$ (bulk),~~$J^{\mu}_{(s)}$(surface),~~
                     $J^{\mu}_{(e)}$ (exterior),~~$J^\mu$ (full) \\
    Entropy current & $J^{\mu}_{(b)\text{ent}}$ (bulk),~~$J^{\mu}_{(s)\text{ent}}$ (sur.),~~
                      $J^{\mu}_{(e)\text{ent}}$ (ext.),~~$J^\mu_{\text{ent}}$ (full) \\
    Derivative corrections & $\Pi^{\mu\nu}_{(b/s/e)}$, \quad $\U^{\mu}_{(b/s/e)}$, \quad
                             $\U^\mu_{(b/s/e)\text{ent}}$ \\
    \hline\hline
    \multicolumn{2}{|c|}{\textbf{Equations of Motion}} \\
    \hline
    
    Energy-mom. conservation & $\N_\mu T^{\mu\nu}_{(b/e)} = \cF^{\nu\r}J_\r^{(b/e)}$,
                               $\tilde\N_{\mu} T^{\mu\nu}_{(s)} = \cF^{\nu\r}J_\r^{(s)} + n_\mu
                               (T^{\mu\nu}_{(b)} - T^{\mu\nu}_{(e)})$ \\
    Charge conservation & $\N_\mu J^{\mu}_{(b/e)} = 0$, \quad
                               $\tilde\N_{\mu}J^{\mu}_{(s)} = n_\mu(J^{\mu}_{(b)} - J^{\mu}_{(e)})$
  \\
    Local second law & $\N_\mu J^\mu_{(b/e)\text{ent}} \geq 0$, \quad
                            $\N_\mu J^\mu_{(s)\text{ent}}
                       - n_\mu(J^{\mu}_{(b)\text{ent}} - J^{\mu}_{(e)\text{ent}})  \geq 0$ \\
    \hline
    Young-Laplace equation & $n_\nu \tilde\N_{\mu} T^{\mu\nu}_{(s)} = n_\nu \cF^{\nu\r}J_\r^{(s)}
                             + n_\mu n_\nu (T^{\mu\nu}_{(b)} - T^{\mu\nu}_{(e)})$ \\
    $f$ equation of motion & $u^\mu n_\mu = \g_{diss}$ \\
    Josephson equation & $u^\mu \xi_\mu = \mu + \mu_{diss}$ \\
    \hline
  \end{tabular}
  \caption{\label{notation1} Quick reference guide for relativistic superfluid bubbles (Part I).}
  \vspace{3em}
\end{table}
    
\begin{table}[t]
  \renewcommand{\arraystretch}{1.3}
  \centering         
  \begin{tabular}[t]{|l|c|}
    \hline
    \multicolumn{2}{|c|}{\textbf{Constitutive Relations}} \\
    \hline
    \multicolumn{2}{|c|}{$T^{\mu\nu}_{(b)} = (E+P) u^\mu u^\nu + P \cG^{\mu\nu} + F \xi^\mu \xi^\nu
    + \Pi^{\mu\nu}_{(b)}$, \qquad
    $J^\mu_{(b)} = Q u^\mu - F \xi^\mu + \U^\mu_{(b)}$} \\
    \multicolumn{2}{|c|}{$T^{\mu\nu}_{(s)}=(\cE{-}\cC) u^\mu u^\nu - \cC \tilde\cG^{\mu\nu}
    + \cF \tilde\xi^\mu \tilde\xi^\nu + 2 \cU u^{(\mu} \bar n^{\nu)}
    + \Pi^{\mu\nu}_{(s)} $, ~
    $J^\mu_{(s)} = \cQ u^\mu - \cF \tilde\xi^\mu + \cV \bar n^\mu + \U^\mu_{(s)}$} \\
    \multicolumn{2}{|c|}{$J^\mu_{(b)\text{ent}} = S u^\mu + \U_{(b)\text{ent}}^\mu$, \quad
    $J^\mu_{(s)\text{ent}} = \cS u^\mu + \frac{1}{T} (\cU{-}\mu \cV) \bar n^\mu + \U_{(s)\text{ent}}^\mu$
    } \\
    \hline
    Pressure and surface tension & $P$ (bulk), \quad
                                   $\cY = - \cC$ (surface) \\
    Bulk densities & $E$ (energy), $Q$ (charge), $S$ (entropy), $F$ (superfluid den.) \\
    Surface densities & $\cE$ (energy), $\cQ$ (charge), $\cS$ (entropy), $\cF$ (superfluid den.) \\
    \hline
    First order even bulk coeff. & $f_1$, \quad $f_2$ \\
    First order odd bulk coeff. & $g_1$,~$g_2$ (3+1), \quad 
                                  $\k_1$,~$\k_2$,~$\k_3$,~$\k_4$ (2+1) \\
    Surface inflow densities (3+1) & $\cU = g_1$, $\cV = g_2$ (3+1), \quad 
                                     $\cU = \k_1$, $\cV = \k_2$ (2+1) \\
    \hline
    Euler relation & $E+P = TS + \mu Q$ (bulk),~~$\cE-\cY = T\cS + \mu\cQ$ (surface) \\ 
    Bulk first law & $\df E = T \df S + \mu \df Q - \half F \df \chi$ \\ 
    Surface first law (3+1) & $\df (\cE{-}\l f_1) = T \df (\cS{-}\frac{\l}{T} (f_1{-}\mu f_2))
                                 + \mu \df (\cQ{-}\l f_2) - \half \cF \df \tilde\chi $ \\
    Surface first law (2+1) & $\df (\cE-\l f_1+\bar\l \k_4)
                                 = T \df (\cS-\frac{\l}{T} (f_1-\mu f_2)+\frac{\bar\l}{T} (
                                 \k_3-\mu \k_4)) $ \\
                              & $~~~~~~~~~~~~~~~~~~~~+ ~\mu \df (\cQ-\l f_2+\bar\l \k_4) -
                                \half \cF \df \tilde\chi $ \\
    \hline
  \end{tabular}
  \caption{\label{notation2} Quick reference guide for relativistic superfluid bubbles (Part II).}
\end{table}

\subsection*{\emph{Relativistic Superfluids}}

We have given a list of useful definitions and relations for relativistic superfluids in tables
\ref{notation1} and \ref{notation2}. In equilibrium, the metric and gauge field can be dimensionally
reduced in a Kaluza-Klein framework as
\begin{equation}
  \cG_{\mu\nu} =
  \begin{pmatrix}
    - \E{2\sigma} & - \E{2\sigma} a_j \\
    - \E{2\sigma} a_i &- \E{2\sigma} a_i a_j + g_{ij}
  \end{pmatrix}, \qquad
  \cG^{\mu\nu} =
  \begin{pmatrix}
    - \E{-2\sigma} + a^k a_k & - a^j \\
    - a^i & g^{ij}
  \end{pmatrix},
\end{equation}
\begin{equation}
  \cA_\mu =
  \begin{pmatrix}
    A_0 \\ A_i + a_i A_0 
  \end{pmatrix}, \qquad
  \cA^\mu =   \begin{pmatrix}
    - \E{-2\sigma} A_0 - a_j A^j \\ A^i
  \end{pmatrix}.
\end{equation}
Let $W$ be a partition function for relativistic superfluids with a surface, written as a functional
of the background fields $\sigma$, $a_i$, $A_0$, $A_i$, $g_{ij}$, the Goldstone boson $\f$ and the
shape-field $f$. Varying it, we can read out the components of the energy-momentum tensor and charge
current as,
\begin{equation}\nn
  T_{00}  = -{T_0 e^{\sigma} \over \sqrt{g}}{\delta W \over \delta \sigma},\qquad
  T^i_{\ 0} =  {T_0 e^{-\sigma} \over \sqrt{g}}\lb {\delta W \over \delta a_i} - A_0
  {\delta W \over \delta A_i} \rb,\qquad  
  T^{ij}= 2 {T_0 e^{-\sigma} \over \sqrt{g}} {\delta W \over \delta g_{ij}},
\end{equation}
\begin{equation}\label{rel_variation}
  J_{0}  = -{T_0 e^{\sigma} \over \sqrt{g}}{\delta W \over \delta A_0}, \qquad
  J^i = {T_0 \E{-\sigma} \over \sqrt{g}}{\delta W \over \delta A_i}.
\end{equation}
Here $T_0$ is the inverse radius of the Euclidean time circle.

\subsection*{\emph{Null/Galilean Superfluids}}

In a similar spirit (see \cite{Banerjee:2016qxf} for details), let $W$ be a partition function for
null superfluids with a surface, written in terms of the background fields $\sigma$, $a_i$, $A_0$,
$A_i$, $B_0$, $B_i$, $g_{ij}$, the Goldstone boson $\f$ and the shape-field $f$. Upon variation, it
gives various components of the energy-momentum tensor and charge current respectively as
\begin{equation}\nn
  T_{--} = \frac{T_0}{\sqrt{g}} \frac{\d W}{\d B_0}, \qquad
  T_{0-} = - \frac{T_0}{\sqrt{g}} \lb \frac{\d W}{\d \galsigma} + B_0 \frac{\d W}{\d B_0} \rb,  
\end{equation}
\begin{equation}\nn
  T^i_{\ -} = - \frac{T_0\E{-\galsigma}}{\sqrt{g}} \lb \frac{\d W}{\d B_i} - \frac{\d W}{\d
    A_i} \rb, \quad
  T^i_{\ 0} = \frac{T_0\E{-\galsigma}}{\sqrt{g}} \lb \frac{\d W}{\d a_i} - A_0 \frac{\d W}{\d
    A_i} \rb, \quad
  T^{ij} = \frac{2T_0 \E{-\galsigma}}{\sqrt{g}} \frac{\d W}{\d g_{ij}},
\end{equation}
\begin{equation}\label{null_variational}
  J_- = -\frac{T_0}{\sqrt{g_3}} \frac{\d W}{\d A_0}, \qquad
  J^i = \frac{T_0\E{-\galsigma}}{\sqrt{g_3}} \frac{\d W}{\d A_i}.
\end{equation}
Note that the components $T_{00}$ and $J_0$ is not determined by the partition function. In fact,
these two components are ``unphysical'' as they do not enter the respective conservation laws. The
formulae \eqref{null_variational} can also be recasted directly into a null reduced Galilean language
\begin{equation}\nn
    \r = \frac{T_0}{\sqrt{g}} \frac{\d W}{\d B_0}, \qquad
  \r^i = \frac{T_0\E{-\galsigma}}{\sqrt{g}} \lb \frac{\d W}{\d B_i} - \frac{\d W}{\d
    A_i} \rb, \qquad
  t^{ij} = \frac{2 T_0\E{-\galsigma}}{\sqrt{g}} \frac{\d W}{\d g_{ij}},
\end{equation}
\begin{equation}\nn
  \e = - \frac{T_0\E{-\galsigma}}{\sqrt{g}} \frac{\d W}{\d \galsigma}, \qquad
  \e^i = \frac{T_0\E{-2\galsigma}}{\sqrt{g}} \lb - \frac{\d W}{\d a_i} + (A_0 - B_0) \frac{\d W}{\d
    A_i} + B_0 \frac{\d W}{\d B_i} \rb,
\end{equation}
\begin{equation}\label{eqbPF_varformula}
  q = \frac{T_0}{\sqrt{g}} \frac{\d W}{\d A_0}, \qquad
  q^i = \frac{T_0\E{-\galsigma}}{\sqrt{g}} \frac{\d W}{\d A_i}.
\end{equation}


\bibliographystyle{JHEP}
\bibliography{supersurface}

\end{document}